\documentclass[elsart12,eqsecnum,graphics,cite,nofootinbib]
{revtex4-2}
\usepackage{epsfig,epsf}
\usepackage{graphicx}
\usepackage{grffile}
\usepackage{amsmath,amsfonts,amssymb,amsthm,nccmath,latexsym,mathtools}
\usepackage[mathscr]{euscript}
\usepackage{color}
\usepackage{hyperref}
\newcommand{\beq}{\begin{eqnarray}}
\newcommand{\eeq}{\end{eqnarray}}
\newcommand{\be}{\begin{eqnarray*}}
\newcommand{\ee}{\end{eqnarray*}}

\def\by{\bar{y}}

\def\nn{\nonumber}

\begin{document}

\title{Single inclusive particle production at next-to-leading order in proton-nucleus collisions at forward rapidities: hybrid approach meets TMD factorization}
\author{Tolga Altinoluk$^1$, N\'estor Armesto$^2$, Alexander Kovner$^{3}$, Michael Lublinsky$^4$}
\affiliation{$^1$ Theoretical Physics Division, National Centre for Nuclear Research, Pasteura 7, Warsaw 02-093,
Poland\\
$^2$ Instituto Galego de F\'{\i}sica de Altas Enerx\'{\i}as IGFAE, Universidade de Santiago de Compostela, 15782 Santiago de Compostela, Galicia-Spain\\
$^3$ Physics Department, University of Connecticut, 2152 Hillside Road, Storrs, CT 06269, USA\\
$^4$ Department of Physics, Ben-Gurion University of the Negev, Beer-Sheva 84105, Israel\\
}
%\date{\today}

\begin{abstract}
We revisit the calculation of the cross section for forward inclusive single hadron production in $pA$ collisions within the hybrid approach. We show that the proper framework to perform this calculation beyond leading order is not the collinear factorization, as has been  assumed so far, but the TMD factorized framework. Within the TMD factorized approach we show that all the large transverse logarithms appearing in the fixed order calculation, are resummed into the evolution of the TMD PDFs and TMD FFs with factorization scale. The resulting expressions, when written in terms of TMDs evolved to the appropriate, physically well understood factorization scale, contain no additional large logarithms. The absence of any large logarithms in the resummed result should ensure positivity of the cross section and eradicate the persistent problem that have plagued the previous attempts at calculating this observable in the hybrid approach.

\end{abstract}

\maketitle

\tableofcontents

%%%%%%%%%%%%%%%%%%%%%%%%%%%%%%%%%%%%%%%%%%%%%%%%%%%%%%%%%%

\section{Introduction}
\label{sec:intro}

Understanding the high-energy limit of Quantum Chromodynamics~\cite{Kovchegov:2012mbw} is one central issue in particle physics, see the recent review~\cite{Hentschinski:2022xnd}. At high energies or, alternatively, when exploring the region of the structure of hadrons and nuclei where their parton constituents carry small fractions $x$ of the total momentum, new phenomena are predicted that differ from those found at larger values of $x$. Specifically, fixed order perturbation theory is expected to fail and demand the resummation of large logarithms $\ln{1/x}$, and scattering amplitudes become close to their unitarity limit and non-linear phenomena leading to non-linear evolution equations and saturation of parton densities, further enhanced by the size of the hadron or nucleus, are expected. The latter is usually studied in the weak coupling but non-perturbative in density framework of the Color Glass Condensate (CGC) effective theory~\cite{Kovchegov:2012mbw,Gelis:2010nm}.

Charged particle production in the forward rapidity region in hadronic collisions at high energies, studied at the Relativistic Heavy Ion Collider (RHIC) at the Brookhaven National Laboratory and the Large Hadron Collider (LHC) at CERN, is sensitive to the small-$x$ parton distribution of the backward-going hadron or nucleus (target), while the large-$x$ structure of the forward-going hadron (projectile) is probed. Measurements performed in $d$Au collisions at RHIC~\cite{BRAHMS:2004xry,STAR:2006dgg} show a strong suppression with respect to scaled $pp$ yields as expected in saturation, but they lie close to the kinematic limit which complicates their interpretation. Measurements in $pp$ and $p$Pb at the LHC~\cite{ALICE:2012mj,ATLAS:2016xpn,LHCb:2021abm,LHCb:2021vww} are compatible with the suppression expected in saturation but also, within the large present uncertainties, with predictions based on standard fixed-order collinear approaches. Therefore, both a determination of parton structure of hadrons and nuclei at small $x$ through different observables and a reduction of the uncertainties in the theoretical calculations become mandatory in order to elucidate the dynamical origin of the observed suppression.

Single inclusive particle production in the forward region in the CGC framework is usually addressed in the hybrid approach. In this approach, the projectile is described through standard collinear parton densities. The partons in the projectile then scatter on the color field of the target, considered as a dense object and described through target averages of Wilson lines, to produce the final parton which then hadronizes through standard collinear fragmentation functions. The hybrid approach was formulated at the lowest order (LO) in~\cite{Dumitru:2005gt}. A partial next-to-leading order (NLO) calculation was performed in~\cite{Altinoluk:2011qy}, while the full NLO result appeared in~\cite{Chirilli:2011km,Chirilli:2012jd}. In this result, a collinear subtraction resulted in the Dokshitzer--Gribov--Lipatov--Altarelli--Parisi (DGLAP) evolution~\cite{Gribov:1972ri,Altarelli:1977zs,Dokshitzer:1977sg} of parton densities and fragmentation functions, while a rapidity cut-off to regularize the soft divergencies was required that led to the Jalilian-Marian--Iancu--McLerran--Weigert--Leonidov--Kovner (JIMWLK) evolution equation~\cite{Balitsky:1995ub,Jalilian-Marian:1997ubg,Kovner:2000pt,Kovner:1999bj,Weigert:2000gi,Iancu:2000hn,Ferreiro:2001qy,Balitsky:2007feb,Kovner:2013ona,Kovner:2014lca,Lublinsky:2016meo}, or its mean field version (the Balitsky--Kovchegov equation, BK)~\cite{Balitsky:1998kc,Balitsky:1998ya,Kovchegov:1999ua}, for the target averages of Wilson lines.

A numerical implementation of the NLO results in~\cite{Stasto:2013cha,Stasto:2014sea,Zaslavsky:2014asa}, while successful in describing the existing experimental data at lower transverse momentum, suffered from a peculiar instability. The cross section seemed to drop very fast with increasing transverse momentum and rather quickly became negative and thus unphysical. This implementation made a specific choice of the rapidity scale, see discussions in~\cite{Kang:2014lha,Xiao:2014uba,Iancu:2016vyg}, to which the target averages of Wilson lines were evolved through the BK equation.

Later, in~\cite{Altinoluk:2014eka} a restriction on the lifetime of the fluctuations of the projectile wavefunction (Ioffe time) was introduced that provided a soft cut-off and made evident the existence of additional NLO terms (resembling those in the BK evolution equation). Such additional terms were also found in~\cite{Watanabe:2015tja} (coming from kinematic considerations that are equivalent to the Ioffe time restriction but with a different scale choice). When implemented numerically, they alleviated the negativity problem but did not solve it completely.

Further developments have been the use of a different regularization scheme for the rapidity divergence~\cite{Liu:2019iml} within Soft-Collinear Effective Theory (SCET), and attempts at threshold~\cite{Liu:2020mpy,Xiao:2018zxf,Shi:2021hwx} and Sudakov~\cite{Shi:2021hwx} resummations. While these modifications resulted in  improved fits  to data, there is no guarantee that they ensure positivity of the cross section at large transverse momentum\footnote{Additionally, the calculations have been extended to single jet~\cite{Liu:2022ijp,Wang:2022zdu}, trijet~\cite{Iancu:2018hwa} and the real part of dijet~\cite{Iancu:2020mos} production in $p$A, and dijet~\cite{Caucal:2021ent,Taels:2022tza,Caucal:2022ulg,Caucal:2023nci}, single hadron~\cite{Bergabo:2022zhe} and dihadron~\cite{Bergabo:2022tcu,Iancu:2022gpw,Bergabo:2023wed} production in electron-nucleus collisions.}. Principally, the unsettling thing about these developments, is that they are rather ad hoc and  there does not seem to be a commonality of approaches between them that would address and try to rectify a well defined physics point. One is thus left with the impression that we still do not understand the proper way of calculating single particle inclusive production at NLO within the multiple scattering low-$x$ approach.

Negative cross sections at higher order in perturbation theory appear in many calculations. They are usually taken as a signal of the failure of fixed order perturbation theory due to the existence of large logarithms. However the fact that in this calculation they are present at large transverse momentum looks rather peculiar. At low transverse momentum one in principle can expect large contributions not properly calculable in perturbation theory, however at high momentum a properly resummed perturbative calculation should be free from large logarithms and yield sensible results.

 "Properly resummed" is the key phrase in the previous sentence. It is the purpose of the present paper to show that indeed the perturbative resummations performed in the previous works all have one basic flaw in common. Specifically they assume that all large transverse logarithms can be resummed  within the collinear factorization approach, as first formulated in~\cite{Dumitru:2005gt}. We will show instead that the proper framework for resummation is the Transverse Momentum Dependent (TMD) factorization and not the collinear factorization scheme. Once this fact is realized, and the TMD resummation is performed, instabilities should disappear and the NLO calculation should yield a positive physical result.

Our approach to the problem in large measure is motivated by the early analysis of~\cite{Altinoluk:2011qy}. That paper showed that the source of large transverse momentum hadrons (with transverse momenta much larger than the saturation scale in the target) is twofold.  The large momentum of the produced hadron can originate either from the large transverse momentum exchange with the target (this contribution to hadroproduction was named elastic in~\cite{Altinoluk:2011qy} since it is driven by an elastic scattering of a valence parton)
 or from  the large transverse momentum perturbative splittings in the projectile wave function followed by soft scattering with the target (this was named inelastic in~\cite{Altinoluk:2011qy} as at NLO it is driven by  the inelastic scattering of the quark-gluon pair). The elastic production mechanism  contributes both at LO and  NLO and is affected by virtual corrections, while the inelastic contribution
is a pure real NLO effect. The elastic contribution is very sensitive to the high momentum component of the target fields. In many models (such as Golec-Biernat--W\"usthoff, GBW~\cite{Golec-Biernat:1998zce}) it is greatly suppressed.
In these models the large transverse momentum region at NLO is entirely dominated by the inelastic contribution. Even in models where the elastic mechanism does not disappear exponentially at high momentum, the two sources for $p_T\gg Q_s$ are of equal importance.  
The inelastic contribution to production, being
a real NLO squared contribution, is always positive. Obviously this contribution cannot be properly taken into account within the collinear factorization approach as the hard final state momentum is not acquired due to hard scattering. Instead, it is natural to think about this contribution in terms of TMD factorization, with the produced hadron arising from the high $k_T$ quark coming directly from the quark TMD parton distribution function (PDF).

With this intuition it becomes clear that there is another potential source to production at high $p_T$, i.e., a process where a low $k_T$ parton scatters with low momentum transfer but subsequently fragments into a high $p_T$ hadron. This would correspond to the hadron arising from a TMD fragmentation function (FF). The fragmentation was not considered in~\cite{Altinoluk:2011qy}. It was included in~\cite{Chirilli:2011km,Chirilli:2012jd} in the framework of collinear factorization, and its proper treatment in the TMD factorization framework is performed in this paper for the first time.

In addition to large transverse logarithms, the NLO calculation encounters large soft logarithms. These logarithms have to be resummed into the energy evolution of the Wilson lines -- the scattering amplitudes of the projectile partons on a dense target. Originally in~\cite{Chirilli:2011km,Chirilli:2012jd} this resummation was performed by explicitly subtracting a large logarithmic term from the cross section and attributing it to the BK-like evolution of the Wilson lines (or dipole amplitudes). At large $p_T$ this procedure over subtracted a large contribution, and, coupled with the incomplete treatment of the inelastic contribution, led to a negative cross section at large transverse momentum. 

In terms of the resummation of soft logarithms we follow the approach of~\cite{Altinoluk:2014eka}. 
We perform the calculation in a frame where most of the energy is carried by the target. The life time of fluctuations in the projectile wave function is limited by the Ioffe time constraint. As a result, in this frame the projectile wave function does not contain many soft gluons and no large soft logarithms appear explicitly in the calculation. All such logarithms have been implicitly resummed in  the dipole scattering amplitude on a highly evolved target by the choice of the convenient frame.

As for the transverse logarithms, we do not employ collinear subtraction in an attempt to resum such logarithms into   the collinear parton densities and fragmentation functions. Instead we show that such logarithms are naturally resummed into the  transverse momentum dependent (TMD) PDFs and FFs.  We show that in this TMD factorized framework after evolving TMDs to the naturally chosen resolution scale  only genuinely small NLO terms remain which do not contain large logarithms. These genuinely perturbative corrections are understood as NLO corrections to the "hard part" - i.e. production probability of a low $k_T$ parton via a hard scattering from the target.  

In the remaining part of the introduction we present a sketch of our approach.
As in~\cite{Altinoluk:2014eka}, we require the production of a parton with longitudinal momentum $k^+$, at a forward rapidity (in the projectile-going direction) and with a sizable transverse momentum. %$k_\perp$. 
This parton then fragments into a hadron of momentum $p^+=\zeta\, k^+$, $0<\zeta<1$ at a forward rapidity $\eta$. 
By definition, one has
\begin{equation}\label{eta}
\eta=\frac{1}{2}\ln \frac{p^+}{p^-}\ .
\end{equation}
Let us define the fractions $x_p$ and $x_F$ of the light-cone momentum $P^+$ of the projectile carried by the produced parton and hadron respectively, as
\begin{equation}
x_p=\frac{k^+}{P^+}   \qquad \textrm{and} \qquad x_F=\frac{p^+}{P^+}\, .
\end{equation}
Notice that  the standard Feynman-$x$ variable $x_F=x_p \zeta$. 

\begin{figure}[h]
\vskip -1.5cm
\begin{center}
 \includegraphics[width=0.7\textwidth]{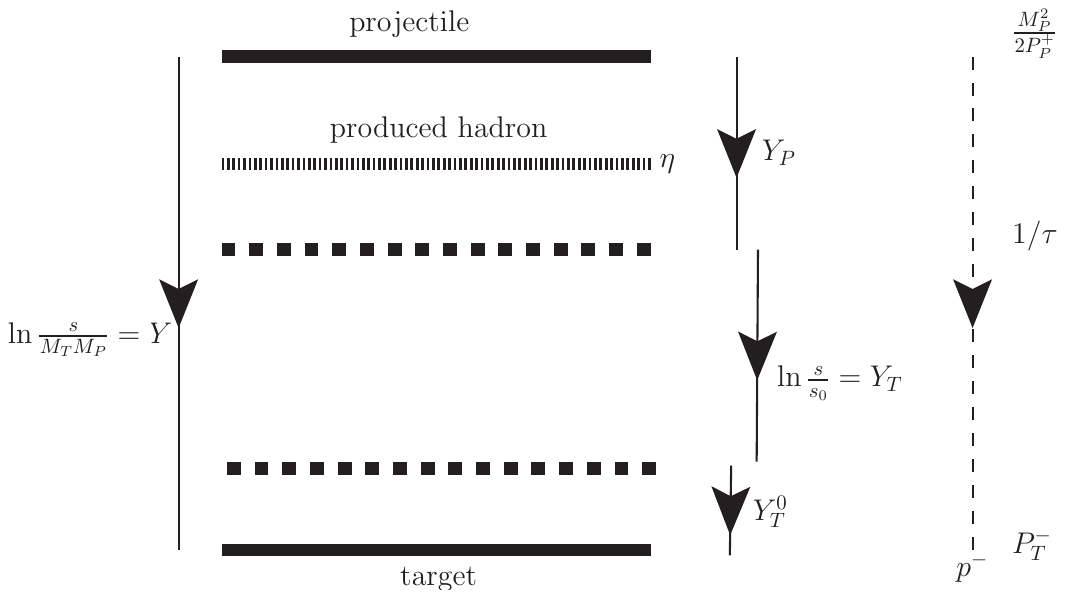}
 \end{center}
 \vskip -9.5cm
 \caption{Illustration of the different rapidity and momentum scales in our setup, taken from~\cite{Altinoluk:2014eka}. Here $P_i,M_i$, $i,j=P,T$ denote the momentum and mass of the projectile and target.}
 \label{fig1}
\end{figure}

In Fig.~\ref{fig1} we show the different rapidity and momentum scales in our setup. As discussed at length in Sections II.A and III.A in~\cite{Altinoluk:2014eka}, we work in a frame where the ensembles of Wilson lines representing the target have been evolved to rapidity $Y_T=\ln(s/s_0)< 1/\alpha_s$, for which there is no need of additional rapidity evolution. In this frame, the projectile momentum is $P^+\equiv P_P^+=\frac{M_P}{\sqrt{2}}\, e^{Y_P}$ and the target momentum $P^-_T=\frac{M_T}{\sqrt{2}}\, e^{Y_T}$,
while the total energy of the process is $s=2P^+P_T^-$.
At a given energy $s$, the choice of parameter $s_0$ is equivalent to the choice of Lorentz frame in which the calculation is performed. 
Any  variation of $s_0$ must be accompanied by the change of the dipole scattering amplitudes $s(p_T)$ (or Wilson line averages) that enter the scattering probability. As was shown in~\cite{Altinoluk:2014eka}, the change of $s(p_T)$ must be given by the BK evolution, where the evolution parameter  is proportional to $\ln s_0$. The Ioffe time $\tau$ is simply related to $s_0$:  $P^+/\tau=s_0/2$. We explicitly verify here, that in the TMD factorized framework  we pursue, such a change of the choice of $s_0$ leaves physical observables unaffected, as should be the case for a choice of frame\footnote{ This derivation is presented in Appendix~\ref{app:BK}.}.

In the calculation we assume that the highly evolved target is characterized by a large saturation scale $Q_s^2\gg\Lambda_{QCD}$, and the hadron is produced with a large transverse momentum $p_T^2\gg \Lambda^2_{QCD}$. Since both at RHIC and LHC the saturation momentum of the nucleus is at most $1-2$ GeV, while the available values of transverse momenta can be much higher, phenomenologically one is mostly interested in the situation where $p_T^2\gg Q_s^2$. Nevertheless our TMD factorized  approach is also valid for transverse momenta $Q_s^2>p_T^2\gg\Lambda_{QCD}^2$.

\begin{figure}[h]
\begin{center}
 \includegraphics[width=0.7\textwidth]{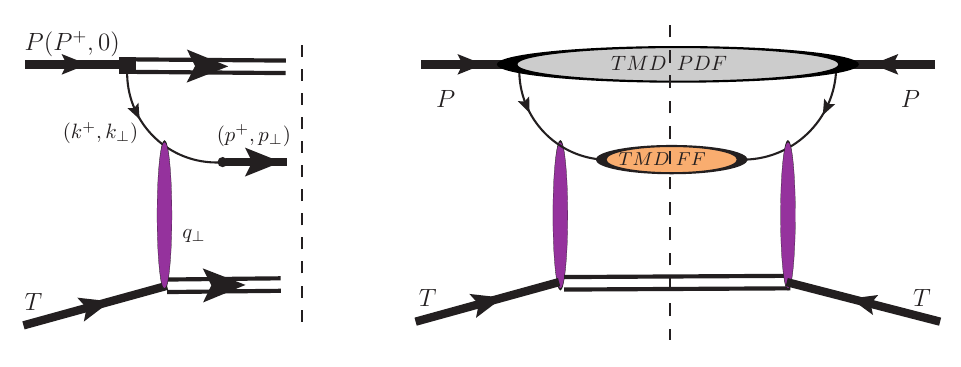}
 \end{center}
 \vskip -0.6cm
 \caption{Illustration of the parton-model expression~\eqref{parton}, with the cut (left) and the squared (right) diagrams.}
 \label{fig2}
\end{figure}

Our physical motivation, as discussed above, relies on the naive  transverse momentum dependent parton model. The particle production process in this model is illustrated in Fig.~\ref{fig2}, and it suggests the following simple expression for inclusive single particle production cross section (we assume a single quark species for the sake of the argument in this section):
\beq\label{parton}
\int\frac{d\zeta}{\zeta^2}\int d^2k_\perp d^2q_\perp\,\mathcal{T}\left(\frac{x_F}{\zeta},k_\perp;\mu^2_T\right)P(k_\perp,q_\perp)\,\mathcal{F}\left(\zeta; p_\perp,(k_\perp+q_\perp);\mu^2_F\right).
\eeq
Here $\mathcal{T}$ is the initial TMD parton distribution function (PDF), $\mathcal{F}$ the final parton TMD fragmentation function (FF), corresponding to a parton with transverse momentum $k_\perp+q_\perp$ fragmenting into a hadron with transverse momentum $p_\perp$, and $P(k_\perp,q_\perp)$  is the differential probability to produce a parton with momentum  $k_\perp+q_\perp$ from a parton with momentum $k_\perp$ due to scattering off the target\footnote{Here we have assumed for simplicity that there is no longitudinal momentum transfer during scattering and therefore the probability $P$ depends only on transverse momentum. This is true in the leading order where the scattering is eikonal. As we will see below this is not quite true in general and finite NLO terms do involve finite longitudinal momentum transfer. We ignore this in the qualitative discussion in this section for simplicity.}.
Our main goal in this paper is to show explicitly that all large logarithms at NLO can be resummed into perturbative evolution of the TMD PDF and FF with the resolution scale precisely in the form of eq.~\eqref{parton}. Thus eq.~\eqref{parton} is not just a cartoon, but is indeed the correct theoretical framework for performing this calculation.

Note, that eq.~\eqref{parton} is not quite what is usually called TMD factorization in hadronic collisions. In other words the picture of the process is not that one draws a parton from the TMD PDF of the projectile and another parton from the TMD PDF of the target, and perturbatively collides the two with subsequent fragmentation. Instead we have only the parton that arises from the TMD PDF of the projectile, which scatters (eikonally) on the nonperturbative fields of the target. The target here is not described in terms of TMD, like for example in the $k_T$- factorized approach to particle production at mid rapidity. This reflects the hybrid nature of eq.~\eqref{parton} in the spirit of the original proposal~\cite{Dumitru:2005gt}.

The correct value of the factorization scales for TMD PDF and FF, is of course, very important. Again, our naive expectation based on simple arguments below (which is born out by the explicit calculations in subsequent sections) is
\begin{eqnarray}\label{thescales11}
&& \mu^2_T={\rm max}\left\{k_\perp^2,q_\perp^2,Q_s^2,\left(\frac{p}{\zeta}\right)^2\right\}\approx{\rm max}\left\{ (k_\perp+q_\perp)^2, Q_s^2,\left(\frac{p}{\zeta}\right)^2\right\}, \\
&& \mu^2_F=[(q_\perp+k_\perp)-p_\perp/\zeta]^2\approx {\rm max}\left\{(q_\perp+k_\perp)^2, (p_\perp/\zeta)^2\right\}.\nonumber
\end{eqnarray}
Qualitatively this is understood in the following way. For the initial parton production, if $k_\perp$ is the largest scale then clearly the TMD is taken at this resolution scale, since the scale has to be at least $k_\perp^2$ in order to resolve the parton, and no larger scale is available. On the other hand if the  momentum transfer from the target, $q_T$ is larger than the momentum of produced parton, then it is this momentum transfer, and not the final momentum that provides the highest resolution and defines the factorization scale. It is also possible that both $k_\perp$ and $q_\perp$ are smaller than $Q_s$. That would mean that in most likelihood, the partner of the incoming quark (or partners, depending on the structure of the dressed quark state) is scattered with momentum of order $Q_s$, as this is the typical scale for scattering off the target. The resolution scale then is determined by $Q_s$ which resolves the scattered quark from the rest of the wave function. In case that neither $k_\perp$ nor $q_\perp$ are of the order of the final momentum, the momentum $p/\zeta$ is acquired during the fragmentation. Then the fragmentation momentum scale provides the relevant resolution, since it resolves the members of a pair of the size up to the inverse of this momentum scale, which emerge from the scattering.

On the other hand for the fragmentation we reason as follows. The fragmentation process proceeds in two steps - first,  the quark with momentum $(p^+, (k_\perp+q_\perp))$ fragments perturbatively into a quark with momentum $(p^+, p_\perp/\zeta)$ \footnote{Strictly speaking the longitudinal momentum of the quark changes a little in the fragmentation process, so that it is $(1-\xi)p^+$, but we neglect this in this discussion.}. In the second step the quark fragments nonperturbatively collinearly into a hadron with momentum $(p^+\zeta,p_\perp)$. It is natural to take the factorization scale for the fragmentation function as the component of momentum of the parton produced in the first step of fragmentation perpendicular to the direction of motion of the fragmenting parton. Assuming dominance of longitudinal (plus) momenta, $p^+\gg p_T$,  and at leading order gives then the value in~\eqref{thescales11}.

The main contribution to particle production arises from the kinematic regions $(k_\perp+q_\perp)^2\sim (p/\zeta)^2$ and $(k_\perp+q_\perp)^2\sim Q_s^2$.
Thus  both factorization scales become approximately $Q_s$ and $|p_\perp|/\zeta$ for $|p_\perp|\ll Q_s$ and $|p_\perp|\gg Q_s$, respectively. Therefore we expect the physically relevant choice to be
\beq\label{thescales1}
\mu_T^2=\mu^2_F=\mu^2={\rm max}\left\{Q_s^2,\left(p_\perp/\zeta\right)^2\right\}.
\eeq

Note that the production probability $P$ may depend on two momenta - the momentum transfer from the target as well as that of the incoming quark. Perhaps naively one would think that only the momentum transfer matters, since one can always boost to a frame where the quark has vanishing transverse momentum. However, since in actuality the quark is not alone, but is a member of a quark-gluon pair, such boost affects the gluon momentum as well. As a result in the frame where the quark has no transverse momentum, the gluon can still be in different momentum states which is the imprint of the initial momentum of the quark, and this value of gluon momentum affects the quark production probability.

This  paper is devoted to showing that~\eqref{parton} beyond being a naive expectation, is in fact the correct framework to resum all large logarithms that appear in the NLO calculation, and deriving the correct form of the production probability $P$.  The paper is structured as follows. In Section~\ref{sec:tmds} we discuss the definition of TMD distributions and their evolution with the transverse resolution scale that are used in this paper. In Section~\ref{eq:qtqth} we concentrate for simplicity of presentation of our approach  on a single channel, $q\to q\to H$. This is the process where the incoming quark produces the final state hadron via either elastic scattering, or inelastic scattering into a quark with subsequent quark fragmentation. The other channels, i.e., those either initiated by a gluon, or producing a gluon in the inelastic scattering of an initial quark,  are discussed in Section~\ref{sec:otherch}.  Section~\ref{sec:conclu} presents our conclusions. Technical details are contained in the Appendices.

\section{The transverse momentum dependent distributions}
\label{sec:tmds}

In this Section to set the stage we discuss the definition of TMD PDFs and FFs that we use in this paper. We note that several definitions of TMDs are used in the literature, differing predominantly in the way one treats the soft resolution scale, see, e.g.,~\cite{Collins:2011zzd,Boussarie:2023izj} and references therein. This is in addition to the process dependence of TMDs which requires, for different observables, the inclusion of different Wilson line factors in their definition. 
It is not our intention here to go in depth into differences and similarities, as well as subtle questions arising in these different definitions. Additionally, we are not interested in the nonperturbative aspects of TMDs, as we only require the high momentum  perturbative regime in the application in this paper. We will therefore use the simple intuitive perturbative definition, which is perfectly adequate for our purposes. It resembles that used in the parton branching method~\cite{Hautmann:2017fcj,Martinez:2023azt}.

\subsection{The no-gluon TMDs}
\label{seciia}
 In order not to crowd expressions we first provide the definitions for a single parton species (quark) and later generalize them by including gluons. Hereafter we also drop the $\perp$ subscript for transverse momenta.

The unpolarized quark TMD PDF ($f_1$ in standard notations) is defined perturbatively through its relation with the collinear PDFs as
\beq\label{eq:tmdd1pp}
x{\cal T}_q(x,k^2;k^2; \xi_0)=\frac{g^2}{(2\pi)^3}\frac{N_c}{2}\int_{\xi_0}^{1}d\xi\frac{1+(1-\xi)^2}{\xi}\, \frac{x}{1-\xi}\,f^q_{k^2}\left(\frac{x}{1-\xi}\right)\frac{1}{k^2}\ ,
\eeq
This  closely resembles the known perturbative relation between TMD and collinear PDFs in the large $k$ region~\cite{Collins:2011zzd,Boussarie:2023izj,Collins:1981uw,Collins:1984kg}. Here $\xi$ denotes the momentum fraction taken by the emitted gluon. The soft divergence in the gluon emission is regulated by the cutoff $\xi_0$, which has therefore the meaning of the resolution in the longitudinal momentum fraction.

The third argument in the TMD is the transverse resolution (factorization) scale.
eq.~\eqref{eq:tmdd1pp} is intuitively very simple. It states that partons with high transverse momentum are produced  from partons with lower transverse momentum by DGLAP splittings. The transverse resolution scale in these splittings is simply equal to the transverse momentum of the parton in question, $\mu^2=k^2$.

The factorization scale dependence of the TMD PDFs is then given by the DGLAP-like equation
\beq
\label{eq:pdftmdmu}
x{\cal T}_q(x,k^2;\mu^2; \xi_0)&&=\theta(\mu^2-k^2)\, \Bigg[x{\cal T}_q(x,k^2;k^2; \xi_0)\nonumber \\
&& \hskip 2.2cm -\frac{g^2}{(2\pi)^3}\frac{N_c}{2}\int_{k^2}^{\mu^2}\frac{\pi dl^2}{l^2}\,\int_{\xi_0}^{1}d\xi\frac{1+(1-\xi)^2} {\xi}x\,{\cal T}_q\left(x,k^2;l^2; \xi_0\right)\Bigg].
\eeq
Again, this is easy to understand. Increasing the transverse resolution means that the number of quarks at a fixed transverse momentum decreases due to DGLAP splittings into quark-gluon pairs with higher longitudinal momentum given by the resolution scale. 

As noted above, in~\eqref{eq:tmdd1pp} and \eqref{eq:pdftmdmu} we regulate the soft divergence in gluon emissions by introducing the cut-off on momentum fraction, $\xi_0$. Such regularization is standard in the TMD literature, although details of its implementations vary, see~\cite{Collins:2011zzd,Boussarie:2023izj} for discussions on the different implementations of such cut-off and the cut-off independence of physical observables.
The definition of the longitudinal cutoff we use follows our earlier approach~\cite{Altinoluk:2014eka}, where we have limited the life time of the fluctuations by the Ioffe time cutoff. The resolution $\xi_0$ then depends on the virtuality $l$ of the gluon in the splitting in~\eqref{eq:pdftmdmu} as $\xi_0(l)=l^2/(xs_0)$, where $s_0$ is the Ioffe cutoff parameter. Since in~\eqref{eq:pdftmdmu} the cutoff  appears under the integral over $l$, at the end of the day  the soft regulator effectively  depends both on the momentum $k^2$ and the transverse resolution $\mu^2$ of the TMD. Thus, it is better to label the longitudinal resolution by $s_0$ rather than $\xi_0$, although for most of our calculations we will stick to the above simplified notations.

With these definitions, the collinear quark PDF, related to the quark TMD PDF via (see~\cite{Qiu:2000hf,Berger:2002ut,Bacchetta:2013pqa,Ebert:2022cku})
\beq
\label{eq:reltmdpdf}
xf^q_{\mu^2}(x)=\int_0^{\mu^2} \pi dk^2 \,x{\cal T}_q(x,k^2;\mu^2; \xi_0),
\eeq
satisfies the DGLAP evolution equations\footnote{For $g(x)$ sufficiently smooth at $x=0$, we define $\int_0^1dx [f(x)]_+\ g(x)=\int_0^1dx f(x)[g(x)-g(0)]$.  To relate eq.~\eqref{eq:dglap} with a more standard expression of the quark-to-quark splitting function, note that $\left[\frac{1+(1-\xi)^2}{\xi}\right]_+=\frac{1+(1-\xi)^2}{[\xi]_+}+\frac{3}{2}\delta(\xi)$.}:
\beq\label{eq:dglap}
\frac{dxf^q_{\mu^2}(x)}{d\mu^2}&=&\pi x{\cal T}_q(x,\mu^2;\mu^2; \xi_0)+\int_0^{\mu^2} \pi dk^2 \,\frac{d}{d\mu^2} x{\cal T}_q(x,k^2;\mu^2; \xi_0)\nonumber\\
&=&\pi\,\frac{g^2}{(2\pi)^3}\frac{N_c}{2}\int_{\xi_0}^{1}d\xi\frac{1+(1-\xi)^2}{\xi}\, \frac{x}{1-\xi}\,f^q_{\mu^2}\left(\frac{x}{1-\xi}\right)\frac{1}{\mu^2}\nonumber \\
&-& \pi\,\frac{g^2}{(2\pi)^3}\frac{N_c}{2}\int_{\xi_0}^{1}d\xi\frac{1+(1-\xi)^2}{\xi}\, x\frac{1}{\mu^2}\int_0^{\mu^2} \pi dk^2 \,{\cal T}_q\left(x,k^2;\mu^2; \xi_0\right)\nonumber \\
&=&\ \pi\,\frac{g^2}{(2\pi)^3}\frac{N_c}{2}\int_{\xi_0}^{1}d\xi\left[\frac{1+(1-\xi)^2}{\xi}\right]_+\, \frac{x}{1-\xi}\,f^q_{\mu^2}\left(\frac{x}{1-\xi}\right)\frac{1}{\mu^2},
\eeq
where we can take the limit $\xi_0 \to 0$ without encountering any obstacles\footnote{Concerning the $\xi_0$-independence of the collinear PDFs, note that using~\eqref{eq:tmdd1pp} and \eqref{eq:pdftmdmu} we get
$$
%\label{eq:reltmdpdf2}
xf^q_{\mu^2}(x)=\frac{g^2}{(2\pi)^3}\frac{N_c}{2}\int_0^{\mu^2} \frac{\pi dk^2}{k^2}\int_{\xi_0}^{1}d\xi\left[\frac{1+(1-\xi)^2}{\xi}\right]_+   \,\frac{x}{1-\xi} \,f_{k^2}^q\left(\frac{x}{1-\xi}\right),
%\nonumber
$$
where it is evident that we can take $\xi_0\to 0$. This definition is therefore sound and independent of the choice of $\xi_0$ as long as  $\xi_0\ll 1$ as we implicitly assume.}.

The evolution equations for these TMD PDFs with respect to the transverse and longitudinal resolution scales are given in Appendix~\ref{appendixA}. They are easily obtained from the definitions given here and have in general similar structure to the evolution equations for more standard TMDs~\cite{Collins:2011zzd,Boussarie:2023izj,Hautmann:2017fcj,Collins:1981uk,Collins:1981uw,Aybat:2011zv,Becher:2010tm,Becher:2011xn,Echevarria:2011epo,Chiu:2012ir,Echevarria:2012js,Becher:2012yn,Echevarria:2014rua,Ebert:2019tvc}. We have not scrutinized more closely  the correspondence between these differently defined TMDs although we feel that such a  study is warranted in future.

Similarly, for TMD FFs ($D_1$ in standard notations)\footnote{At LO the DGLAP evolution kernels for PDFs and FFs, i.e., for space-like and time-like evolution, coincide~\cite{Collins:2011zzd}.},
\beq\label{tmdffk}
\mathcal{F}_H^q(\zeta,k^2;k^2,\xi_0)=\frac{g^2}{(2\pi)^3}\frac{N_c}{2}\int_{\xi_0}^{1}d\xi\,\frac{1+(1-\xi)^2}{\xi}\, \frac{1}{1-\xi}\,D_{H,k^2}^q\left(\frac{\zeta}{1-\xi}\right)\,\frac{1}{k^2}\ , 
\eeq
and
\beq
\mathcal{F}_H^q(x,k^2;\mu^2; \xi_0)&&=\theta(\mu^2-k^2)\, \Bigg[\mathcal{F}_H^q(x,k^2;k^2; \xi_0)\nonumber \\
&& \hskip 2.2cm -\frac{g^2}{(2\pi)^3}\frac{N_c}{2}\int_{k^2}^{\mu^2}\frac{\pi dl^2}{l^2}\int_{\xi_0}^{1}d\xi\frac{1+(1-\xi)^2} {\xi}\,\mathcal{F}_H^q\left(x,k^2;l^2; \xi_0\right)\Bigg],
\eeq
where $D_{H,\mu^2}^q(x)$ is the collinear fragmentation function giving the projection of parton $q$ onto a hadron $H$.
Analogously to eq.~\eqref{eq:reltmdpdf} we have
\beq
\label{eq:reltmdff1}
D_{H,\mu^2}^q(x)=\int_0^{\mu^2} \pi dk^2 \,\mathcal{F}_H^q(x,k^2;\mu^2; \xi_0).
\eeq

In the following, in order to identify the logarithms to be resummed we will need  perturbative expressions for the TMDs to order $g^2$.
Expanding eq.~\eqref{eq:pdftmdmu} to first order we have 
\beq
\label{eq:1storder}
x{\cal T}_q(x,k^2;\mu^2; \xi_0)&&=\theta(\mu^2-k^2)\, x{\cal T}_q(x,k^2;k^2; \xi_0)\Bigg[1
-\frac{g^2}{(2\pi)^3}\frac{N_c}{2}\int_{k^2}^{\mu^2}\frac{\pi dl^2}{l^2}\int_{\xi_0}^{1}d\xi\frac{1+(1-\xi)^2} {\xi}\Bigg],
\eeq
and similarly for $\mathcal{F}_H$.

\subsection{Including the gluons}
\label{seciib}
We now generalize the previous expressions by including the gluons and also allowing for  $n_f$ massless quark species. It is these TMDs that will actually appear in our final expressions for the particle production. The generalization is straightforward and the following expressions should be self explanatory:
\beq\label{eq:tmdd1pfull}
x{\cal T}_q(x,k^2;k^2; \xi_0)&=&\frac{g^2}{(2\pi)^3}\frac{N_c}{2}\int_{\xi_0}^{1}d\xi\frac{1+(1-\xi)^2}{\xi}\, \frac{x}{1-\xi}\,f^q_{k^2}\left(\frac{x}{1-\xi}\right)\frac{1}{k^2}\nonumber \\
&&+\frac{g^2}{(2\pi)^3}\frac{1}{2}\int_{\xi_0}^{1}d\xi \left[\xi^2+(1-\xi)^2\right]\frac{x}{1-\xi}\,f^g_{k^2}\left(\frac{x}{1-\xi}\right)\frac{1}{k^2} \ ,
\eeq
\beq
\label{eq:pdftmdmufull}
x{\cal T}_q(x,k^2;\mu^2; \xi_0)&&=\theta(\mu^2-k^2)\, \Bigg[x{\cal T}_q(x,k^2;k^2; \xi_0)\nonumber \\
&& \hskip 2.2cm -\frac{g^2}{(2\pi)^3}\frac{N_c}{2}\int_{k^2}^{\mu^2}\frac{\pi dl^2}{l^2}\,\int_{\xi_0}^{1}d\xi\frac{1+(1-\xi)^2} {\xi}x\,{\cal T}_q\left(x,k^2;l^2; \xi_0\right)\Bigg]
\eeq
 and analogously for $\bar q$, and
\beq\label{eq:tmdg1pfull}
x{\cal T}_g(x,k^2;k^2; \xi_0)&=&\frac{g^2}{(2\pi)^3}\,2N_c\int_{\xi_0}^{1}d\xi\left[\frac{1-\xi}{\xi}+\frac{\xi}{1-\xi}+\xi(1-\xi)\right]\, \frac{x}{1-\xi}\,f^g_{k^2}\left(\frac{x}{1-\xi}\right)\frac{1}{k^2}\nonumber \\
&&+\frac{g^2}{(2\pi)^3}\frac{N_c}{2}\sum_q\int_{\xi_0}^{1}d\xi \,\frac{1+\xi^2}{1-\xi}\,\frac{x}{1-\xi}\left[f^q_{k^2}\left(\frac{x}{1-\xi}\right)+f^{\bar q}_{k^2}\left(\frac{x}{1-\xi}\right)\right]\frac{1}{k^2} 
\eeq
where the sum runs over quark flavors. The evolution of the gluon TMD with the transverse resolution scale is given by
\beq
\label{eq:pdftmdgmufull}
x{\cal T}_g(x,k^2;\mu^2; \xi_0)&&=\theta(\mu^2-k^2)\, \Bigg[x{\cal T}_g(x,k^2;k^2; \xi_0)\nonumber \\
&& \hskip 2.2cm -\frac{g^2}{(2\pi)^3}\, N_c\int_{k^2}^{\mu^2}\frac{\pi dl^2}{l^2}\,\int_{\xi_0}^{1}d\xi\left[\frac{1-\xi}{\xi}+\frac{\xi}{1-\xi}+\xi(1-\xi)\right]\,x\,{\cal T}_g\left(x,k^2;l^2; \xi_0\right)\nonumber \\
&&\hskip 2.2cm -\frac{g^2}{(2\pi)^3}\, \frac{n_f}{2}\int_{k^2}^{\mu^2}\frac{\pi dl^2}{l^2}\,\int_{\xi_0}^{1}d\xi\,[\xi^2+(1-\xi)^2]\,x\,{\cal T}_g\left(x,k^2;l^2; \xi_0\right)
\Bigg].
\eeq

Using~\eqref{eq:reltmdpdf} for the quark collinear PDF and an analogous expression
\beq
\label{eq:reltmdgpdf}
xf^g_{\mu^2}(x)=\int_0^{\mu^2} \pi dk^2 \,x{\cal T}_g(x,k^2;\mu^2; \xi_0),
\eeq
for the gluon collinear PDF, 
yields the standard DGLAP evolution equations for collinear PDFs, see Appendix~\eqref{appendixA}.

The TMD FFs generalize to
\beq\label{tmdffkp}
\mathcal{F}_H^q(\zeta,k^2;k^2,\xi_0)&=&\frac{g^2}{(2\pi)^3}\frac{N_c}{2}\int_{\xi_0}^{1}d\xi\,\frac{1+(1-\xi)^2}{\xi}\, \frac{1}{1-\xi}\,D_{H,k^2}^q\left(\frac{\zeta}{1-\xi}\right)\,\frac{1}{k^2}\nonumber \\
&&+\frac{g^2}{(2\pi)^3}\frac{N_c}{2}\int_{\xi_0}^{1}d\xi\,\frac{1+\xi^2}{1-\xi}\, \frac{1}{1-\xi}\,D_{H,k^2}^g\left(\frac{\zeta}{1-\xi}\right)\,\frac{1}{k^2}\ , 
\eeq
with
\beq
\mathcal{F}_H^q(x,k^2;\mu^2; \xi_0)&&=\theta(\mu^2-k^2)\, \Bigg[\mathcal{F}_H^q(x,k^2;k^2; \xi_0)\nonumber \\
&& \hskip 2.2cm -\frac{g^2}{(2\pi)^3}\frac{N_c}{2}\int_{k^2}^{\mu^2}\frac{\pi dl^2}{l^2}\int_{\xi_0}^{1}d\xi\frac{1+(1-\xi)^2} {\xi}\,\mathcal{F}_H^q\left(x,k^2;l^2; \xi_0\right)\Bigg],
\eeq
with analogous expressions for $\bar q$. The gluon TMD FF is given by
\beq\label{eq:tmdgff1pfull}
{\cal F}_H^g(x,k^2;k^2; \xi_0)&=&\frac{g^2}{(2\pi)^3}\,2N_c\int_{\xi_0}^{1}d\xi\left[\frac{1-\xi}{\xi}+\frac{\xi}{1-\xi}+\xi(1-\xi)\right]\, \frac{1}{1-\xi}\,D_{H,k^2}^g\left(\frac{x}{1-\xi}\right)\frac{1}{k^2}\nonumber \\
&&+\frac{g^2}{(2\pi)^3}\frac{N_c}{2}\sum_q\int_{\xi_0}^{1}d\xi \,\left[\xi^2+(1-\xi)^2\right]\,\frac{1}{1-\xi}\left[D^q_{H,k^2}\left(\frac{x}{1-\xi}\right)+D^{\bar q}_{H,k^2}\left(\frac{x}{1-\xi}\right)\right]\frac{1}{k^2} \,,
\eeq
\beq
\label{eq:pdftmdgffmufull}
{\cal F}_H^g(x,k^2;\mu^2; \xi_0)&&=\theta(\mu^2-k^2)\, \Bigg[{\cal F}_H^g(x,k^2;k^2; \xi_0)\nonumber \\
&& \hskip 2.2cm -\frac{g^2}{(2\pi)^3}\, N_c\int_{k^2}^{\mu^2}\frac{\pi dl^2}{l^2}\,\int_{\xi_0}^{1}d\xi\left[\frac{1-\xi}{\xi}+\frac{\xi}{1-\xi}+\xi(1-\xi)\right]\,{\cal F}_H^g\left(x,k^2;l^2; \xi_0\right)\nonumber \\
&&\hskip 2.2cm -\frac{g^2}{(2\pi)^3}\, \frac{n_f}{2}\int_{k^2}^{\mu^2}\frac{\pi dl^2}{l^2}\,\int_{\xi_0}^{1}d\xi\,[\xi^2+(1-\xi)^2]\,{\cal F}_H^g\left(x,k^2;l^2; \xi_0\right)
\Bigg].
\eeq
With~\eqref{eq:reltmdff1} and the definition
\beq
\label{eq:reltmdgff}
D_{H,\mu^2}^g(x)=\int_0^{\mu^2} \pi dk^2 \,\mathcal{F}_H^g(x,k^2;\mu^2; \xi_0),
\eeq
the DGLAP evolution equations for the collinear FFs are recovered in full analogy to~\eqref{eq:dglapq} and \eqref{eq:dglapg} for the collinear PDFs.

Note that all the evolution equations in $\mu^2$ for TMDs are diagonal in parton species and the longitudinal momentum fraction~\cite{Collins:2011zzd,Boussarie:2023izj,Collins:1981uk,Collins:1981uw,Aybat:2011zv,Becher:2010tm,Becher:2011xn,Echevarria:2011epo,Chiu:2012ir,Echevarria:2012js,Becher:2012yn,Echevarria:2014rua,Ebert:2019tvc}. This is  a direct reflection of the fact that evolution in $\mu^2$ proceeds because of  disappearance of partons of a given species via  DGLAP splitting  into pairs of partons with higher transverse momentum.  On the other hand  the "initial condition" for this evolution, i.e., the TMD at $\mu^2=k^2$, involves  a sum over all parton species. Note also that the imposition of the Ioffe time cutoff for collinear PDFs and FFs leads to  $\xi_0\propto \mu^2$, as advocated in~\cite{Qiu:2000hf,Berger:2002ut,Bacchetta:2013pqa,Ebert:2022cku}.

\section{The $q\to q\to H$ channel}
\label{eq:qtqth}

We start our discussion of hadron production by considering a simplified setup, where the process is initiated by a valence quark and proceeds via fragmentation of the scattered quark into the hadron.
We will include all other channels in the next section. 

Just like in \cite{Altinoluk:2011qy,Chirilli:2011km,Chirilli:2012jd} we assume that the incoming "valence" partons have small transverse momentum,  $k^2< \mu^2_0\sim \Lambda^2_{QCD}$. In previous work this momentum was taken to vanish, but this of course should not be understood literally, but rather only in the sense that the typical momentum in the hadronic wave function is of order of the nonperturbative soft scale.  This same nonperturbative scale is the natural value to choose for the factorization scale (both in PDF and FF) at LO.

\subsection{The starting point - \cite{Altinoluk:2014eka}}

\begin{figure}[h]
\begin{center}
 \includegraphics[width=0.3\textwidth]{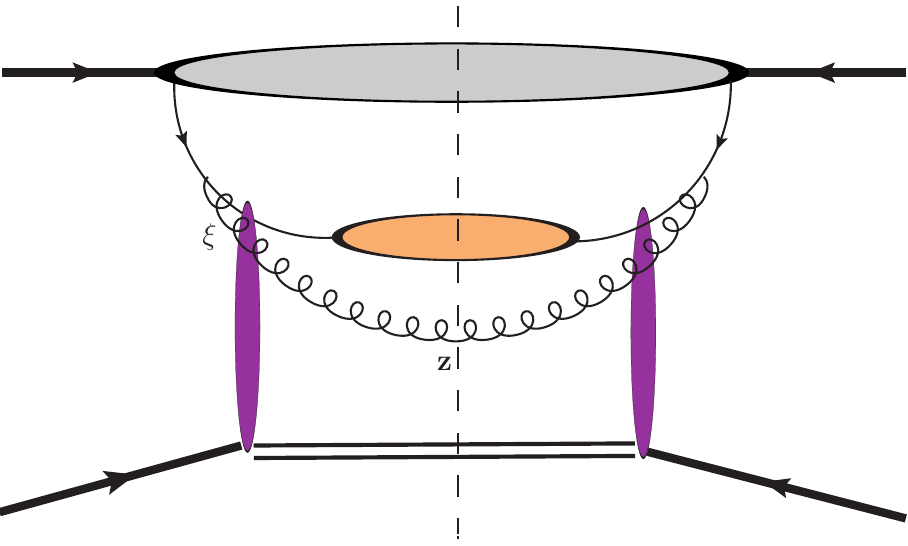} \hfill \includegraphics[width=0.3\textwidth]{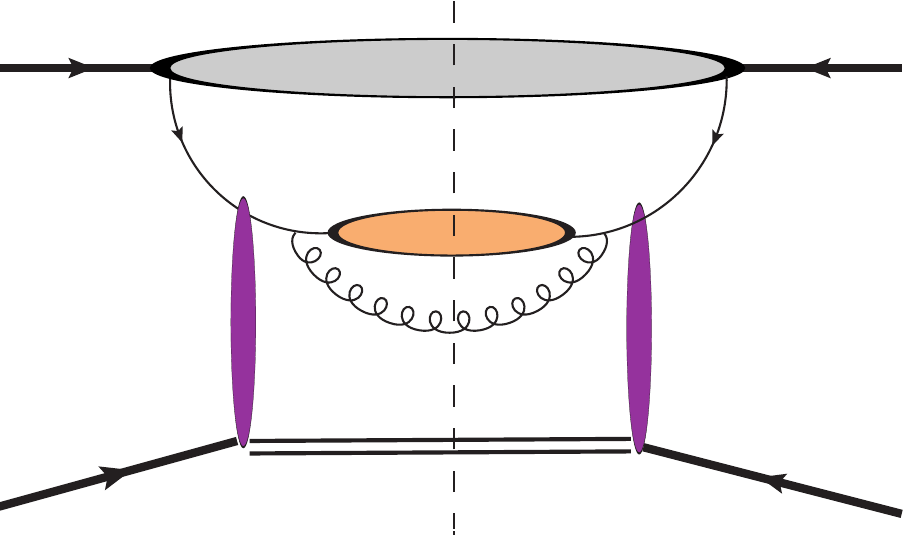}\hfill\includegraphics[width=0.3\textwidth]{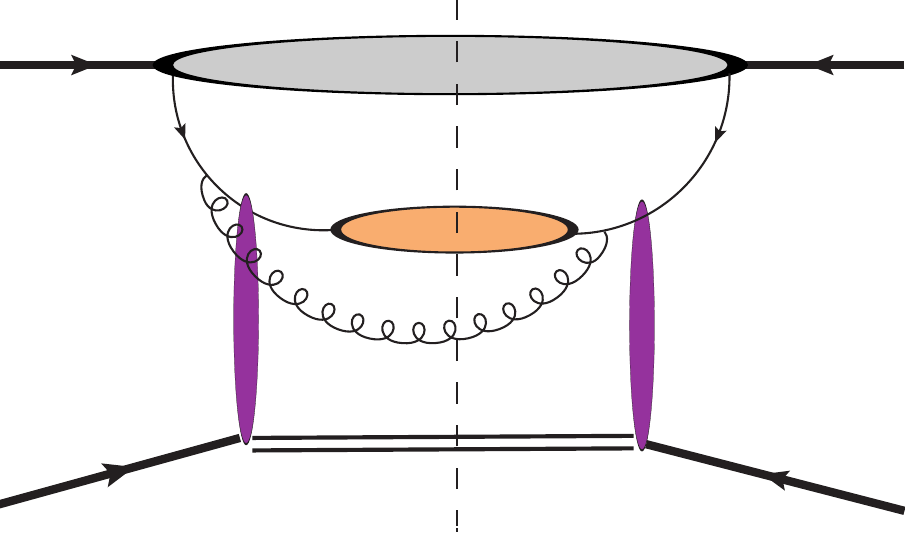}
 \end{center}
 \vskip -0.6cm
 \caption{Examples of the real diagrams included in eq.~\eqref{sigmafinal1r} with the interaction with the target nucleus after (left), before (middle) and the corresponding interference term (right).}
 \label{fig3}
\end{figure}

\begin{figure}[h]
\begin{center}
 \includegraphics[width=0.3\textwidth]{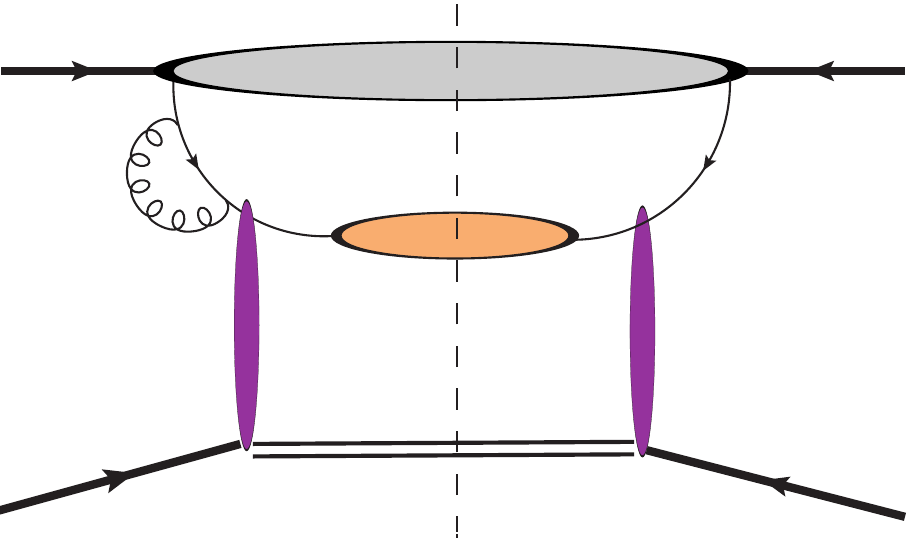} \hskip 1cm\includegraphics[width=0.3\textwidth]{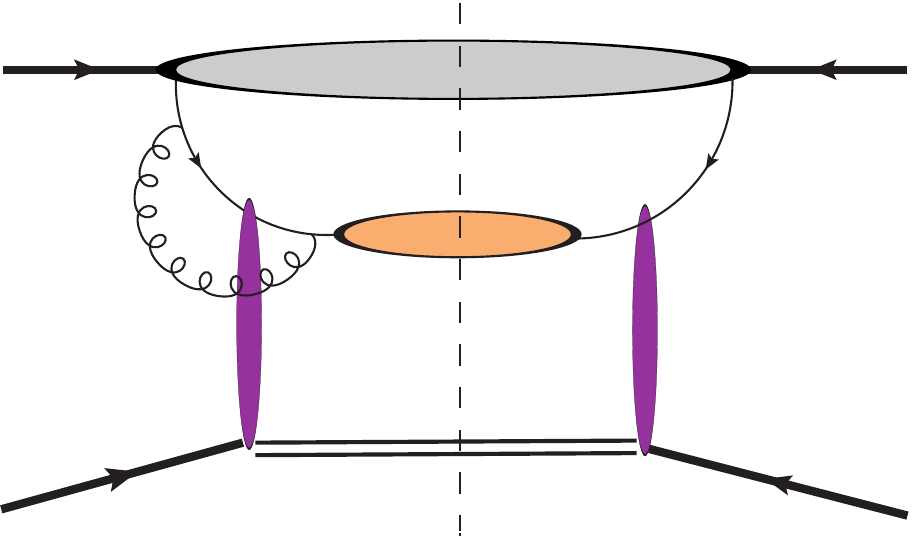}
 \end{center}
 \vskip -0.6cm
 \caption{Examples of the virtual diagrams included in eq.~\eqref{sigmafinal1r} with the interaction with the target nucleus after (left) or through (right) the loop.}
 \label{fig4}
\end{figure}

We use freely the results of \cite{Altinoluk:2014eka}.  However, in contradistinction to \cite{Altinoluk:2014eka} we do not perform explicitly any collinear subtractions.
We start from eqs. (A.11) and (A.13) there. These expressions do not contain fragmentation effects, and we will include those a little later. The production cross section is written as the sum of three terms:
\beq\label{sigmafinal}
\frac{d\bar{\sigma}^{q\to q}}{d^2k d\eta}(k, x_p)=\frac{d\bar{\sigma}_0^{q\to q}}{d^2k d\eta}(k, x_p)+\frac{d\bar{\sigma}_{1,r}^{q\to q}}{d^2k d\eta}(k, x_p)+\frac{d\bar{\sigma}_{1,v}^{q\to q}}{d^2k d\eta}(k, x_p),
\eeq
with the LO term
\beq\label{sigmafinal0}
\frac{d\bar{\sigma}_0^{q\to q}}{d^2k d\eta}(k, x_p)=\frac{1}{(2\pi)^2}\,x_p f^q_{\mu_0^2}(x_p)\int_{y,\by}e^{ik\cdot(y-\by)}s[y-\by],
\eeq
the "real" NLO term (see examples of diagrams in Fig.~\ref{fig3})
\beq\label{sigmafinal1r}
&&\frac{d\bar{\sigma}_{1,r}^{q\to q}}{d^2k d\eta}(k, x_p)=\frac{g^2}{(2\pi)^3}
\int_0^{1} d\xi  \int_{y,\bar y,z} e^{ik\cdot(y-\bar y)}\frac{1+(1-\xi)^2}{\xi} \\
 &&\hskip 0.3cm\times \Bigg[\frac{x_p}{1-\xi}f^q_{\mu_0^2}\left(\frac{x_p} {1-\xi}\right)\;A_\xi^i(y-z)A_\xi^i(\bar y-z)\Bigg\{  C_F  \, \Big[  s[y-\by] +
 s\big[(1-\xi)(y-\bar y) \big] \Big] \nonumber\\
&& \hskip 0.9cm-\frac{N_c}{2}\Big[s\big[ (1-\xi)(y-z)\big] \, s[\by-z] + s\big[ (1-\xi)(\by-z)\big] \, s[y-z] \Big]\Bigg\}\Bigg]\nonumber
\eeq
and the "virtual" NLO term (see examples of diagrams in Fig.~\ref{fig4})
\beq\label{sigmafinal1v}
&&\frac{d\bar{\sigma}_{1,v}^{q\to q}}{d^2k d\eta}(k, x_p)=-\frac{g^2}{(2\pi)^3}
\int_0^1 d\xi  \int_{y,\bar y,z} e^{ik\cdot (y-\bar y)}\frac{1+(1-\xi)^2}{\xi} \\
&&\hskip 0.3cm\times \Bigg[x_p\: f^{q}_{\mu_0^2}(x_p) \; \Bigg\{  C_F \Big[A_{\xi,x_p}^i(y-z)A_{\xi,x_p}^i(y-z) + A_{\xi,x_p}^i(\bar y-z)A_{\xi,x_p}^i(\bar y-z) \Big]s[y-\by]
\nonumber\\
&&\hskip 2.8cm-\;\bigg\{
A_{\xi,x_p}^i(y-z)A_{\xi,x_p}^i(y-z) \left[ \frac{N_c}{2} \, s\Big[ y-z \Big] \, s\Big[ (z-\bar y)+\xi(y-z)\Big]\right]\nonumber \\
&&\hskip 3.4cm + A_{\xi,x_p}^i(\bar y-z)A_{\xi,x_p}^i(\bar y-z)  \frac{N_c}{2} \, s\Big[ z-\bar y \Big] \, s\Big[ (y-z)-\xi(\bar y-z)\Big]  \bigg\}\Bigg\}\Bigg].
\nonumber\eeq
In the above, as always, $s(x)\equiv \frac{1}{N_c}<{\rm Tr}[U^\dagger (0)U(x)>_T$ is the target averaged scattering amplitude of the projectile fundamental dipole. The superscript indices $i$ refer to transverse components. As discussed in the Introduction, this dipole amplitude has to be averaged over the target color fields evolved with rapidity by the amount $Y_T=\ln s/s_0$ via the JIMWLK or BK equation. Here  $s$ is the total energy of the process, while $s_0$ is a high, but not very high hadronic scale chosen so that at energy $s_0$ one can already use eikonal approximation for scattering, but the energy evolution from the rest frame to energy $s_0$ does not yield a significant change in the dipole amplitude. As explained in~\cite{Altinoluk:2014eka}, the scale $s_0$ determines the Ioffe time cutoff on the life time of the fluctuations in the projectile wave function,
\beq
\frac{(1-\xi)\xi x_p}{q^2}>s_0^{-1}\ ,
\eeq
resulting in the longitudinal resolution of the TMD discussed in the previous section.
Here $\xi$ and $q$ are the longitudinal momentum fraction and transverse momentum of the emitted gluon\footnote{Note that if we change $s_0 \to s$ (the total squared center-of-mass energy of the collision), this restriction coincides with the kinematic constraint used in~\cite{Watanabe:2015tja}.}, respectively.

The Ioffe time constraint enters the above expressions explicitly via the modified Weizs\"acker-Williams (WW) fields
\beq
\label{eq:ww}
A_{\xi,x_p}^i(y-z)&\equiv &-i\int_{l^2<\xi(1-\xi) x_ps_0}
\frac{d^2l}{(2\pi)^2} \frac{l^i}{l^2} e^{-il\cdot (y-z)}\\
&=&-\frac{1}{2\pi} \; \frac{(y-z)^i}{(y-z)^2}\; \left[1- \textrm{J}_0\left(|y-z| \sqrt{\xi(1-\xi)x_ps_0}\right)\right]\, ,\nonumber\\
A_{\xi}^i(y-z)&\equiv &A_{\xi,x_p/(1-\xi)}^i(y-z)\, .\label{eq:ww2}
\eeq

In eq.~\eqref{sigmafinal1r} we have put the upper limit of the integration over $\xi$ to unity, relying on the fact that the PDF vanishes if the fraction of the longitudinal momentum in its argument is greater than one\footnote{We have also dropped the prefactor $(1-\xi)^2$ in the second term in~\eqref{sigmafinal1r} compared to eq. (A.11) in~\cite{Altinoluk:2014eka}. This factor in~\cite{Altinoluk:2014eka} is incorrect, and arose due to an overcourageus treatment of a divergent integral.  Our original expressions (before rescaling in eq. (A.10) in~\cite{Altinoluk:2014eka}) reduce to the expression given here directly if we assume translational invariance of the dipoles. We thank Yair Mulian for a confirmation on this point.}.
Hereafter, $\int_r=\int d^2r$, $\int_{k}=\int d^2k$.

In the above expressions we assumed the large $N_c$ factorization of the dipole amplitudes, and translational invariance of the target ensemble, approximations which are invariably employed in numerical implementations. We have also neglected the $N_c$ suppressed terms. This last approximation can only be relaxed if one also relaxes the factorization hypothesis, since some of the nonfactorisabe terms are of the same order as the explicit $1/N_c$ suppressed terms in the expressions of \cite{Altinoluk:2014eka}.

Fragmentation can be accounted for in a straightforward way by modifying the expression for cross section to
\beq\label{sigmafinalfr}
\frac{d{\sigma}^{q\to q\to H}}{d^2p d\eta}=\int_{x_F}^1 \frac{d\zeta}{\zeta^2}\,D_{H,\mu_0^2}^q(\zeta)\ \frac{d\bar{\sigma}^{q\to q}}{d^2k d\eta}\left(\frac{p}{\zeta},\frac{x_F}{\zeta}\right).
\eeq

\subsection{Transforming into momentum space}

We now transform the above expressions into momentum space.
We define, as in~\cite{Shi:2021hwx}, the Fourier transform of the dipole 
\beq
\label{eq:dip_FT}
s(k)=\int_r \frac{1}{(2\pi)^2} \ e^{-ik\cdot r} s(r)\Longrightarrow s(r)=\int_{l}e^{il \cdot r} s(l).
\eeq
Note that with this definition $s(r=0)=1= \int_{l} s(l)$.

The LO expression reads then
\beq\label{sigmafinal0ft}
\frac{d{\sigma}_0^{q\to q\to H}}{d^2p d\eta}=S_\perp \int_{x_F}^1 \frac{d\zeta}{\zeta^2}\,D_{H,\mu_0^2}^q(\zeta)\ \frac{x_F}{\zeta} f_{\mu_0^2}^q\left(\frac{x_F}{\zeta}\right) s\left(\frac{p}{\zeta}\right),
\eeq
with $S_\perp$ the overlap area of projectile and target that comes from the integration over the impact parameter $b=\frac{y+\bar y}{2}$ of the dipole under our translational invariance assumption.

The real part of the NLO piece reads (with $C_F\to N_c/2$ at large $N_c$)
\beq\label{sigmafinal1rftp}
\frac{d{\sigma}_{1,r}^{q\to q\to H}}{d^2p d\eta}&=&S_\perp\frac{g^2}{(2\pi)^3}\frac{N_c}{2}
\int_{x_F}^1 \frac{d\zeta}{\zeta^2}\,D_{H,\mu_0^2}^q(\zeta)\ \int_{k,q} \int_{\zeta k^2/(x_Fs_0)}^{1} d\xi  \,\frac{1+(1-\xi)^2}{\xi} \ \frac{x_F}{(1-\xi) \zeta} f_{\mu_0^2}^q\left(\frac{x_F}{(1-\xi) \zeta}\right)\noindent \\
&& \times \ s(k)s(q)\,\frac{[k-(1-\xi)q]^2}{\left(\frac{p}{\zeta}-k\right)^2\left(\frac{p}{\zeta}-(1-\xi)q\right)^2}\,.\nonumber
\eeq

The virtual term~\eqref{sigmafinal1v} in the Fourier space reads\footnote{Note that in the regime $k^2/(x_F s_0)\ll 1$ where our approach is valid, $A_{\xi}^i(y-z)\simeq A_{\xi,x_p}^i(y-z)$, see~\eqref{eq:ww} and \eqref{eq:ww2}. Note also that we integrate $k$ for $k^2>\mu_0^2$ because partons with smaller $k^2$ are already included in the LO term, as discussed above.}
\begin{eqnarray}\
\label{eq:sigmafinal1vft}
&&\frac{d{\sigma}_{1,v}^{q\to q\to H}}{d^2p d\eta} =-2\frac{g^2}{(2\pi)^3} S_\perp \frac{N_c}{2}\int_{x_F}^1 \frac{d\zeta}{\zeta^2}\, D_{H,\mu_0^2}^q(\zeta) \int_{k^2>\mu_0^2} \int_{k^2\zeta/(x_Fs_0)}^1 d\xi \,\frac{x_F}{\zeta} \,f_{\mu_0^2}^q\left( \frac{x_F}{\zeta}\right)\, \frac{1+(1-\xi)^2}{\xi} \nonumber \\
&& \hskip 3cm \times \int_q s\left(\frac{p}{\zeta}\right)\,s(q)\,\left[\frac{\frac{p}{\zeta}(1-\xi)-q-k}{(\frac{p}{\zeta}(1-\xi)-q-k)^2}\cdot \frac{k}{k^2}+\frac{1}{k^2}\right].
\end{eqnarray}

\subsection{Resumming the transverse logarithms}
\subsubsection{The real term}
We start with the observation that the real NLO contribution can be algebraically manipulated in the following form: \beq\label{sigmafinal1rft1}
\frac{d{\sigma}_{1,r}^{q\to q\to H}}{d^2p d\eta}&=&S_\perp\frac{g^2}{(2\pi)^3}\frac{N_c}{2}
\int_{x_F}^1 \frac{d\zeta}{\zeta^2}\,D_{H,\mu_0^2}^q(\zeta)\ \int_{k,q} \int_{\zeta k^2/(x_Fs_0)}^{1} d\xi  \,\frac{1+(1-\xi)^2}{\xi} \ \frac{x_F}{(1-\xi) \zeta} f_{\mu_0^2}^q\left(\frac{x_F}{(1-\xi) \zeta}\right)\noindent \\
&& \times\  s(k)s(q)\,\left\{\left[\frac{\frac{p}{\zeta}-k}{\left(\frac{p}{\zeta}-k\right)^2}-\frac{\frac{p}{\zeta}-(1-\xi)k}{\left(\frac{p}{\zeta}-(1-\xi)k\right)^2}\right]\cdot\left[\frac{\frac{p}{\zeta}-q}{\left(\frac{p}{\zeta}-q\right)^2}-\frac{\frac{p}{\zeta}-(1-\xi)q}{\left(\frac{p}{\zeta}-(1-\xi)q\right)^2}\right]\right.\nonumber\\
&&\hskip 2cm \left.+\frac{1}{2}\left[\frac{(k-q)^2}{\left(\frac{p}{\zeta}-k\right)^2\left(\frac{p}{\zeta}-q\right)^2}+\frac{(1-\xi)^2(k-q)^2}{\left(\frac{p}{\zeta}-(1-\xi)k\right)^2\left(\frac{p}{\zeta}-(1-\xi)q\right)^2}\right]\right\}.\nonumber
\eeq
It is obvious that the first term on the right hand side contains no transverse or longitudinal logarithms, and is a genuine finite NLO correction. Both terms in the last line however contain transverse logarithms coming from the poles in the integrals over $q$ and $k$. We now concentrate on these two terms. 

With some changes of variables the first term in the last line can be cast into the form
\beq
\int_{k,q} \frac{1}{2}\,s(k)s(q)\frac{(k-q)^2}{\left(\frac{p}{\zeta}-k\right)^2\left(\frac{p}{\zeta}-q\right)^2}=\int_{k,q} \frac{1}{k^2} \,s\left(-k+\frac{p}{\zeta}\right) \left[ 1- \frac{k\cdot q}{q^2}\right] \, s\left(-q+\frac{p}{\zeta}\right).
\label{eq:secondmagic}
\eeq
Now, using~\eqref{eq:tmdd1pp}, we can write this term as
\beq
S_\perp \int_{x_F}^1 \frac{d\zeta}{\zeta^2}\,  \int_{k^2>\mu_0^2}  \frac{x_F}{\zeta}\,D_{H,\mu_0^2}^q(\zeta)  {\cal T}_q\left(\frac{x_F}{\zeta},k^2;k^2,\xi_0=k^2\zeta/(x_Fs_0)\right) \int_qs(-k+p/\zeta)\left[1-\frac{k\cdot q}{q^2}\,\right]s(-q+p/\zeta).
\eeq

In the last term in eq.~\eqref{sigmafinal1rft1} we make the change of variables $\zeta'=\zeta(1-\xi)$ and assuming 
$k^2/(x_F s_0)\ll 1$, obtain (dropping the prime over $\zeta$ for simplicity)
\beq
&&\frac{g^2}{(2\pi)^3} S_\perp \frac{N_c}{2} \int_k \int_{x_F}^{1}\frac{d\zeta}{\zeta^2}\int_{k^2\zeta/(x_F s_0)}^{1} \frac{d\xi}{1-\xi}\, D_{H,\mu_0^2}^q\left(\frac{\zeta}{1-\xi}\right) \,\frac{x_F}{\zeta} \,f_{\mu_0^2}^q\left( \frac{x_F}{\zeta}\right)\, \frac{1+(1-\xi)^2}{\xi} \nonumber\\
&&\hskip 3cm \times\frac{1}{k^2}\, \int_qs\left(-k+\frac{p}{\zeta}\right)\left[1-\frac{k\cdot q}{q^2}\right]\,s\left(-q+\frac{p}{\zeta}\right).
\label{eq:thirdp3}
\eeq
This is easily expressed in terms of the TMD FF using eq.~\eqref{tmdffk}. At the end of the day the real term is written as
\beq
\label{eq:myreal2}
&&\frac{d{\sigma}_{1,r}^{q\to q\to H}}{d^2p d\eta}\\
&&=S_\perp \int_{x_F}^1 \frac{d\zeta}{\zeta^2}\,  \int_{k^2>\mu_0^2}  \frac{x_F}{\zeta}\,\left\{D_{H,\mu_0^2}^q(\zeta)  {\cal T}_q\left(\frac{x_F}{\zeta},k^2;k^2,\xi_0=k^2\zeta/(x_Fs_0)\right)+f_{\mu_0^2}^q\left( \frac{x_F}{\zeta}\right) \, {\cal F}_H^q\left(\zeta,k^2;k^2,\xi_0=k^2\zeta/(x_Fs_0)\right)\right\}\nonumber\\
&&\hskip 3cm \times \int_qs(-k+p/\zeta)\left[1- \frac{k\cdot q}{q^2}\,\right]s(-q+p/\zeta)\nonumber \\
&&\hskip 0.5cm + \frac{g^2}{(2\pi)^3} S_\perp \frac{N_c}{2} \int_{x_F}^1 \frac{d\zeta}{\zeta^2}\, D_{H,\mu_0^2}^q(\zeta) \int_{k^2>\mu_0^2} \int_{k^2\zeta/(x_Fs_0)}^1 d\xi \,\frac{x_F}{\zeta(1-\xi)} \,f_{\mu_0^2}^q\left( \frac{x_F}{\zeta(1-\xi)}\right)\, \frac{1+(1-\xi)^2}{\xi} \nonumber \\
&&\hskip 3cm \times \int_q s(k)s(q) \left[\frac{p/\zeta -k}{(p/\zeta -k)^2}-\frac{p/\zeta -(1-\xi)k}{(p/\zeta -(1-\xi)k)^2}\right]\cdot\left[\frac{p/\zeta -q}{(p/\zeta -q)^2}-\frac{p/\zeta -(1-\xi)q}{(p/\zeta -(1-\xi)q)^2}\right].\nonumber
\eeq
The first term in the second line describes the process where the incoming quark has momentum $k$. It scatters with momentum transfer $-k+p/\zeta$. The outgoing quark with transverse momentum $p/\zeta$ then collinearly fragments into hadron with momentum $p$. 

To understand the second term in the second line it is convenient to change variables a little: $k\rightarrow -q+p/\zeta$, $q\rightarrow -k+p/\zeta$. Then this describes the process where the incoming quark has vanishing momentum, it scatter with momentum transfer $q$ and later fragments into hadron with momentum $p$, via first fragmenting perturbatively into quark with momentum $p/\zeta$.

Note that the integration in $k$ is limited to $k^2 > \mu_0^2$, as the perturbative splitting process produces partons above the non-perturbative scale. Note also that the TMD PDF ${\cal T}_q\left(\frac{x_F}{\zeta},k^2;k^2,\xi_0=k^2\zeta/(x_Fs_0)\right)$ in~\eqref{eq:myreal2} is already of order $\alpha_s$ and, therefore, to this order we can choose the resolution scale in the TMD to be any $\mu^2\ge k^2$. In the kinematics of this term, the momentum transfer from the target is $-k+p/\zeta$, which is always dominated by $Q_s$. Thus $k$ is always either greater (if $p/\zeta\gg Q_s$) or equal (if $p/\zeta\ll Q_s$) to $Q_s$.  We can therefore write in~\eqref{eq:myreal2} for TMD PDF ${\cal T}_q\left(\frac{x_F}{\zeta},k^2;\mu^2,\xi_0=\mu^2\zeta/(x_Fs_0)\right)$ with $\mu_T^2$ as defined in~\eqref{thescales1}. The same is true for the TMD FF term. Here the resolution scale also is set by $\mu_F^2$ in~\eqref{thescales1}, since $k^2=[(p/\zeta-k)-p/\zeta]^2$. We will see below that this choice of scale is best when the virtual correction is included.

Thus we find that the large logarithms in the real contribution are resummed into the TMD PDF and FF. We now move to the virtual contribution.

\subsubsection{The virtual term}

We first rewrite the virtual term as
\begin{eqnarray}\
\label{eq:sigmafinal1v2}
&&\frac{d{\sigma}_{1,v}^{q\to q\to H}}{d^2p d\eta} =-2\frac{g^2}{(2\pi)^3} S_\perp \frac{N_c}{2}\int_{x_F}^1 \frac{d\zeta}{\zeta^2}\, D_{H,\mu_0^2}^q(\zeta) \int_{k^2>\mu_0^2} \int_{k^2\zeta/(x_Fs_0)}^1 d\xi \,\frac{x_F}{\zeta} \,f_{\mu_0^2}^q\left( \frac{x_F}{\zeta}\right)\, \frac{1+(1-\xi)^2}{\xi} \\
&&  \hskip 0.5cm \times \int_q s\left(\frac{p}{\zeta}\right)\,s(q)\,\left\{\left[\frac{\frac{p}{\zeta}-q-k}{\left(\frac{p}{\zeta}-q-k\right)^2}\cdot\frac{k}{k^2}+\frac{1}{k^2}\right] +\left[\frac{\frac{p}{\zeta}(1-\xi)-q-k}{\left(\frac{p}{\zeta}(1-\xi)-q-k\right)^2}-\frac{\frac{p}{\zeta}-q-k}{\left(\frac{p}{\zeta}-q-k\right)^2}\right]\cdot
\frac{k}{k^2}\right\}.\nonumber
\end{eqnarray}
The second term here is analogous to a similar term in the real contribution. It contains no large logarithms, either transverse or longitudinal, and is therefore a small, genuinely perturbative correction.

To understand the physics of the first term we perform the angular integration over the angle of vector $k$ in eq.~\eqref{eq:sigmafinal1vft},
\begin{equation}
\int_{k^2>\mu_0^2}d^2k\left[\frac{\frac{p}{\zeta}-q-k}{(\frac{p}{\zeta}-q-k)^2}\cdot\frac{k}{k^2}+\frac{1}{k^2}\right]=\int_{\mu_0^2}^{\left(q-\frac{1}{\zeta} p\right)^2} \frac{d^2k}{k^2}\ .
\end{equation}
We can now write for the first term in eq.~\eqref{eq:sigmafinal1v2}
\begin{eqnarray} \label{eq:myvirrtual2}
&&-2\frac{g^2}{(2\pi)^3} S_\perp \frac{N_c}{2}\int_{x_F}^1 \frac{d\zeta}{\zeta^2}\, D_{H,\mu_0^2}^q(\zeta) \int_{\mu_0^2}^\infty d^2k \int_{k^2\zeta/(x_Fs_0)}^1 d\xi \,\frac{x_F}{\zeta} \,f_{\mu_0^2}^q\left( \frac{x_F}{\zeta}\right)\, \frac{1+(1-\xi)^2}{\xi} \nonumber \\
&& \hskip 3cm \times \int_q s\left(\frac{p}{\zeta}\right)\,s(q)\,\left[\frac{\frac{p}{\zeta}-q-k}{\left(\frac{p}{\zeta}-q-k\right)^2}\cdot\frac{k}{k^2}+\frac{1}{k^2}\right] \\
&&=-2\frac{g^2}{(2\pi)^3} S_\perp \frac{N_c}{2}\int_{x_F}^1 \frac{d\zeta}{\zeta^2}\, D_{H,\mu_0^2}^q(\zeta)\int_q\int_{\mu_0^2}^{\left(q-\frac{1}{\zeta} p\right)^2} \frac{d^2k}{k^2} \int_{k^2\zeta/(x_Fs_0)}^1 d\xi \,\frac{x_F}{\zeta} \,f_{\mu_0^2}^q\left( \frac{x_F}{\zeta}\right)\, \frac{1+(1-\xi)^2}{\xi}  s\left(\frac{p}{\zeta}\right)\,s(q)\nonumber\\
&&=-2\frac{g^2}{(2\pi)^3} S_\perp \frac{N_c}{2}\int_{x_F}^1 \frac{d\zeta}{\zeta^2}\, D_{H,\mu_0^2}^q(\zeta)\int_q\left[\int_{\mu_0^2}^{\mu^2}+\int_{\mu^2}^{\left(q-\frac{1}{\zeta} p\right)^2}\right] \frac{d^2k}{k^2} \int_{k^2\zeta/(x_Fs_0)}^1 d\xi \,\frac{x_F}{\zeta} \,f_{\mu_0^2}^q\left( \frac{x_F}{\zeta}\right)\, \frac{1+(1-\xi)^2}{\xi}  s\left(\frac{p}{\zeta}\right)\,s(q).\nonumber
\end{eqnarray}
This simple result has a nice interpretation. Recall that the first term in the square brackets in the first line in \eqref{eq:myvirrtual2} originates from the diagrams where the incoming quark splits into a $qg$ pair, which then scatters and recombines after the scattering 
into a quark. This is clearly an NLO correction to the LO  elastic quark scattering, $\delta\sigma_{elastic}$. The second term in the square brackets is just the $qg$ loop on the quark propagator, which occurs either before or after the scattering of the quark -- so the proper virtual diagram, $\sigma_v^{proper}$.

What do we expect from the elastic contribution $\delta\sigma_{elastic}$? If the transverse size of the $qg$ pair is greater than the inverse momentum transfer from the target (or the relative momentum is smaller than the momentum transfer), we expect this contribution to be very small. This is because the scattering will be dominated by a single kick to a single parton, but this clearly cannot be elastic since in the outgoing state the relative momentum between $q$ and $g$ then will be large, while in the elastic state the relative momentum should be small. On the other hand, if the size of the $qg$ pair is much smaller that the inverse momentum transfer, the scattering does not resolve the pair, and there should be no correction to the elastic cross section. In other words $\delta\sigma_{elastic}$ should be cancelled by the NLO correction to the single quark elastic cross section which does not include splitting into $qg$ pair in the intermediate state, i.e., $\sigma_v^{proper}$. Thus for large sizes we expect the sum of the two virtual terms to be simply equal to the "proper" virtual term $\sigma_v^{proper}$, while for small sizes we expect the sum to vanish since the two terms should cancel each other. eq.~\eqref{eq:myvirrtual2} reflects precisely this behavior in a somewhat extreme form. Recall that the integral over $k^2$ in \eqref{eq:myvirrtual2} is precisely the integral over the (inverse) sizes of the $qg$ pair. Also note that the dipole function $s(q)$ should be peaked rather sharply at $q^2\sim Q_s^2$. So for large sizes (or $k^2<\left(q-\frac{1}{\zeta} p\right)^2\sim {\rm max}\left[Q_s^2, \frac{p^2}{\zeta^2}\right]$) the whole contribution in eq.~\eqref{eq:myvirrtual2} is given by the proper virtual term, while for small sizes there indeed is complete cancellation. 

Thus the virtual term essentially tells us that the  $qg$ pairs of large size scatter inelastically, while those of very small size are not resolved and therefore do not contribute a perturbative correction.

In the last equality in eq.~\eqref{eq:myvirrtual2} we have deliberately split the integration interval into two. It is easy to see that the first term (integral up to $\mu^2$) combines with the LO to evolve the resolution scale in the TMDs to $\mu^2$:
\beq\label{virtmod}
&&S_\perp \int_{x_F}^1 \frac{d\zeta}{\zeta^2}\,D_{H,\mu_0^2}^q(\zeta)\ \frac{x_F}{\zeta} f_{\mu_0^2}^q\left(\frac{x_F}{\zeta}\right) s\left(\frac{p}{\zeta}\right)\\
&&\hskip 0.3cm -2\frac{g^2}{(2\pi)^3} S_\perp \frac{N_c}{2}\int_{x_F}^1 \frac{d\zeta}{\zeta^2}\, D_{H,\mu_0^2}^q(\zeta)\int_{\mu_0^2}^{\mu^2} \frac{d^2k}{k^2} \int_{k^2\zeta/(x_Fs_0)}^1 d\xi \,\frac{x_F}{\zeta} \,f_{\mu_0^2}^q\left( \frac{x_F}{\zeta}\right)\, \frac{1+(1-\xi)^2}{\xi}  s\left(\frac{p}{\zeta}\right)\,\nonumber\\
&&\approx S_\perp\int_{x_F}^1 \frac{d\zeta}{\zeta^2}\int_q \int_0^{\mu_0^2} d^2l \int_0^{\mu_0^2}d^2k\nonumber\\
&&\hskip 1cm\times {\cal F}_H^q\left(\zeta,l^2;\mu^2;\xi_0=\frac{\zeta \mu^2}{x_F s_0}\right)\,\frac{x_F}{\zeta } {\cal T}_q\left(\frac{x_F}{\zeta},k^2;\mu^2; \xi_0=\frac{\zeta \mu^2}{x_F s_0}\right)s\left(-(k+l)+\frac{p}{\zeta}\right)\left[1-\frac{(k+l)\cdot q}{q^2}\right]s\left(-q+\frac{p}{\zeta}\right),\nonumber
\eeq
We did two things to arrive at the last equality. First, we have evolved the factorization scale in the collinear PDF and FF up to $\mu^2$, but kept the integral over the momentum up to the low factorization scale:
\begin{equation}\label{evolvdf}
D_{H,\mu_0^2}^q(\zeta)\rightarrow\int_0^{\mu_0^2} d^2l{\cal F}_H^q\left(\zeta,l^2;\mu^2;\xi_0=\frac{\zeta \mu^2}{x_F s_0}\right); \ \ \ \  f_{\mu_0^2}^q\left(\frac{x_F}{\zeta}\right)\rightarrow \int_0^{\mu_0^2} d^2k{\cal T}_q\left(\frac{x_F}{\zeta},k^2;\mu^2; \xi_0=\frac{\zeta \mu^2}{x_F s_0}\right).
\end{equation}
This introduces a term ${\cal O}\left(\alpha_s^2\right)$ and is therefore legitimate in our order $\alpha_s$ calculation. In addition we have altered the scattering amplitude:
\beq
\label{eq:discard}
s\left(\frac{p}{\zeta}\right)\rightarrow 
\int_qs\left(-(k+l)+\frac{p}{\zeta}\right)\left[1-\frac{(k+l)\cdot q}{q^2}\right]s\left(-q+\frac{p}{\zeta}\right).
\eeq
This is legitimate since in eq.~\eqref{virtmod},  $|k+l|^2\lesssim \mu_0^2\ll \frac{p^2}{\zeta^2}$, while the momentum $q$ is dominated by the region where the argument of the second dipole is of order $Q_s$, and thus $q^2\sim \rm{max}\left[Q_s^2,\frac{p^2}{\zeta^2}\right]$. We also recall that $\int_qs(q)=1$. In all, this modification only adds subleading power-correction terms of order $\mu_0^2/Q_s^2$ and therefore are beyond the accuracy of our calculation.
The utility in these modifications is that they allow us to put the virtual and real terms together in a simple way.

Now going back to eq.~\eqref{eq:myvirrtual2}, we note that the second contribution comes only from the pairs of the transverse size close to the resolution provided by the target. 
We show in  Appendix \ref{appb} that this term is a small perturbative correction to the elastic scattering probability and does not contain large logarithms as long as our choice of the Ioffe time parameter $s_0$ is close enough to the factorization scale $\mu^2$, so that $\ln \left(s_0/\mu^2\right)$ is not large.

\subsection{Putting it all together}

We can now put together the real and virtual pieces. To do that we use eq.~\eqref{evolvdf} in the first term in eq.~\eqref{eq:myreal2}, which again is legitimate within the accuracy of our calculation.
Then the first term in  eq.~\eqref{eq:myreal2}, up to order $\alpha_s^2$ corrections, can be cast in the form of the first term in~\eqref{virtmod} with the difference of the domain integration in $l$ and $k$. The real and virtual contributions  can be combined into the following expression, which now does not contain any large logarithms apart from those that are resummed into the TMD PDF and TMD FF:
\beq
\label{eq:mytotal2}
&&\frac{d{\sigma}^{q\to q\to H}}{d^2p d\eta}\\
&&=S_\perp\int_{x_F}^1 \frac{d\zeta}{\zeta^2}\int_q \int d^2l \int d^2k\nonumber \\
&&\hskip 1cm\times {\cal F}_H^q\left(\zeta,l^2;\mu^2;\xi_0=\frac{\zeta \mu^2}{x_F s_0}\right)\,\frac{x_F}{\zeta } {\cal T}_q\left(\frac{x_F}{\zeta},k^2;\mu^2; \xi_0=\frac{\zeta \mu^2}{x_F s_0}\right)s\left(-(k+l)+\frac{p}{\zeta}\right)\left[1-\frac{(k+l)\cdot q}{q^2}\right]s\left(-q+\frac{p}{\zeta}\right)\nonumber \\
&&\hskip 0.5cm - 2\frac{g^2}{(2\pi)^3} S_\perp \frac{N_c}{2}\int_{x_F}^1 \frac{d\zeta}{\zeta^2}\, D_{H,\mu_0^2}^q(\zeta)  \frac{x_F}{\zeta} \,f_{\mu_0^2}^q\left( \frac{x_F}{\zeta}\right)\int_q \int_{\mu^2}^{\left(q-\frac{1}{\zeta}p\right)^2}\frac{d^2k}{k^2}\int_{k^2\zeta/(x_Fs_0)}^1 d\xi \,\, \frac{1+(1-\xi)^2}{\xi}  
s\left(\frac{p}{\zeta}\right)\,s\left(q\right)\nonumber \\
&&\hskip 0.5cm+\frac{g^2}{(2\pi)^3} S_\perp \frac{N_c}{2} \int_{x_F}^1 \frac{d\zeta}{\zeta^2}\, D_{H,\mu_0^2}^q(\zeta) \int_{k} \int_{k^2\zeta/(x_Fs_0)}^1 d\xi \,\frac{x_F}{\zeta(1-\xi)} \,f_{\mu_0^2}^q\left( \frac{x_F}{\zeta(1-\xi)}\right)\, \frac{1+(1-\xi)^2}{\xi} \nonumber \\
&&\hskip 3cm \times \int_q s(k)s(q) \left[\frac{p/\zeta -k}{(p/\zeta -k)^2}-\frac{p/\zeta -(1-\xi)k}{(p/\zeta -(1-\xi)k)^2}\right]\cdot\left[\frac{p/\zeta -q}{(p/\zeta -q)^2}-\frac{p/\zeta -(1-\xi)q}{(p/\zeta -(1-\xi)q)^2}\right].\nonumber\\
&&\hskip 0.5cm-2\frac{g^2}{(2\pi)^3} S_\perp \frac{N_c}{2}\int_{x_F}^1 \frac{d\zeta}{\zeta^2}\, D_{H,\mu_0^2}^q(\zeta) \int_{k^2>\mu_0^2} \int_{k^2\zeta/(x_Fs_0)}^1 d\xi \,\frac{x_F}{\zeta} \,f_{\mu_0^2}^q\left( \frac{x_F}{\zeta}\right)\, \frac{1+(1-\xi)^2}{\xi}\nonumber \\
&& \hskip 3cm \times \int_q s\left(\frac{p}{\zeta}\right)\,s(q)\,\left[\frac{\frac{p}{\zeta}(1-\xi)-q-k}{(\frac{p}{\zeta}(1-\xi)-q-k)^2}-\frac{\frac{p}{\zeta}-q-k}{(\frac{p}{\zeta}-q-k)^2}\right]\cdot
\frac{k}{k^2}\ .\nonumber
\eeq

Now, making a change of notation $k\to m$ in the genuine NLO corrections,
we can reorganize~\eqref{eq:mytotal2} to read
\beq
\label{eq:mytotal3}
&&\frac{d{\sigma}^{q\to q\to H}}{d^2p d\eta}=
S_\perp\int_{x_F}^1 \frac{d\zeta}{\zeta^2}\int d\xi \int d^2l \int d^2k\ 
{\cal F}_H^q\left(\zeta,l^2;\mu^2;\xi_0=\frac{\zeta \mu^2}{x_F s_0}\right)\\
&&\hskip 3.3cm \times\frac{x_F}{\zeta (1-\xi)} {\cal T}_q\left(\frac{x_F}{\zeta(1-\xi)},k^2;\mu^2; \xi_0=\frac{\zeta \mu^2}{x_F s_0}\right)\, \tilde{\cal P}(\xi,\zeta;k+l;p,s_0,\mu^2,\mu_0^2),\nonumber
\eeq
with the production probability 
\beq
\label{eq:hardfactor}
&&\tilde{\cal P}(\xi,\zeta;k+l;p,s_0,\mu^2,\mu_0^2)=\delta(\xi)\Bigg\{\int_qs\left(-(k+l)+\frac{p}{\zeta}\right)\left[1-\frac{(k+l)\cdot q}{q^2}\right]s\left(-q+\frac{p}{\zeta}\right)\\
&& 
\nonumber\hskip 0.8cm -2\frac{g^2}{(2\pi)^3} S_\perp \frac{N_c}{2}\theta\left[\mu_0^2-k^2\right]\theta\left[\mu_0^2-l^2\right]
\Bigg(\int_q \int_{\mu^2}^{\left(q-\frac{1}{\zeta}p\right)^2}\frac{d^2m}{m^2}\int_{m^2\zeta/(x_Fs_0)}^1 d\lambda \,\, \frac{1+(1-\lambda)^2}{\lambda} 
s\left(\frac{p}{\zeta}\right)\,s\left(q\right)\\
&&+\int_{m^2>\mu_0^2} \int_{m^2\zeta/(x_Fs_0)}^1 d\lambda \, \frac{1+(1-\lambda)^2}{\lambda}  \int_qs\left(\frac{p}{\zeta}\right)\,s(q)\,\left[\frac{\frac{p}{\zeta}(1-\lambda)-q-m}{(\frac{p}{\zeta}(1-\lambda)-q-m)^2}-\frac{\frac{p}{\zeta}-q-m}{(\frac{p}{\zeta}-q-m)^2}\right]\cdot
\frac{m}{m^2}\Bigg)\Bigg\}\nonumber\\
&&\hskip 0.3cm+\frac{g^2}{(2\pi)^3}\,\frac{N_c}{2}\theta\left[\mu_0^2-k^2\right]\theta\left[\mu_0^2-l^2\right]\int_m \frac{1+(1-\xi)^2}{\xi} \theta\left(\xi-\frac{m^2\zeta}{x_Fs_0}\right)\nonumber \\
&&\hskip 0.6cm\times \int_q s(m)s(q) \left[\frac{p/\zeta -m}{(p/\zeta -m)^2}-\frac{p/\zeta -(1-\xi)m}{(p/\zeta -(1-\xi)m)^2}\right]\cdot\left[\frac{p/\zeta -q}{(p/\zeta -q)^2}-\frac{p/\zeta -(1-\xi)q}{(p/\zeta -(1-\xi)q)^2}\right].\nonumber 
\eeq
In this expression, the first term has the form of a production probability discussed above, while the remaining factors correspond to genuine NLO contributions without any logarithmic enhancement for the choice of scales in~\eqref{thescales1}. The production probability in the first term has a natural interpretation: a quark with momentum $k+l$ should be scattered with momentum transfer $-(k+l)+p/\zeta$ in order to emerge with momentum $p/\zeta$. The unity in the square bracket would be the probability for such scattering if the quark would scatter independently of the rest of the spectators. The second term, $-\frac{(k+l)\cdot q}{q^2}$, corrects this by  taking into account that the quark has to decohere from the gluon with which it is correlated in the incoming wave function, in order to be actually produced.

We note that the change $\mu_0^2\to \mu^2$ in the collinear PDFs and FFs only affects the expression at ${\cal O}(\alpha^2_s)$. We can therefore  replace $\mu_0^2$ by $\mu^2$ as the factorization scale in the TMD, and also drop the constraint on the momenta $k^2$ and $l^2$ as being smaller than $\mu_0^2$ in the second and fourth terms in \eqref{eq:hardfactor}. 
Beyond the formal argument of this change being a higher order correction, physically it is clear that for any parton with intrinsic transverse momentum smaller than $p$, the correction to the production probability should not depend on this momentum and should be the same as we have calculated for $k^2<\mu_0^2$. The same is true for the momentum in the fragmentation function. Thus replacing $\mu_0^2$ by $\mu^2$ is physically well motivated.
Finally, we can simplify our expression somewhat further 
 by setting the lower limit of the integration over $\lambda$ and $m$
 in the third term in~\eqref{eq:hardfactor},  to zero. Since the integrand in this term is regular both at $m^2\rightarrow 0$ and $\lambda\rightarrow 0$, this amounts to dropping power corrections in $\mu_0^2/s_0$, which is indeed a small number.
 
All in all we can reorganize~\eqref{eq:mytotal3} to read,
\beq
\label{eq:mytotal4}
&&\frac{d{\sigma}^{q\to q\to H}}{d^2p d\eta}=
S_\perp\int_{x_F}^1 \frac{d\zeta}{\zeta^2}\int d\xi \int d^2l \int d^2k\ 
{\cal F}_H^q\left(\zeta,l^2;\mu^2;\xi_0=\frac{\zeta \mu^2}{x_F s_0}\right)\\
&&\hskip 3.3cm \times\frac{x_F}{\zeta (1-\xi)} {\cal T}_q\left(\frac{x_F}{\zeta(1-\xi)},k^2;\mu^2; \xi_0=\frac{\zeta \mu^2}{x_F s_0}\right)\, {\cal P}(\xi,\zeta;k+l;p,s_0,\mu^2),\nonumber
\eeq
with the production probability 
\beq
\label{eq:hardfactor4}
&&{\cal P}(\xi,\zeta;k+l;p,s_0,\mu^2)=\delta(\xi)\Bigg\{\int_qs\left(-(k+l)+\frac{p}{\zeta}\right)\left[1-\frac{(k+l)\cdot q}{q^2}\right]s\left(-q+\frac{p}{\zeta}\right)\\
&& 
\nonumber\hskip 0.8cm -2\frac{g^2}{(2\pi)^3} S_\perp \frac{N_c}{2}
\Bigg(\int_q \int_{\mu^2}^{\left(q-\frac{1}{\zeta}p\right)^2}\frac{d^2m}{m^2}\int_{m^2\zeta/(x_Fs_0)}^1 d\lambda \,\, \frac{1+(1-\lambda)^2}{\lambda} 
s\left(\frac{p}{\zeta}\right)\,s\left(q\right)\\
&&+\int_{m} \int_{0}^1 d\lambda \, \frac{1+(1-\lambda)^2}{\lambda}  \int_qs\left(\frac{p}{\zeta}\right)\,s(q)\,\left[\frac{\frac{p}{\zeta}(1-\lambda)-q-m}{(\frac{p}{\zeta}(1-\lambda)-q-m)^2}-\frac{\frac{p}{\zeta}-q-m}{(\frac{p}{\zeta}-q-m)^2}\right]\cdot
\frac{m}{m^2}\Bigg)\Bigg\}\nonumber\\
&&\hskip 0.3cm+\frac{g^2}{(2\pi)^3}\,\frac{N_c}{2}\int_m \frac{1+(1-\xi)^2}{\xi} \theta\left(\xi-\frac{m^2\zeta}{x_Fs_0}\right)\nonumber \\
&&\hskip 0.6cm\times \int_q s(m)s(q) \left[\frac{p/\zeta -m}{(p/\zeta -m)^2}-\frac{p/\zeta -(1-\xi)m}{(p/\zeta -(1-\xi)m)^2}\right]\cdot\left[\frac{p/\zeta -q}{(p/\zeta -q)^2}-\frac{p/\zeta -(1-\xi)q}{(p/\zeta -(1-\xi)q)^2}\right].\nonumber 
\eeq
Note that in the $l$- and $k$-independent terms in ${\cal P}$, the integral over these momentum variables in \eqref{eq:mytotal4} simply turns the appropriate TMD into the collinear PDF. These "genuine NLO" correction terms therefore have the same collinear structure as the naive leading order expression in the hybrid approach, with the production probability having an NLO correction which is not necessarily eikonal, since it involves a nontrivial $\xi$ dependence.

Eqs.~\eqref{eq:mytotal4} and~\eqref{eq:hardfactor4} are our final result for the hadron produced from the projectile quark. 

In addition we need of course to account for all available channels. Those include the quark initiated channel, which produces a gluon which eventually fragments into the hadron, as well as all gluon initiated channels. The detailed calculation of these processes is presented in Appendix~\ref{appc}. In the next section we summarize the results of these calculations.
 
\section{The other channels}
\label{sec:otherch}
In the previous section we showed in detail how the transverse logarithms are resummed into TMD PDF and FF in the case when the projectile quark scatters on the target and eventually fragments into the observed hadron. In this exercise we used the simplified evolution equations for the TMDs, given in Sec.~\ref{seciia} which did not involved the gluon contributions. The details of the full calculation, including all channels is presented in Appendix~\ref{appc}. The results are qualitatively the same. All large transverse logarithms can be absorbed into the quark and gluon TMD PDFs and TMD FFs, whose definitions and evolutions are given in Sec.~\ref{seciib}. The rest of the terms are finite and are genuine NLO corrections. We summarize the results here.

Consider the channels where the leading order contribution is given by the small transverse momentum gluon scattering off the target. At NLO this channel gets several real contribution. The gluon can split  either into a $q\bar q$ pair or into a $gg$ pair. The splitting can either occur in the projectile wave function before scattering, or in the final state after scattering on the target.
The large logarithms arising from the former contribution are summed into the TMD PDFs of the member of the pair ($q$ or $\bar q$, or one of the gluons) which produces the hadron after scattering. In the later case the logarithms are summed into the TMD FF of the gluon. 

There are of course also virtual contribution to all these channels. Just like for the quark, they ensure that the optimal choice of the transverse resolution scale in the TMDs is $\mu^2$ discussed in the previous section. 

After resummation of the logarithms, both real and virtual diagrams leave a finite remainder which is not enhanced by any logarithms, and we refer to it as the genuine NLO correction. In all cases this genuine NLO correction can be reabsorbed into the appropriate production probability as in eq.~\eqref{eq:hardfactor4}.

In addition there is a contribution arising from the incoming quark like in the previous section but where the hadron is produced from the gluon which splits off the quark. The contribution due to splitting before scattering completes the relevant piece in the gluon TMD PDF, while splitting after scattering completes the quark TMD FF, completing all the terms in the formulae of Sec.~\ref{seciib}.

All said and done, the final result has the naive TMD factorized form like in the previous section augmented by "genuine NLO" terms:
\beq
&&\frac{d{\sigma}}{d^2p d\eta}=S_\perp\int_{x_F}^1 \frac{d\zeta}{\zeta^2}\int_q \int d^2l \int d^2k\nonumber \\
&&\hskip 0.5cm\times \Bigg\{{\cal F}_H^q\left(\zeta,l^2;\mu^2;\xi_0=\frac{\zeta \mu^2}{x_F s_0}\right)\,\frac{x_F}{\zeta } {\cal T}_q\left(\frac{x_F}{\zeta},k^2;\mu^2; \xi_0=\frac{\zeta \mu^2}{x_F s_0}\right)s\left(-(k+l)+\frac{p}{\zeta}\right)\left[1-\frac{(k+l)\cdot q}{q^2}\right]s\left(-q+\frac{p}{\zeta}\right)\nonumber \\
&&
\hspace{1.cm}
+\, \int_t{\cal F}_g\left(\zeta,l^2;\mu^2;\xi_0=\frac{\zeta \mu^2}{x_F s_0}\right)\frac{x_F}{\zeta} {\cal T}_g\left( \frac{x_F}{\zeta},k^2;\mu^2;\xi_0=\frac{\zeta \mu^2}{x_F s_0}\right)\nonumber \\
&&\hskip 4cm \times
s(t)\, s\left(\frac{p}{\zeta}-t+(k+l)\right) \bigg[1-\frac{(k+l)\cdot q}{q^2}\bigg]\, s\left(\frac{p}{\zeta}-t+q\right)\Bigg\}\nonumber\\
&&\hskip 0.6cm +\sum_i({\rm Gen.\, NLO})_i\ ,
\eeq
with the resolution scale $\mu^2\approx Q_s^2+p^2/\zeta^2$.

The finite terms have the meaning of NLO corrections to production probability of the appropriate parton when interacting with the target. All of these terms have a collinearly factorized structure, i.e., they describe a parton with low transverse momentum coming from the projectile, producing another parton (either of identical or different species) with high transverse momentum due to scattering from the target, and then fragmenting collinearly into the final state hadron.

These terms come in two basic varieties. The first type is corrections to eikonal production, which do not involve longitudinal momentum transfer during interaction with the target, where the outgoing parton has the same longitudinal momentum as the incoming one. These are of the type of the terms proportional to $\delta(\xi)$ in eq.~\eqref{eq:hardfactor4}. The incoming and the outgoing parton here is always of the same species. In the nomenclature of~\cite{Altinoluk:2011qy} these are the genuine NLO corrections to the elastic production probability.

The second type are terms that involve the change of longitudinal momentum during production, like the last term in eq.~\eqref{eq:hardfactor4}. These corrections can be thought of as arising from the process where the incoming parton splits into a pair in close proximity to the target, and the two members of the pair are produced due to scattering. These are the NLO corrections to the inelastic production probability. In cases where the produced parton is of a different variety than the incoming one, these terms are the leading contributions since at ${\cal O}(1)$ such processes are absent in the naive leading order collinearly factorized expression in the hybrid approximation.

We present the expressions for the finite terms below. We use a notation that stresses the interpretation discussed above. For example $({\rm Gen.\, NLO})^{q}_{el}$ denotes the genuine NLO contribution to hadron production via elastic scattering of a quark, while $({\rm Gen.\, NLO})^{q\rightarrow g}_{inel}$ means the contribution of the inelastic channel where an incoming quark produces a gluon, which later fragments into the hadron.
The results derived in the Appendix~\ref{appc} are the following (for completeness we also list here the results derived in the previous section):
 
\noindent $\bullet$ The elastic corrections are:
\beq
&&({\rm Gen.\, NLO})^q_{el}=\\
&&\hskip 1cm- 2\frac{g^2}{(2\pi)^3} S_\perp \frac{N_c}{2}\int_{x_F}^1 \frac{d\zeta}{\zeta^2}\, D_{H,\mu^2}^q(\zeta)  \frac{x_F}{\zeta} \,f_{\mu^2}^q\left( \frac{x_F}{\zeta}\right)\int_q \int_{\mu^2}^{\left(q-\frac{1}{\zeta}p\right)^2}\frac{d^2k}{k^2}\int_{\xi_0=k^2\zeta/(x_Fs_0)}^1 d\xi \,\, \frac{1+(1-\xi)^2}{\xi}  
s\left(\frac{p}{\zeta}\right)\,s\left(q\right)\nonumber \\
&&\hskip 1cm -2\frac{g^2}{(2\pi)^3} S_\perp \frac{N_c}{2}\int_{x_F}^1 \frac{d\zeta}{\zeta^2}\, D_{H,\mu^2}^q(\zeta) \int_{k^2} \int_{0}^1 d\xi \,\frac{x_F}{\zeta} \,f_{\mu^2}^q\left( \frac{x_F}{\zeta}\right)\, \frac{1+(1-\xi)^2}{\xi}\nonumber \\
&& \hskip 3cm \times \int_q s\left(\frac{p}{\zeta}\right)\,s(q)\,\left[\frac{\frac{p}{\zeta}(1-\xi)-q-k}{(\frac{p}{\zeta}(1-\xi)-q-k)^2}-\frac{\frac{p}{\zeta}-q-k}{(\frac{p}{\zeta}-q-k)^2}\right]^i
\frac{k^i}{k^2}\ ,\nonumber
\eeq
\beq
({\rm Gen. NLO})^{g}_{el}&=& - 2 \frac{g^2N_c}{(2\pi)^3}\, S_\perp 
\int_{x_F}^1\frac{d\zeta}{\zeta^2}D^g_{\mu^2}(\zeta) \int_{q,t} \,  \int_{\mu^2}^{[q+t-p/\zeta]^2}\frac{d^2l}{l^2}
\frac{x_F}{\zeta} \, f^g_{\mu^2}\left(\frac{x_F}{\zeta}\right) \, \\
&&\times\,
\, \int_{\xi_0=l^2\zeta/(x_Fs_0)}^{1-\xi_0}d\xi \ \bigg[ \frac{1-\xi}{\xi}+\frac{\xi}{1-\xi}+\xi(1-\xi)\bigg]
s(q)s(t-p/\zeta)s(t)\nonumber \\
& -&\, \frac{g^2}{(2\pi)^3}\, S_\perp \int_{x_F}^1\frac{d\zeta}{\zeta^2} D^g_{\mu^2}(\zeta)
\int_t 2\int_{\xi_0=l^2\zeta/(x_Fs_0)}^1 d\xi\int_{\mu^2}^{[t-(1-\xi)p/\zeta]^2} \frac{d^2l}{l^2}
 \frac{x_F}{\zeta}f^g_{\mu^2}\left(\frac{x_F}{\zeta}\right)\, 
\nonumber\\
&&\times\, 
\frac{1}{2}\, \Big[\xi^2+(1-\xi)^2\Big]\,  s(t)s(t-p/\zeta)\nonumber\\
 &-&2\frac{g^2N_c}{(2\pi)^3}\, S_\perp 
\int_{x_F}^1\frac{d\zeta}{\zeta^2}D^g_{\mu^2}(\zeta)\frac{x_F}{\zeta}f^g_{\mu^2}\left(\frac{x_F}{\zeta}\right) \, \, \int_{l^2}\int_{0}^1d\xi \bigg[ \frac{1-\xi}{\xi}+\frac{\xi}{1-\xi}+\xi(1-\xi)\bigg]
\nonumber\\
&&\times\,
\int_{q,t}s(q)s(t-p/\zeta)s(t)
 \left[\frac{(1-\xi)p/\zeta-l-q-t}{[(1-\xi)p/\zeta-l-q-t]^2}-\frac{p/\zeta-l-q-t}{[p/\zeta-l-q-t]^2}\right]^i
\frac{l^i}{l^2}\ ,\nonumber
\eeq
Here in the last term the cancellation of the $\xi=0$ pole is explicit. To see that the $\xi=1$ pole also cancels we note that under the transformation $t\rightarrow -t+p/\zeta,\ q\rightarrow -q,\ l\rightarrow -l$, we have $\frac{l^i}{l^2}\frac{[-l-q-t]^i}{[-l-q-t]^2}\rightarrow \frac{l^i}{l^2}\frac{[p/\zeta-l-q-t]^i}{[p/\zeta-l-q-t]^2}$. Thus under the assumption $s(k)=s(-k)$ which we use throughout, the cancellation indeed occurs.

\noindent $\bullet$ The inelastic corrections are:
\beq
({\rm Gen.\, NLO})^{q\rightarrow q}_{inel}&=&\frac{g^2}{(2\pi)^3} S_\perp \frac{N_c}{2} \int_{x_F}^1 \frac{d\zeta}{\zeta^2}\, D_{H,\mu^2}^q(\zeta) \int_{k} \int_{0}^1 d\xi \,\frac{x_F}{\zeta(1-\xi)} \,f_{\mu^2}^q\left( \frac{x_F}{\zeta(1-\xi)}\right)\, \frac{1+(1-\xi)^2}{\xi} \nonumber \\
&& \times \int_q s(k)s(q) \left[\frac{p/\zeta -k}{(p/\zeta -k)^2}-\frac{p/\zeta -(1-\xi)k}{(p/\zeta -(1-\xi)k)^2}\right]^i\left[\frac{p/\zeta -q}{(p/\zeta -q)^2}-\frac{p/\zeta -(1-\xi)q}{(p/\zeta -(1-\xi)q)^2}\right]^i,
\eeq
\beq
({\rm Gen.\, NLO})^{g\rightarrow q}_{inel}&=& \frac{g^2}{(2\pi)^3}\, S_{\perp} \int_{x_F}^1\frac{d\zeta}{\zeta^2}\, D^q_{\mu^2}(\zeta)\int_{k^2,q^2}\int_t \int_{0}^1d\xi
\, \frac{x_F}{\zeta(1-\xi)}f^g_{\mu^2}\left(\frac{x_F}{\zeta(1-\xi)}\right)\frac{1}{2}\Big[\xi^2+(1-\xi)^2\Big] \nonumber\\
&&\times s(k)s(q)s(t)\bigg\{ 
\frac{(p/\zeta-q)^i}{(p/\zeta-q)^2}\bigg[\frac{p/\zeta-k}{(p/\zeta-k)^2} -\frac{p/\zeta-(1-\xi)(q-k)}{[p/\zeta-(1-\xi)(q-k)]^2}\bigg]^i
\nonumber\\
&&
\hspace{2.6cm}
+
\frac{[p/\zeta-(1-\xi)(q-k)]^i}{[p/\zeta-(1-\xi)(q-k)]^2}\bigg[ \frac{p/\zeta-(1-\xi)(q-t)}{[p/\zeta-(1-\xi)(q-t)]^2}-\frac{p/\zeta-q}{(p/\zeta-q)^2}
\bigg]^i\bigg\},
\eeq
\beq
&&
({\rm Gen.\, NLO})^{q\rightarrow g}_{inel}= \frac{g^2}{(2\pi)^3}\, S_\perp \int_{x_F}^1 \frac{d\zeta}{\zeta^2} \, D^g_{\mu^2}(\zeta)\int_{k^2,q^2} \int_t \int_{0}^1d\xi \, \frac{x_F}{\zeta\xi}\, f^q_{\mu^2}\left(\frac{x_F}{\zeta\xi}\right)\frac{N_c}{2}\bigg[\frac{1+(1-\xi)^2}{\xi}\bigg]
s(k)s(q)s(t)\nonumber\\
&&
\hspace{-0.5cm}
\times
\bigg\{ \frac{[p/\zeta-(q-k)]^i}{[p/\zeta-(q-k)]^2}\bigg[\frac{p/\zeta-(q-t)}{[p/\zeta-(q-t)]^2}-\frac{p/\zeta-\xi q}{(p/\zeta-\xi q)^2}\bigg]^i+
 \frac{(p/\zeta-\xi q)^i}{(p/\zeta-\xi q)^2}\bigg[ \frac{p/\zeta-\xi k}{(p/\zeta-\xi k)^2}-\frac{p/\zeta-(q-k)}{[p/\zeta-(q-k)]^2}\bigg]^i\bigg\},
\eeq
\beq
&&
({\rm Gen.\, NLO})^{g\rightarrow g}_{inel}=\frac{g^2}{(2\pi)^3}\, S_\perp
\int_{x_F}^1\frac{d\zeta}{\zeta^2}\, D^g_{\mu^2}(\zeta) \\
&& \hskip 2cm \times \int_{k^2}
\int_{0}^1d\xi \frac{x_F}{\zeta(1-\xi)}f^g_{\mu^2}\left(\frac{x_F}{\zeta(1-\xi)}\right)\, 2N_c \bigg[ \frac{1-\xi}{\xi}+\frac{\xi}{1-\xi}+\xi(1-\xi)\bigg]
\nonumber\\
&&\times
\int_{q,t}s(k) \, s(q)\, s(t)\bigg[
\frac{p/\zeta-(q-t)}{[p/\zeta-(q-t)]^2}-\frac{p/\zeta-(1-\xi)(q-t)}{[p/\zeta-(1-\xi)(q-t)]^2}\bigg]^i
\bigg[
\frac{p/\zeta-(q-k)}{[p/\zeta-(q-k)]^2}-\frac{p/\zeta-(1-\xi)(q-k)}{[p/\zeta-(1-\xi)(q-k)]^2}\bigg]^i. \nonumber
\eeq

In addition to all contributions listed above, there are similar contributions associated with antiquarks. Their functional form is identical to the  appropriate terms involving quarks, with the substitution of antiquark TMDs for the quark TMDs. We are not listing those explicitly, but they of course have to be added in, in any numerical calculation.

\section{Discussion and conclusions}
\label{sec:conclu}

In this paper we have revisited the calculation of single inclusive hadron production in $pA$ at forward rapidities at NLO within the hybrid approximation. We have shown that beyond leading order the collinear resummation paradigm does not hold in the hybrid approximation. In order to properly resum  large transverse logarithms at NLO one needs to work within the TMD factorization framework. 

The need to introduce TMDs is in fact quite  clear intuitively.
At high transverse momentum of the observed hadron, the naive parton model  picture of a low transverse momentum projectile parton that scatters with high momentum transfer off the target breaks down. We showed that this is only one of the mechanisms for producing the high momentum hadron in the final state, but not the only one. 

Another mechanism has been discussed a long time ago in~\cite{Altinoluk:2011qy}. It amounts to the projectile parton acquiring large transverse momentum due to perturbative splittings in the projectile wave function, and undergoing only soft scattering with the target with typical momentum transfer of order $Q_s$\footnote{This was dubbed "inelastic scattering" in~\cite{Altinoluk:2011qy}, since it can be viewed as  the production of the observed hadron from the inelastic scattering of the low $p_T$ valence parton, which produces two, and not one, partons in the final state.}. It is clear that collinear factorization is physically inappropriate for the description of such a process.  We have shown in the present paper that the proper way to account for it, in the sense of proper resummation of  large transverse logarithms associated with perturbative splittings in the projectile wave function, is to view the parton as coming from the projectile TMD PDF with large transverse momentum. 

The importance of this contribution was first recognized in~\cite{Altinoluk:2011qy}. Although fragmentation was not included in that analysis, the basic physics discussed there is correct and we recap the argument slightly adjusted by our results. Consider for simplicity the quark channel production cross section. It is given by eq.~\eqref{eq:mytotal4}. The integral over the momentum in the TMD PDF $k$ has two distinct regions, $|k|\ll |p|/\zeta$ and $|k|\approx |p|/\zeta$. The first region corresponds to the collinear regime, while the second to a large transverse momentum quark coming directly from the TMD. In the collinear regime the production cross section is very sensitive to the high momentum behavior of the dipole amplitude $s(p/\zeta)$. For example, in the GBW model~\cite{Golec-Biernat:1998zce} $s(p)$ is strongly suppressed at high momentum and the collinear contribution to production is very small. On the other hand, the large $k$ contribution probes the bulk of the scattering amplitude $s(|k-p/\zeta|\sim Q_s)$ and is not suppressed. Its high momentum behavior is determined by the TMD PDF, which, at least perturbatively falls only as $1/k^2$. Thus, for a GBW-like dipole amplitude the collinear contribution is completely negligible at $p^2\gg Q_s^2$ and the production probability is dominated by high-$k$ quarks originating from splittings in the wave function. In general one does not expect such a steep drop of $s(p)$ with momentum. The simple perturbative expectation  $s(p)\sim 1/p^4$ leads to comparable contributions from the two mechanisms, with the collinear contribution slightly larger due to saturation in the target wave function for momenta below $Q_s$~\cite{Altinoluk:2011qy}.

To summarize, the high-$k$ contribution to the production cross section is distinct from the collinear channel, and should lead to a significant numerical effect, especially at high transverse momentum, as discussed in~\cite{Altinoluk:2011qy} . Such an effect indeed has been observed in an early numerical work~\cite{Jalilian-Marian:2011tvq}, albeit the framework of that study was the same as~\cite{Altinoluk:2011qy} and therefore not complete.

The present work suggests also  importance of an additional mechanism of production, which has not been discussed either in~\cite{Altinoluk:2011qy}, or in any of the subsequent works. Referring back to eq.~\eqref{eq:mytotal4}, we observe that the integration region also contains the region $|k|\ll |p|$, $|l|\approx |p|/\zeta$. This is the hard perturbative fragmentation of a low transverse momentum parton emerging from the scattering with the target. The naive collinear framework assumes that no hard momentum is produced in the fragmentation process. We showed here that this is not the case for the observable at hand, and 
that this process needs to be taken into account in order to resum the large logarithms. Indeed, the same estimate as above suggest that the high momentum behavior of this channel is determined by the TMD FF, which perturbatively decays as $1/l^2$. Thus there is no reason to expect that this
 process is suppressed, and the  large transverse momentum TMD FF should contribute to particle production on par with the other two mechanisms discussed above.

The previous discussion mostly pertains to the situation when $p^2\gg Q_s^2$ since phenomenologically this is the relevant kinematic region at RHIC and LHC.  However, in principle our derivation did not assume this and one could consider a hypothetical situation when $Q_s^2\gg p^2\gg \Lambda_{QCD}^2$. What happens in this case? As discussed in the introduction, the main difference is that the transverse resolution scale on TMDs is now set by $Q_s$ rather than $p$. Superficially this is similar to the "threshold resummation" considered in~\cite{Xiao:2018zxf,Shi:2021hwx} 
where the resolution scale on the {\it collinear} PDF and FF was taken to be the higher of the scales $Q_s^2$ and $p^2$ (up to a small correction induced by the running of the coupling). However, the similarity is only superficial. 

First, the authors of~\cite{Shi:2021hwx} were not able to resum all the logarithms into the variation of the resolutions scale but in addition required a Sudakov resummation of part of the remaining logarithms\footnote{Our understanding is that even with the putative Sudakov resummation, the final result of~\cite{Shi:2021hwx} still contains a remaining large transverse logarithm which the authors declared to be a small NLO correction.} which we do not find necessary. Second, physically the TMD and collinear factorizations are very different even for the same choice of the resolution scale.
In the collinear picture, for $Q_s^2\gg p^2$, choosing $\mu^2\approx Q_s^2$ amounts to the assumption that all the incoming projectile partons with transverse momenta up to $Q_s^2$ scatter on the target with the same amplitude $s(p)$. On the other hand, in the TMD picture a projectile parton enters the interaction with arbitrary momentum $k$ and scatters with amplitude $s(p-k)$ to produce the outgoing parton with momentum $p$ (we disregard  fragmentation in this qualitative discussion).
 In the TMD picture the scattering is therefore, in principle, sensitive to the behavior of the dipole amplitude at all momenta rather than just at momentum of the final state particle. In practice, of course, momenta $(k-p)^2\gg Q_s^2$ do not contribute significantly, since $s(k-p)$ is small in this region. However, any variation of $s(k-p)$ for $(k-p)^2\lesssim Q_s^2$ will affect the result by a factor of order unity, which is not a small power correction. For example, with the naive dependence of the TMD PDF on momentum, ${\cal T}(k)\sim \Lambda_{QCD}^2/k^2$, the logarithmic integral over $k$ is dominated by  momenta of order $k^2\sim \sqrt{Q_s^2\Lambda_{QCD}^2}$. The momentum transfer in the scattering amplitude  is then dominated by the typical values $(k-p)^2\sim p^2+\sqrt{Q_s^2\Lambda^2_{QCD}}$ and  depends nontrivially on the relative value of momentum $p$ and the semihard scale $\sqrt{Q_s^2\Lambda^2_{QCD}}$. The collinear limit is only recovered if one assumes that $s(k-p)=const.$ for $(k-p)^2<Q_s^2$.

We note that we have used here an entirely perturbative definition of the TMDs since only their high momentum structure is pertinent for the physics at hand. These are  also commonly used in perturbative parton branching calculations~\cite{Hautmann:2017fcj,Martinez:2023azt}\footnote{We thank Peter Taels for pointing this out to us.}.

Finally the important question is whether the resulting cross section we obtained is positive for all transverse momenta. The crucial point here is that in our final result all the large logarithms have been resummed into the scale dependence of the TMDs, and none of the remaining genuine perturbative corrections are suspiciously large. Although the sign of some of these corrections are not obvious to determine without explicit calculation, there is no reason to expect that  they will give rise to unnaturally large negative contributions. We are thus confident that the problem of negative production cross section should not arise. Of course to put this beyond any doubt, a numerical implementation of our results is necessary. We leave all these aspects for future work.

\section*{Acknowledgements}

 Special thanks are due to Guillaume Beuf who participated in early stages of this work. We also thank Jamal Jalilian-Marian,Tuomas Lappi, Yair Mulian, Pieter Taels and other participants of the ECT* workshop {\it Color Glass Condensate at the Electron-Ion Collider}, and Cyrille Marquet and Bowen Xiao, for useful discussions. AK thanks the Physics Departments of the Ben Gurion University of the Negev and Tel Aviv University for hospitality while this work was being completed. 
NA has received financial support from Xunta de Galicia (Centro singular de investigaci\'on de Galicia
accreditation 2019-2022), from the European Union ERDF, and from the Spanish Research State Agency under project PID2020-119632GB-I00.
AK is supported by the NSF Nuclear Theory grant 2208387. 
ML is supported by the Binational Science Foundation grant \#2021789.
This material is based upon work supported by the U.S. Department of
Energy, Office  of Science, Office of Nuclear Physics through the Saturated Glue (SURGE)
Topical Collaboration.
This work has been performed in the framework
of the European Research Council project ERC-2018-ADG-835105 YoctoLHC and the MSCA RISE 823947 "Heavy
ion collisions: collectivity and precision in saturation physics" (HIEIC), and has received funding from the European
Union's Horizon 2020 research and innovation program under grant agreement No. 824093.

\appendix

\section{Evolution equations}
\label{appendixA}

\subsection{Evolution equations for the TMDs}

The definition of the TMDs we use in this paper is purely perturbative, since the nonperturbative region of small momenta in our calculation is not important and it is the high momentum region of the TMDs that gives the contribution not previously accounted for. In this appendix we list the evolution equations satisfied by these perturbative TMDs.

In our setup, \eqref{eq:pdftmdmu} defines the evolution of the quark TMD with respect to the transverse resolution parameter $\mu^2$, which can be obtained simply by differentiating the equation. From \eqref{eq:pdftmdmu} 
\beq
\label{eq:evolmu2}
\frac{\partial x{\cal T}_q(x,k^2;\mu^2; \xi_0)}{\partial \ln \mu^2}=\mu^2\delta\left(\mu^2-k^2\right)x{\cal T}_q(x,k^2;k^2; \xi_0) -
\frac{\alpha_s}{2\pi}\frac{N_c}{2}\int_{\xi_0}^{1}d\xi\frac{1+(1-\xi)^2} {\xi}\,x\,{\cal T}_q\left(x,k^2;\mu^2; \xi_0\right),
\eeq
where the delta function acts on the virtuality, not on the transverse momentum. 
The longitudinal resolution scale in the RHS of  eq.~\eqref{eq:evolmu2} has been taken independent of $\mu^2$ but in reality it should be taken as $\xi_0=\mu^2/(xs_0)$, and the evolutions in $\mu^2$ and $\xi_0$ would therefore become interlinked. Note that 
 $\int_{\xi_0}^{1}d\xi\frac{1+(1-\xi)^2} {\xi}=2\ln\frac{1}{\xi_0}-\frac{\xi_0^2}{2}+2\xi_0-\frac{3}{2}$.

In addition it is straightforward to write down the evolution with respect to the longitudinal resolution parameter. The evolution equation is akin to the Balitsky--Fadin--Kuraev--Lipatov (BFKL) equation~\cite{Kuraev:1977fs,Balitsky:1978ic}, although for the quark TMD it is actually a little simpler. 
One considers  the probability for a quark with transverse momentum $k$ to emit a gluon with longitudinal momentum fraction $\xi_0$ and transverse momentum $p$ with the usual collinear splitting function.
Allowing additional such emissions with the longitudinal fraction between $\xi_0$ and $\xi_0(1-\epsilon)$ and using \eqref{eq:tmdd1pp}, \eqref{eq:pdftmdmu} and \eqref{eq:reltmdpdf} we obtain
\beq
\label{eq:evolxi0}
\frac{\partial x{\cal T}_q(x,k^2;\mu^2; \xi_0)}{\partial \ln \frac{1}{\xi_0}}&=& \frac{g^2}{(2\pi)^3}\frac{N_c}{2}\,\left[1+(1-\xi_0)^2\right]\Bigg\{-\int_{k^2}^{\mu^2}\frac{\pi dl^2}{l^2}\,x\,{\cal T}_q\left(x,k^2;l^2; \xi_0\right)\nonumber \\
&&\hskip 4.1cm+\frac{1}{k^2}\,\frac{x}{1-\xi_0} \,f^q_{k^2}\left(\frac{x}{1-\xi_0}\right)\Bigg\}\nonumber\\
&&+\frac{\alpha_s}{2\pi}
\frac{N_c}{2}\int_{k^2}^{\mu^2}\frac{\pi dl^2}{l^2}\int_{\xi_0}^{1}d\xi\frac{1+(1-\xi)^2} {\xi}\,\frac{\partial x{\cal T}_q(x,k^2;l^2; \xi_0)}{\partial \ln \frac{1}{\xi_0}}\nonumber \\
&=& \frac{g^2\pi}{(2\pi)^3}\frac{N_c}{2}\,\left[1+(1-\xi_0)^2\right]\Bigg\{-\int_{k^2}^{\mu^2}\frac{dl^2}{l^2}\,x\,{\cal T}_q\left(x,k^2;l^2; \xi_0\right)\nonumber \\
&&\hskip 4.1cm+\frac{1}{k^2}\int_0^{k^2} dl^2 \,\frac{x}{1-\xi_0} \,{\cal T}_q\left(\frac{x}{1-\xi_0},l^2;k^2;\xi_0\right)\Bigg\}\nonumber\\
&&+\frac{g^2}{(2\pi)^3}\frac{N_c}{2}\int_{k^2}^{\mu^2}\frac{\pi dl^2}{l^2}\int_{\xi_0}^{1}d\xi\frac{1+(1-\xi)^2} {\xi}\,\frac{\partial x{\cal T}_q(x,k^2;l^2; \xi_0)}{\partial \ln \frac{1}{\xi_0}}\,.
\eeq
To order $\alpha_s$ the last term should be ignored. In addition, since we are only interested in $\xi_0\ll 1$
we set $1-\xi_0\to 1$ everywhere except in the argument of the TMD,  to get
\beq
\label{eq:evolxi01st}
\frac{\partial x{\cal T}_q(x,k^2;\mu^2; \xi_0)}{\partial \ln \frac{1}{\xi_0}}&=& 2\,\frac{g^2\pi}{(2\pi)^3}\frac{N_c}{2}\Bigg\{-\int_{k^2}^{\mu^2}\frac{dl^2}{l^2}\,x\,{\cal T}_q\left(x,k^2;l^2; \xi_0\right)\nonumber \\
&&\hskip 2.0cm+\frac{1}{k^2}\int_0^{k^2} dl^2 \,x {\cal T}_q\left(x,l^2;k^2;\xi_0\right)\Bigg\}.
\eeq

The evolution equations for the quark TMD FF and gluon TMDs can be derived in a similar manner.

\subsection{DGLAP evolution equations of the collinear distributions}

Given the evolution of the TMD PDFs we can explicitly check that the collinear PDF defined as eq.~\eqref{eq:reltmdpdf} satisfies the DGLAP evolution equation. We use
~\eqref{eq:reltmdpdf},~\eqref{eq:reltmdgpdf},~\eqref{eq:tmdd1pfull},~\eqref{eq:pdftmdmufull},~\eqref{eq:tmdg1pfull} and~\eqref{eq:pdftmdgmufull}, to obtain for the quark and gluon PDF
\beq
\label{eq:dglapq}
\frac{dxf^q_{\mu^2}(x)}{d\mu^2}&=&\pi x{\cal T}_q(x,\mu^2;\mu^2; \xi_0)+\int_0^{\mu^2} \pi dk^2 \,\frac{d}{d\mu^2} x{\cal T}_q(x,k^2;\mu^2; \xi_0)\nonumber\\
&=&\pi\,\frac{g^2}{(2\pi)^3}\frac{N_c}{2}\int_{\xi_0}^{1}d\xi\frac{1+(1-\xi)^2}{\xi}\, \frac{x}{1-\xi}\,f^q_{\mu^2}\left(\frac{x}{1-\xi}\right)\frac{1}{\mu^2}\nonumber \\
&&+\pi\,\frac{g^2}{(2\pi)^3}\frac{1}{2}\int_{\xi_0}^{1}d\xi \left[\xi^2+(1-\xi)^2\right]\frac{x}{1-\xi}\,f^g_{\mu^2}\left(\frac{x}{1-\xi}\right)\frac{1}{\mu^2}\nonumber\\
&&- \pi\,\frac{g^2}{(2\pi)^3}\frac{N_c}{2}\int_{\xi_0}^{1}d\xi\frac{1+(1-\xi)^2}{\xi}\, x\frac{1}{\mu^2}\int_0^{\mu^2} \pi dk^2 \,{\cal T}_q\left(x,k^2;\mu^2; \xi_0\right)\nonumber \\
&=&\ \pi\,\frac{g^2}{(2\pi)^3}\frac{N_c}{2}\int_{\xi_0}^{1}d\xi\left[\frac{1+(1-\xi)^2}{\xi}\right]_+\, \frac{x}{1-\xi}\,f^q_{\mu^2}\left(\frac{x}{1-\xi}\right)\frac{1}{\mu^2}\nonumber \\
&&+\pi\,\frac{g^2}{(2\pi)^3}\frac{1}{2}\int_{\xi_0}^{1}d\xi \left[\xi^2+(1-\xi)^2\right]\frac{x}{1-\xi}\,f^g_{\mu^2}\left(\frac{x}{1-\xi}\right)\frac{1}{\mu^2}
\eeq
and
\beq
\label{eq:dglapg}
\frac{dxf^g_{\mu^2}(x)}{d\mu^2}&=&\pi x{\cal T}_g(x,\mu^2;\mu^2; \xi_0)+\int_0^{\mu^2} \pi dk^2 \,\frac{d}{d\mu^2} x{\cal T}_g(x,k^2;\mu^2; \xi_0)\nonumber\\
&=&\frac{g^2}{(2\pi)^3}\,2N_c\int_{\xi_0}^{1}d\xi\left[\frac{1-\xi}{\xi}+\frac{\xi}{1-\xi}+\xi(1-\xi)\right]\, \frac{x}{1-\xi}\,f^g_{\mu^2}\left(\frac{x}{1-\xi}\right)\frac{1}{\mu^2}\nonumber \\
&&+\frac{g^2}{(2\pi)^3}\frac{N_c}{2}\sum_q \int_{\xi_0}^{1}d\xi \,\frac{1+\xi^2}{1-\xi}\,\frac{x}{1-\xi}\left[f^q_{\mu^2}\left(\frac{x}{1-\xi}\right)+f^{\bar q}_{\mu^2}\left(\frac{x}{1-\xi}\right)\right]\frac{1}{\mu^2} \nonumber \\
&&- \pi \frac{g^2}{(2\pi)^3}\, N_c\int_{\xi_0}^{1}d\xi\left[\frac{1-\xi}{\xi}+\frac{\xi}{1-\xi}+\xi(1-\xi)\right]\,x\frac{1}{\mu^2}\,\int_0^{\mu^2} \pi dk^2 \,{\cal T}_g\left(x,k^2;\mu^2; \xi_0\right)\nonumber \\
&&- \pi\frac{g^2}{(2\pi)^3}\, \frac{n_f}{2}\int_{\xi_0}^{1}d\xi\,[\xi^2+(1-\xi)^2]\,x\,\frac{1}{\mu^2}\int_0^{\mu^2} \pi dk^2 \,{\cal T}_g\left(x,k^2;\mu^2; \xi_0\right)\nonumber \\
&=&\frac{g^2}{(2\pi)^3}\,2N_c\int_{\xi_0}^{1}d\xi\left[\left\{\frac{1-\xi}{\xi}+\frac{1}{2}\xi(1-\xi)\right\}_+ +\frac{\xi}{1-\xi}+\frac{1}{2}\xi(1-\xi)\right]\, \frac{x}{1-\xi}\,f^g_{\mu^2}\left(\frac{x}{1-\xi}\right)\frac{1}{\mu^2}\nonumber \\
&&- \pi\frac{g^2}{(2\pi)^3}\, \frac{n_f}{2}\int_{\xi_0}^{1}d\xi\,[\xi^2+(1-\xi)^2]\,xf_{\mu^2}^g(x)\,\frac{1}{\mu^2}\nonumber \\
&&+\frac{g^2}{(2\pi)^3}\frac{N_c}{2}\sum_q\int_{\xi_0}^{1}d\xi \,\frac{1+\xi^2}{1-\xi}\,\frac{x}{1-\xi}\left[f^q_{\mu^2}\left(\frac{x}{1-\xi}\right)+f^{\bar q}_{\mu^2}\left(\frac{x}{1-\xi}\right)\right]\frac{1}{\mu^2}\,.
\eeq
Here we have employed a somewhat nonstandard definition of the $+$ prescription. Using  
\begin{equation}2 N_c \left[\left\{\frac{1-\xi}{\xi}+\frac{1}{2}\xi(1-\xi)\right\}_+ +\frac{\xi}{1-\xi}+\frac{1}{2}\xi(1-\xi)\right]=2 N_c \left[\frac{1-\xi}{[\xi]_+}+\frac{\xi}{1-\xi}+\xi(1-\xi)\right]+\frac{11N_c}{6}\delta(\xi)
\end{equation}
and $\int_{0}^{1}d\xi\,[\xi^2+(1-\xi)^2]=\frac{2}{3}$ one can verify that our expression is identical to  the standard gluon-to-gluon splitting function. The equations for collinear FFs are identical to this order. This demonstrates that the definition of TMDs we use here are consistent with the standard DGLAP evolution of collinear PDFs and FFs.

\section{$q\to q\to H$ channel}
\label{appb}
In this Appendix we give more details of the derivations in Section III.
\subsection{The real part}
In the real part we have the rational function of momenta multiplying an expression symmetric under the interchange $k\leftrightarrow q$ and integrated over both momenta. 
The crucial observation is that we can use the symmetry $k\leftrightarrow q$ and renaming momenta in part of the terms, we can write
\begin{eqnarray}
\label{awesome}
\frac{[k-(1-\xi)q]^2}{\left(\frac{p}{\zeta}-k\right)^2\left(\frac{p}{\zeta}-(1-\xi)q\right)^2}&\rightarrow&\left[\frac{\frac{p}{\zeta}-k}{\left(\frac{p}{\zeta}-k\right)^2}-\frac{\frac{p}{\zeta}-(1-\xi)q}{\left(\frac{p}{\zeta}-(1-\xi)q\right)^2}\right]^2\\
&\rightarrow&\left[\frac{\frac{p}{\zeta}-k}{\left(\frac{p}{\zeta}-k\right)^2}-\frac{\frac{p}{\zeta}-(1-\xi)k}{\left(\frac{p}{\zeta}-(1-\xi)k\right)^2}\right]\left[\frac{\frac{p}{\zeta}-q}{\left(\frac{p}{\zeta}-q\right)^2}-\frac{\frac{p}{\zeta}-(1-\xi)q}{\left(\frac{p}{\zeta}-(1-\xi)q\right)^2}\right]\nonumber\\
&&+\frac{1}{2}\left[\frac{\frac{p}{\zeta}-k}{\left(\frac{p}{\zeta}-k\right)^2}-\frac{\frac{p}{\zeta}-q}{\left(\frac{p}{\zeta}-q\right)^2}\right]\left[\frac{\frac{p}{\zeta}-k}{\left(\frac{p}{\zeta}-k\right)^2}-\frac{\frac{p}{\zeta}-q}{\left(\frac{p}{\zeta}-q\right)^2}\right]\nonumber\\
&&+\frac{1}{2}\left[\frac{\frac{p}{\zeta}-(1-\xi)k}{\left(\frac{p}{\zeta}-(1-\xi)k\right)^2}-\frac{\frac{p}{\zeta}-(1-\xi)q}{\left(\frac{p}{\zeta}-(1-\xi)q\right)^2}\right]\left[\frac{\frac{p}{\zeta}-(1-\xi)k}{\left(\frac{p}{\zeta}-(1-\xi)k\right)^2}-\frac{\frac{p}{\zeta}-(1-\xi)q}{\left(\frac{p}{\zeta}-(1-\xi)q\right)^2}\right]\nonumber\\
&\rightarrow&\left[\frac{\frac{p}{\zeta}-k}{\left(\frac{p}{\zeta}-k\right)^2}-\frac{\frac{p}{\zeta}-(1-\xi)k}{\left(\frac{p}{\zeta}-(1-\xi)k\right)^2}\right]\left[\frac{\frac{p}{\zeta}-q}{\left(\frac{p}{\zeta}-q\right)^2}-\frac{\frac{p}{\zeta}-(1-\xi)q}{\left(\frac{p}{\zeta}-(1-\xi)q\right)^2}\right]\nonumber\\
&&+\frac{1}{2}\left[\frac{(k-q)^2}{\left(\frac{p}{\zeta}-k\right)^2\left(\frac{p}{\zeta}-q\right)^2}+\frac{(1-\xi)^2(k-q)^2}{\left(\frac{p}{\zeta}-(1-\xi)k\right)^2\left(\frac{p}{\zeta}-(1-\xi)q\right)^2}\right].\nonumber
\end{eqnarray}
Then the real NLO term can be written in the form
\beq\label{sigmafinal1rft}
\frac{d{\sigma}_{1,r}^{q\to q\to H}}{d^2p d\eta}&=&S_\perp\frac{g^2}{(2\pi)^3}\frac{N_c}{2}
\int_{x_F}^1 \frac{d\zeta}{\zeta^2}\,D_{H,\mu_0^2}^q(\zeta)\ \int_{k,q} \int_{\zeta k^2/(x_Fs_0)}^{1} d\xi  \,\frac{1+(1-\xi)^2}{\xi} \ \frac{x_F}{(1-\xi) \zeta} f_{\mu_0^2}^q\left(\frac{x_F}{(1-\xi) \zeta}\right)\noindent \\
&& \times\  s(k)s(q)\,\left\{\left[\frac{\frac{p}{\zeta}-k}{\left(\frac{p}{\zeta}-k\right)^2}-\frac{\frac{p}{\zeta}-(1-\xi)k}{\left(\frac{p}{\zeta}-(1-\xi)k\right)^2}\right]\left[\frac{\frac{p}{\zeta}-q}{\left(\frac{p}{\zeta}-q\right)^2}-\frac{\frac{p}{\zeta}-(1-\xi)q}{\left(\frac{p}{\zeta}-(1-\xi)q\right)^2}\right]\right.\nonumber\\
&&\hskip 2cm \left.+\frac{1}{2}\left[\frac{(k-q)^2}{\left(\frac{p}{\zeta}-k\right)^2\left(\frac{p}{\zeta}-q\right)^2}+\frac{(1-\xi)^2(k-q)^2}{\left(\frac{p}{\zeta}-(1-\xi)k\right)^2\left(\frac{p}{\zeta}-(1-\xi)q\right)^2}\right]\right\}.\nonumber
\eeq

With some changes of variables, and taking $p \to -p$\footnote{Here we assume  parity invariance of the dipole amplitudes $s(q)=s(-q)$, which we have already used before.}, the first term in the third line in~\eqref{sigmafinal1rft} reads
\beq
\int_{k,q}s(k)s(q) \frac{1}{2}\,\frac{(k-q)^2}{\left(\frac{p}{\zeta}-k\right)^2\left(\frac{p}{\zeta}-q\right)^2}=\int_{k,q} \frac{1}{k^2} \,s\left(-k+\frac{p}{\zeta}\right) \left[ 1-\frac{k\cdot q}{q^2} \right]\, s\left(-q+\frac{p}{\zeta}\right).
\eeq

The second term in the third line in~\eqref{sigmafinal1rft} can be written
\beq
&&\frac{g^2}{(2\pi)^3} S_\perp \frac{N_c}{2} \int_{x_F}^1 \frac{d\zeta}{\zeta^2}\, D_{H,\mu^2}^q(\zeta) \int_{k} \int_{k^2\zeta/(x_Fs_0)}^1 d\xi \,\frac{x_F}{\zeta(1-\xi)} \,f_{\mu^2}^q\left( \frac{x_F}{\zeta(1-\xi)}\right)\, \frac{1+(1-\xi)^2}{\xi} \nonumber\\
&&\hskip 3cm \times\frac{1}{(1-\xi)^2}\,\frac{1}{k^2}\, s\left(-k+\frac{p}{\zeta(1-\xi)}\right)\left[1-\int_q \frac{k\cdot q}{q^2}\,s\left(-q+\frac{p}{\zeta(1-\xi)}\right)\right]. \label{eq:third}
\eeq
Noting that
\beq
\int_{x_F}^1d\zeta \int_{k^2\zeta/(x_Fs_0)}^1 d\xi = \int_{k^2/s_0}^1 d\xi \int_{x_F}^{\mathrm{min}[x_Fs_0\xi/k^2,1]} d\zeta
\eeq
and making the change of variables $\zeta^\prime=\zeta(1-\xi)$ we get
\beq
&&\frac{g^2}{(2\pi)^3} S_\perp \frac{N_c}{2} \int_k \int_{k^2/s_0}^1 \frac{d\xi}{1-\xi} \int_{x_F(1-\xi)}^{\mathrm{min}[x_Fs_0\xi(1-\xi)/k^2,1-\xi]} \frac{d\zeta}{\zeta^2}\, D_{H,\mu^2}^q\left(\frac{\zeta}{1-\xi}\right) \,\frac{x_F}{\zeta} \,f_{\mu^2}^q\left( \frac{x_F}{\zeta}\right)\, \frac{1+(1-\xi)^2}{\xi} \nonumber\\
&&\hskip 3cm \times\int_q\frac{1}{k^2}\, s\left(-k+\frac{p}{\zeta}\right)\left[1- \frac{k\cdot q}{q^2}\right]\,s\left(-q+\frac{p}{\zeta}\right). \label{eq:thirdp1}
\eeq
We now change the order of integrations. For $k^2/(x_F s_0)\ll 1$ where  our approach holds, this leads to 
\beq
\int_{k^2/s_0}^1 \frac{d\xi}{1-\xi} \int_{x_F(1-\xi)}^{\mathrm{min}[x_Fs_0\xi(1-\xi)/k^2,1-\xi]} \frac{d\zeta}{\zeta^2}[\cdots]=\int_{0}^{1-k^2/(x_F s_0)}\frac{d\zeta}{\zeta^2}\int_{k^2\zeta/(x_F s_0)}^{1-\zeta} \frac{d\xi}{1-\xi}[\cdots]\ .
\eeq
When this change of integration order is performed in eq.~\eqref{eq:thirdp1}, due to finite support of $f_{\mu^2}^q\left( \frac{x_F}{\zeta}\right)$, the lower limit on the  $\zeta$ integration becomes $x_F$.
Therefore~\eqref{eq:thirdp1} becomes
\beq
&&\frac{g^2}{(2\pi)^3} S_\perp \frac{N_c}{2} \int_k \int_{x_F}^{1}\frac{d\zeta}{\zeta^2}\int_{k^2\zeta/(x_F s_0)}^{1-\zeta} \frac{d\xi}{1-\xi}\, D_{H,\mu^2}^q\left(\frac{\zeta}{1-\xi}\right) \,\frac{x_F}{\zeta} \,f_{\mu^2}^q\left( \frac{x_F}{\zeta}\right)\, \frac{1+(1-\xi)^2}{\xi} \nonumber\\
&&\hskip 3cm \times\frac{1}{k^2}\, s\left(-k+\frac{p}{\zeta}\right)\left[1-\int_q \frac{k\cdot q}{q^2}\,s\left(-q+\frac{p}{\zeta}\right)\right] \nonumber \\
&&=\frac{g^2}{(2\pi)^3} S_\perp \frac{N_c}{2} \int_k \int_{x_F}^{1}\frac{d\zeta}{\zeta^2}\int_{k^2\zeta/(x_F s_0)}^{1} \frac{d\xi}{1-\xi}\, D_{H,\mu^2}^q\left(\frac{\zeta}{1-\xi}\right) \,\frac{x_F}{\zeta} \,f_{\mu^2}^q\left( \frac{x_F}{\zeta}\right)\, \frac{1+(1-\xi)^2}{\xi} \nonumber\\
&&\hskip 3cm \times\frac{1}{k^2}\, s\left(-k+\frac{p}{\zeta}\right)\left[1-\int_q \frac{k\cdot q}{q^2}\,s\left(-q+\frac{p}{\zeta}\right)\right] ,\label{eq:thirdp2}
\eeq
where in the equality we have used the fact that the upper limit of the integral can be set to 1 due to the support of the collinear FF.

Now, using~\eqref{eq:tmdd1pp}, we can write~\eqref{sigmafinal1rft} as~\eqref{eq:myreal2}.

\subsection{The virtual part}
Here we only consider the term that potentially contains logarithms, i.e.,  eq.~\eqref{eq:myvirrtual2}.

First, eq.~\eqref{eq:myvirrtual2}  tells us that the relevant factorization scale is determined by the values of momentum $q$ that dominate the integral.
Those are indeed easy to understand. The integral is dominated by the values of $q$ for which the argument of the dipole should be close to $Q_s$. This means that if $p/\zeta\gg Q_s$, on average $q\sim p/\zeta$, while if $p/\zeta\ll Q_s$ on average $q\sim Q_s$. This is precisely our conjectured values of the factorization scale~\eqref{thescales1}. 
The second logarithm in~\eqref{eq:myvirrtual2} is never large. It does not contain neither transverse nor longitudinal logarithms, just like the similar term in the real contribution. So it is simply a genuine small perturbative correction which does not need to be resummed.

As shown in Sec.~\ref{eq:qtqth} the first term in  eq.~\eqref{eq:myvirrtual2} contributes to the evolution of the TMDs with the transverse resoluton scale, and disappears when this scale is set to $\mu^2$. The second term still remains.
The natural question to ask is whether this contribution can be treated as a small perturbative correction, or it does contain a large logarithm. It is clear that there is no large transverse logarithm in the game. But what about longitudinal?
The integral we have to analyze is
\beq
\int_q\int_{\mu^2}^{(q-\frac{1}{\zeta} p)^2} \frac{d^2k}{k^2} \int_{k^2\zeta/(x_Fs_0)}^1 d\xi \, \frac{1+(1-\xi)^2}{\xi}  \,s(q).
\eeq

We observe that the integration domain $\{\mu_0^2<\mu^2<k^2<(q-\frac{1}{\zeta} p)^2;\ k^2/\bar s_0 <\xi <1\}$ is equivalent to $\{\mu^2/\bar s_0 <\xi <1,\mu_0^2<\mu^2<k^2<\min [(q-\frac{1}{\zeta} p)^2,\xi \bar s_0]\}$, giving
\beq\label{suspicious}
&&\int_q \int_{\mu^2/s_0}^1 d\xi \, \frac{1+(1-\xi)^2}{\xi}  \int_{\mu^2}^{\min [(q-\frac{1}{\zeta} p)^2,\xi \bar s_0]}
 \frac{d^2k}{k^2}\,s(q)\\
 &&=\pi \int_{\mu^2/\bar s_0}^1 d\xi \,\, \frac{1+(1-\xi)^2}{\xi}  \left[\int_{\mu_0^2}^{\xi \bar s_0} d^2q \ln\frac{q^2}{\mu^2}\,s(q+\bar p)+\int_{\xi\bar s_0}^{\infty} d^2q \ln\frac{\xi \bar s_0}{\mu^2}\,s(q+\bar p)
\right],\nonumber
 \eeq
where
$\bar p=p/\zeta$ and $\bar s_0=x_F s_0/\zeta$.
Assuming that the dipole decreases fast at momenta larger than $Q_s$, and $s(Q_s)\sim 1/Q_s^2$, we see that the integral over $q$ is dominated by the values $q^2\sim Q_s^2+\bar p^2$. Thus the transverse momentum integral is 
\beq
&&\int_{\mu_0^2}^{\xi \bar s_0} d^2q \ln\frac{q^2}{\mu^2}\,s(q+\bar p)+\int_{\xi\bar s_0}^{\infty} d^2q \ln\frac{\xi \bar s_0}{\mu^2}\,s(q+\bar p)\nonumber\\
&&\approx\ln\frac{Q_s^2+\bar p^2}{\mu^2}\theta\Big(\xi\bar s_0-(Q_s^2+\bar p^2)\Big)+\ln\frac{\xi\bar s_0}{\mu^2}\theta\Big((Q_s^2+\bar p^2)-\xi\bar s_0\Big).
\eeq
Now, integrating over $\xi$\footnote{Here we make the low $\xi$ approximation, since only the small values of $\xi$ can potentially lead to a logarithmic integral.} we obtain for the integral in 
eq.~\eqref{suspicious}
\beq 
\ln^2\frac{Q_s^2+\bar p^2}{\mu^2}+\ln\frac{s_0}{Q_s^2+\bar p^2}\ln\frac{Q_s^2+\bar p^2}{\mu^2}\ .
\eeq
The transverse logarithm is not large, since we have chosen $\mu^2$ to be close to $Q_s^2+\bar p^2$. In fact, at least in the approximation considered here we can choose $\mu^2$ such that it vanishes. However even if we do not recourse to such fine tuning, we can see that eq.~\eqref{suspicious} is not dangerously large. The only question here is about the longitudinal logarithm $\ln\frac{s_0}{Q_s^2+\bar p^2}$. Recall that our choice of  $s_0$ is such that although the ratio $\frac{s_0}{Q_s^2+\bar p^2}$ is large, its logarithm is not a large number. If that is the case, this logarithm is also under control. In fact we can always change $s_0$ by evolving the dipole $s(p)$ through a larger or smaller rapidity interval, see Appendix~\ref{app:BK}. The only reason we do not choose $s_0\sim Q_s^2+\bar p^2$, is that then we will have in our projectile wave function gluons with rather small longitudinal momentum, for which we will not be able to use the eikonal scattering approximation. Thus our choice of $s_0$ is the most appropriate, as it does not leave any large logarithms (after resummations discussed here) and also allows us to use eikonal approximation for the partonic scattering amplitude.

We conclude therefore that the second term in the last line of eq.~\eqref{eq:myvirrtual2} is not large with our choice of scales, and should be considered as a small genuine NLO correction.

\section{Energy evolution of the dipole}
\label{app:BK}

\subsection{Balitsky-Kovchegov evolution for $s(k)$}
First we derive the BK equation~\cite{Balitsky:1998kc,Balitsky:1998ya,Kovchegov:1999ua} for $s(k)$ in the translational invariant approximation 
(it is different from the usual momentum space BK evolution  used in  literature - the latter is not formulated for $s(k)$).
We start from
\beq
\frac{ds(x-y)}{dY}=2\frac{g^2}{(2\pi)^3}\frac{N_c}{2}\int_z  \frac{(x-y)^2}{(x-z)^2(z-y)^2}\left[s(x-z)s(z-y)-s(x-y)\right].
\eeq
Using \eqref{eq:dip_FT},
\beq
\label{eq:ffttp}
 \frac{r_i}{r^2}=\int_k \frac{1}{2\pi} e^{-ik\cdot{r}}i\frac{k_{\perp i}}{k_\perp^2}\
\eeq
and
\beq
\frac{(x-y)^2}{(x-z)^2(z-y)^2}=\left[\frac{x-z}{(x-z)^2}+ \frac{z-y}{(z-y)^2}\right]^2,
\eeq
we write
\beq
\int_l e^{il\cdot(x-y)}\frac{ds(l)}{dY}&=&-2\frac{g^2}{(2\pi)^3}\frac{N_c}{2}\int_z\int_{m,n,p,q} \frac{1}{(2\pi)^2}\left[ e^{-im\cdot(x-z)}+e^{-im\cdot (z-y)}\right] \frac{m_in_i}{m^2n^2}\left[ e^{-in\cdot(x-z)}+e^{-in\cdot (z-y)}\right] \nn\\
&\times&s(p) s(q) \left[e^{ip\cdot(x-z)}e^{iq\cdot(z-y)}-e^{ip\cdot(x-y)}\right].
\eeq
Changing variables $p\leftrightarrow q$ and $m\to -m$ in some terms and using our usual assumption of rotational invariance  $s(q)\equiv s(|q|)$, we get
\beq
\int_l e^{il\cdot(x-y)}\frac{ds(l)}{dY}=-4\frac{g^2}{(2\pi)^3}\frac{N_c}{2}\int_{m,p,q} s(p)s(q) \left[e^{i(p-m)\cdot(x-y)}-e^{ip\cdot(x-y)}\right] \left[\frac{m_i(q-p+m)_i}{m^2(q-p+m)^2}-\frac{1}{m^2}\right].
\eeq
Shifting $p\to p+m$, $q\to q+m$ in the terms multiplied by $e^{i(p-m)\cdot(x-y)}$, we get
\beq
\int_l e^{il\cdot(x-y)}\frac{ds(l)}{dY}=-4\frac{g^2}{(2\pi)^3}\frac{N_c}{2}\int_{m,p,q} e^{ip\cdot(x-y)}[s(p+m)s(q+m)-s(p)s(q)]\left[\frac{m_i(q-p+m)_i}{m^2(q-p+m)^2}-\frac{1}{m^2}\right].
\eeq
After renaming the variables, the BK evolution equation for $s(k)$ reads
\beq\
\frac{ds(k)}{dY}=-4\frac{g^2}{(2\pi)^3}\frac{N_c}{2}\int_{p,q} [s(k+p)s(q+p)-s(k)s(q)]\left[\frac{p_i(p+q-k)_i}{p^2(p+q-k)^2}-\frac{1}{p^2}\right].
\label{eq:momBK}
\eeq
Note that in the limits $p\to 0$, as well as  $p,q\to \infty$ the right hand side is finite if $s(k)\to 0$ for $k\to \infty$.

\subsection{Dipole evolution from $s_0$ independence}

Throughout the paper we have worked in the fixed frame, defined by the reference scale $s_0$, when most of the energy evolution is attributed to the target. 
This scale is arbitrary however  and the final result for the cross section must be independent  of $s_0$. The independence on $s_0$ should be achieved by varying the dipole $s(k)$ according to the leading order BK equation. 
We have not explicitly indicated in the body of the paper the dependence of  the dipole amplitude on $s_0$, but such dependence of course is implied throughout. Varying $s_0$ results in the change of the evolution interval for $s(k)$.
The independence on $s_0$ should hold modulo  power- and/or $\alpha_s$-suppressed corrections. The exact independence on $s_0$  may require including some NLO terms in the dipole evolution to account for $\alpha_s$ corrections and also a more exact treatment of power suppressed terms on the TMD factorized side than we have done so far. Our goal in this appendix is to show that modulo these terms, evolving $s(k)$ according to BK equation indeed ensures the independence on $s_0$.

 We note that $d\ln s_0=-d\ln \xi_0=-dY$.
Thus we should be able to derive an evolution  equation for $s(k)$ by imposing invariance of the cross section 
under the change of $s_0$:
\beq
s_0\frac{d}{ds_0}\frac{d{\sigma}^{q\to q\to H}}{d^2p d\eta}\,=\,0\hspace{1cm} \Longrightarrow\hspace{1cm} {ds(k)\over dY}\ .
\eeq
Since in order to arrive to the TMD factorized expressions we have modified some ${\cal O}(\alpha_s^2)$ and power-suppressed terms
(see the discussion below~\eqref{eq:discard}), we should go back to the original expression for the real and virtual pieces in~\eqref{eq:myreal2} and~\eqref{eq:myvirrtual2}, respectively\footnote{For the virtual part we complete the phase space for transverse momentum integration, thus restoring the resolution scale $\mu^2_0$ in the  TMD in the LO term. This actually provides additional confirmation that the virtual NLO remainder is not logarithmically large.
In the real part we use the TMD evolution in $\ln \xi_0$, but disregard the additional $\alpha_s^2$ terms we have added in the product of TMD PDF and FF as explained below eq.~\eqref{eq:hardfactor}.}. Keeping only terms that are ${\cal O}(\alpha_s)$ and that  contribute to logarithmic evolution with respect to $s_0$, we get
\beq
\label{eq:bkevol1}
&&\frac{d}{dY}\frac{d{\sigma}^{q\to q\to H}}{d^2p d\eta}=
S_\perp\int_{x_F}^1 \frac{d\zeta}{\zeta^2} \Biggl\{\int d^2k\Biggl[-\frac{d {\cal F}_H^q\left(\zeta,k^2;k^2;\xi_0=\frac{\zeta k^2}{x_F s_0}\right)}{d\ln\frac{1}{\xi_0}}\ \frac{x_F}{\zeta} f_{\mu_0^2}^q\left(\frac{x_F}{\zeta}\right)\\
&&\hskip 0.8cm -D_{H,\mu_0^2}^q(\zeta)\ \frac{d \frac{x_F}{\zeta} {\cal T}_q\left(\frac{x_F}{\zeta},k^2;k^2; \xi_0=\frac{\zeta k^2}{x_F s_0}\right) }{d\ln\frac{1}{\xi_0}}\Biggl] \int_q s\left(-k+\frac{p}{\zeta}\right)\left[1-\frac{k\cdot q}{q^2}\right]s\left(-q+\frac{p}{\zeta}\right)\nonumber \\
&&\hskip 0.3cm +D_{H,\mu_0^2}^q(\zeta)\frac{x_F}{\zeta}f_{\mu_0^2}^q\left(\frac{x_F}{\zeta}\right) \Biggl[\frac{d s\left(\frac{p}{\zeta}\right)}{dY}\nonumber \\
&&\hskip 2cm  + 2\frac{g^2}{(2\pi)^3} \frac{N_c}{2} \int_{\mu_0^2}^\infty d^2k 
 \int_q s\left(\frac{p}{\zeta}\right)\,s(q)\,\left[\frac{\frac{p}{\zeta}-q-k}{\left(\frac{p}{\zeta}-q-k\right)^2}\cdot\frac{k}{k^2}+\frac{1}{k^2}\right]   \left(1+(1-\lambda_0)^2\right) \Biggl]\Biggl\},\nonumber 
\eeq
with $\lambda_0=k^2\zeta/(x_Fs_0)$. In this equation the two first  lines correspond to the NLO real contributions, the third line to the LO one and the fourth line to the virtual NLO one. In the following we will set $1-\xi_0,1-\lambda_0\to 1$ except in the argument of the TMD distributions. We also consider that in the lower limit of transverse momentum integrals we can set $\mu_0^2\to 0$.

The virtual term reads
\beq
-4\frac{g^2}{(2\pi)^3} \frac{N_c}{2}
 \int_{k,q} s\left(\frac{p}{\zeta}\right)\,s(q)\,\left[\frac{k+q-\frac{p}{\zeta}}{\left(k+q-\frac{p}{\zeta}\right)^2}\cdot\frac{k}{k^2}-\frac{1}{k^2}\right],
\eeq
which is exactly the virtual part of~\eqref{eq:momBK} with the opposite sign.

Now we use~\eqref{eq:evolxi01st} for the evolution of the TMD PDFs and FFs in the real terms in~\eqref{eq:bkevol1}. Note that now the term with the minus sign in~\eqref{eq:evolxi01st} is identically 0. Therefore we get for the first (PDF) real piece
\beq
\label{eq:pppppqqqqq}
&&\int d^2k\Biggl[-D_{H,\mu_0^2}^q(\zeta)\ \frac{d \frac{x_F}{\zeta} {\cal T}_q\left(\frac{x_F}{\zeta},k^2;k^2; \xi_0=\frac{\zeta k^2}{x_F s_0}\right) }{d\ln\frac{1}{\xi_0}}\Biggl] \int_q s\left(-k+\frac{p}{\zeta}\right)\left[1-\frac{k\cdot q}{q^2}\right]s\left(-q+\frac{p}{\zeta}\right)\\
&=&-2\frac{g^2}{(2\pi)^3} \frac{N_c}{2} D_{H,\mu_0^2}^q(\zeta)\int_{k,q}\frac{1}{k^2}\int_0^{k^2}d^2l \ \frac{x_F}{\zeta} {\cal T}_q\left(\frac{x_F}{\zeta},l^2;k^2; \xi_0=\frac{\zeta k^2}{x_F s_0}\right) s\left(-k+\frac{p}{\zeta}\right)\left[1-\frac{k\cdot q}{q^2}\right]s\left(-q+\frac{p}{\zeta}\right)\nonumber \\
&=&2\frac{g^2}{(2\pi)^3} \frac{N_c}{2} D_{H,\mu_0^2}^q(\zeta)\frac{x_F}{\zeta}f_{\mu_0^2}^q\left(\frac{x_F}{\zeta}\right)\int_{k,q} s\left(k+\frac{p}{\zeta}\right)s\left(k+q\right)\left[\frac{k+q-\frac{p}{\zeta}}{\left(k+q-\frac{p}{\zeta}\right)^2}\cdot\frac{k}{k^2}-\frac{1}{k^2}\right], 
\nonumber
\eeq
where to go from the second to the third equality we have changed variables $p\to -p$ and $q+\frac{p}{\zeta}=q^\prime+k$ (and dropped the prime on $q^\prime$), used~\eqref{eq:reltmdpdf} and have taken the virtuality in the collinear PDF to be $\mu_0^2$ (which is justified since further evolution from $\mu_0^2\to k^2>\mu_0^2$ is an ${\cal O}(\alpha_s)$ effect).
 This is exactly one half of the real part of~\eqref{eq:momBK} with the opposite sign. Proceeding in complete analogy with the piece containing the evolution of the TMD FFs in~\eqref{eq:bkevol1}, we get exactly~\eqref{eq:pppppqqqqq} and, therefore, we complete the second half of the real part of~\eqref{eq:momBK} with the opposite sign.
 
 Therefore, by imposing invariance of the cross section under changes of $\ln s_0$ in~\eqref{eq:bkevol1}, we get the BK evolution of the LO dipoles. This constitutes a nontrivial check on the consistency of the evolution equations in $\xi_0$ for the TMDs in our framework, and further supports the absence, in our final results, of large logarithms beyond those contained in the TMD evolution.

\section{All channels - full result}
\label{appc}

\subsection{Real corrections at NLO}

Real corrections at NLO can be written as a sum of four different contributions:
\beq
\label{NLO_real_sch}
\frac{d\sigma}{d^2p\, d\eta}\bigg|^{\rm real}_{\rm NLO}=\frac{d\sigma}{d^2p\, d\eta}\bigg|^{\rm q\to q}_{\rm NLO, r}+
\frac{d\sigma}{d^2p\, d\eta}\bigg|^{\rm g\to q}_{\rm NLO, r}+\frac{d\sigma}{d^2p\, d\eta}\bigg|^{\rm g\to g}_{\rm NLO, r}+
\frac{d\sigma}{d^2p\, d\eta}\bigg|^{\rm q\to g}_{\rm NLO, r}.
\eeq
\subsubsection{Quark initiated quark production}
The first term on the right hand side of eq.~\eqref{NLO_real_sch} corresponds to quark production from quark initiated channel. It is computed in previous sections and it reads
\beq
\label{real_q_q_1}
\frac{d\sigma}{d^2p\, d\eta}\bigg|^{\rm q\to q}_{\rm NLO, r}&=&\frac{g^2}{(2\pi)^3}S_\perp
\int_{x_F}^1 \frac{d\zeta}{\zeta^2} \, D^q_{\mu_0^2}(\zeta) \int_{k^2,q^2>\mu_0^2}
\int_{\xi_0}d\xi \frac{x_F}{\zeta(1-\xi)}f^q_{\mu_0^2}\left(\frac{x_F}{\zeta(1-\xi)}\right)\, 
\frac{N_c}{2} \left[ \frac{1+(1-\xi)^2}{\xi}\right]s(k)s(q)\nonumber\\
&\times&
\bigg\{ \frac{1}{2} \frac{(q-k)^2}{(p/\zeta-k)^2(p/\zeta-q)^2} +\frac{1}{2}\frac{(1-\xi)^2(q-k)^2}{[p/\zeta-(1-\xi)k]^2[p/\zeta-(1-\xi)q]^2}\bigg\}+({\rm Gen. \, NLO})_1\ .
\eeq
The first term in eq.~\eqref{real_q_q_1} contributes to the quark TMD PDF and the second term contributes to the quark TMD FF. The remaining part is referred to as the genuine NLO contribution since it does not involve any large logarithms. Its explicit form reads
\beq
\label{Gen_NLO_1}
({\rm Gen.\;  NLO})_1&=&\frac{g^2}{(2\pi)^3}\, S_\perp\int \frac{d\zeta}{\zeta^2}D^q_{\mu_0^2}(\zeta)\int_{k,q}\int d\xi\, 
\frac{x_F}{\zeta(1-\xi)}\, f^q_{\mu_0^2}\left(\frac{x_F}{\zeta(1-\xi)}\right)\frac{N_c}{2}\left[ \frac{1+(1-\xi)^2}{\xi}\right]s(k)s(q)
\nonumber\\
&&\times\, 
\bigg[ \frac{(p/\zeta-k)^i}{(p/\zeta-k)^2}-\frac{(p/\zeta-(1-\xi) k)^i}{(p/\zeta-(1-\xi) k)^2}\bigg]
\bigg[ \frac{(p/\zeta-q)^i}{(p/\zeta-q)^2}-\frac{(p/\zeta-(1-\xi) q)^i}{(p/\zeta-(1-\xi) q)^2}\bigg].
\eeq
Let us rewrite eq.~\eqref{real_q_q_1} in a more convenient way. In the first term after shifting the transverse momenta, it can be written as
\beq
\label{rewriting}
\frac{1}{2}\int_{k,q}s(k)s(q)\frac{1}{2}\frac{(q-k)^2}{(p/\zeta-k)^2(p/\zeta-q)^2}
&=&
\frac{1}{2}\int_{k,q}s(p/\zeta+k)s(p/\zeta+q)\frac{1}{q^2k^2}(q^2-2k\cdot q+k^2)\nonumber\\
&=&\int_{k,q}\frac{1}{k^2}\, s(-k+p/\zeta)\left[ 1-\frac{k\cdot q}{q^2}\right]\, s(-q+p/\zeta),
\eeq
where we used the symmetry between $k$ and $q$. In the second term in eq.~\eqref{real_q_q_1}, after rescaling $\zeta(1-\xi)\to \zeta$ one can perform the same modifications described in eq.~\eqref{rewriting}. After all, the real correction in to the quark initiated quark production can be written as 
\beq
\label{q-q-f}
\frac{d\sigma}{d^2p\, d\eta}\bigg|^{\rm q\to q}_{\rm NLO, r}&=&\frac{g^2}{(2\pi)^3}S_\perp\int_{x_F}^1\frac{d\zeta}{\zeta^2}\, D^q_{\mu_0^2}(\zeta)
\int_{k^2>\mu_0^2}\int_{\xi_0}d\xi \frac{x_F}{\zeta(1-\xi)}f^q_{\mu_0^2}\left(\frac{x_F}{\zeta(1-\xi)}\right)
\frac{N_c}{2}\bigg[\frac{1+(1-\xi)^2}{\xi}\bigg]\, \frac{1}{k^2}\nonumber\\
&&\times\, \int_q s\big(-k+p/\zeta\big)\, \bigg[1-\frac{k\cdot q}{q^2}\bigg]\, s\big(-q+p/\zeta\big)\nonumber\\
&+&
\frac{g^2}{(2\pi)^3}S_\perp\int_{x_F}^1\frac{d\zeta}{\zeta^2}\frac{1}{(1-\xi)}D^q_{\mu_0^2}\left(\frac{\zeta}{1-\xi}\right)
\int_{k^2>\mu_0^2}\int_{\xi_0}d\xi \, \frac{x_F}{\zeta}f^q_{\mu_0^2}\left(\frac{x_F}{\zeta}\right)
\frac{N_c}{2}\bigg[\frac{1+(1-\xi)^2}{\xi}\bigg]\, \frac{1}{k^2}\nonumber\\
&&\times\, \int_q s\big(-k+p/\zeta\big)\, \bigg[1-\frac{k\cdot q}{q^2}\bigg]\, s\big(-q+p/\zeta\big)\nonumber\\
&+& ({\rm Gen.\;  NLO})_1\ ,
\eeq
where the explicit expression for the genuine NLO correction in this channel is given in eq.~\eqref{Gen_NLO_1}.
\subsubsection{Gluon initiated quark production}

In the large $N_c$ limit, the real contribution to NLO partonic cross section from the gluon initiated quark production reads (eq. (4.8) in Ref.~\cite{Altinoluk:2014eka})
\beq
&&
\frac{d\sigma}{d^2p\, d\eta}\bigg|^{\rm g\to q}_{\rm NLO, r}=\frac{g^2}{(2\pi)^3}\int_{\xi_0}d\xi \, \frac{x_p}{(1-\xi)} \, f^g_{\mu_0^2}\left(\frac{x_p}{1-\xi}\right)\frac{1}{2}[\xi^2+(1-\xi)^2]\int_{y{\bar y}z}e^{ip(y-\bar y)}A^i_{\xi}(y-z)A^i_{\xi}({\bar y}-z)
\\
&&\times
\Big\{ 
s(y,\bar y)+s_A[(1-\xi)y,(1-\xi)\bar y]-s[y, (1-\xi)\bar y+\xi z]s[(1-\xi)\bar y+\xi z, z]- s[z, (1-\xi)y+\xi z]s[(1-\xi)y+\xi z, \bar y]\Big\}. \nonumber
\eeq
Assuming translationally invariant dipoles and using the definitions for the modified WW field (eq.~\eqref{eq:ww}) and for the dipole operators in momentum space (eq.~\eqref{eq:dip_FT}), this contribution can be written as 
\beq
\frac{d\sigma}{d^2p\, d\eta}\bigg|^{\rm g\to q}_{\rm NLO, r}&=&\frac{g^2}{(2\pi)^3}\int_{\xi_0}d\xi \, \frac{x_p}{(1-\xi)} \, f^g_{\mu_0^2}\left(\frac{x_p}{1-\xi}\right)\frac{1}{2}[\xi^2+(1-\xi)^2]\frac{(-1)}{(2\pi)^4}\int_{y{\bar y}z}\int_{l,m>\mu_0}\int_{k,q}\frac{l^i}{l^2}\frac{m^i}{m^2}s(k)s(q) \nonumber\\
&&
\hspace{-1cm}
\times \Big[
e^{iy(p-l+k)}e^{-i\bar y(p+m+k)}e^{iz(l+m)} +e^{iy[p-l+(1-\xi)(k-q)]}e^{-i\bar y[p+m+(1-\xi)(k-q)]}e^{iz(l+m)}\\
&&
\hspace{-1cm}
- e^{iy(p-l+k)}e^{-i\bar y[p+m+(1\xi)(k-q)]}e^{iz[l+m-\xi k-(1-\xi)q]}
-e^{iy[p-l-(1-\xi)(k-q)]}e^{-i\bar y(p+m+q)}e^{iz[l+m+(1-\xi)k+\xi q]}\Big].\nonumber 
\eeq
After performing the integrals, this contribution reads 
\beq
\frac{d\sigma}{d^2p\, d\eta}\bigg|^{\rm g\to q}_{\rm NLO, r}&=&\frac{g^2}{(2\pi)^3}\, S_\perp \int_{\xi_0}d\xi \, \frac{x_p}{(1-\xi)} \, f^g_{\mu_0^2}\left(\frac{x_p}{1-\xi}\right)\frac{1}{2}[\xi^2+(1-\xi)^2]\int_{k,q>\mu_0}s(k)s(q)
\nonumber\\
&&\times\, 
\bigg\{ \frac{1}{(p-q)^2}+\frac{1}{[p-(1-\xi)(q-k)]^2}-2\frac{(p-q)^i}{(p-q)^2}\frac{[p-(1-\xi)(q-k)]^i}{[p-(1-\xi)(q-k)]^2}\bigg\}.
\eeq
At this point we introduce the fragmentation function and reorganize the terms: 
\beq
\label{Real_g_q_1}
&&
\hspace{-1cm}
\frac{d\sigma}{d^2p\, d\eta}\bigg|^{\rm g\to q\to H}_{\rm NLO, r}=\frac{g^2}{(2\pi)^3}\, S_\perp\int_{x_F}^1\frac{d\zeta}{\zeta^2}\, D^q_{\mu_0^2}(\zeta)\int_{k,q>\mu_0}\int_{\xi_0}d\xi \, \frac{x_F}{\zeta(1-\xi)}\, f^g_{\mu_0^2}\left(\frac{x_F}{\zeta(1-\xi)}\right)\frac{1}{2}\Big[\xi^2+(1-\xi)^2\Big]s(k)s(q)
\nonumber\\
&&\times
\bigg\{ \frac{1}{2}\frac{(k-q)^2}{(p/\zeta-k)^2(p/\zeta-q)^2}+\frac{1}{2}\int_ts(t)\frac{(1-\xi)^2[(q-k)-(q-t)]^2}{[p/\zeta-(1-\xi)(q-k)]^2[p/\zeta-(1-\xi)(q-t)]^2}\bigg\}+ ({\rm Gen.\, NLO})_2\ .
\eeq
The first term in eq.~\eqref{Real_g_q_1} is a contribution to the quark TMD PDF, the second term is a contribution to gluon TMD FF and the last term is a genuine NLO correction that do not include any large logarithms. The explicit expression for the genuine NLO contribution is 
\beq
({\rm Gen.\, NLO})_2&=& \frac{g^2}{(2\pi)^3}\, S_{\perp} \int_{x_F}^1\frac{d\zeta}{\zeta^2}\, D^q_{\mu_0^2}(\zeta)\int_{k,q>\mu_0}\int_{\xi_0}d\xi
\, \frac{x_F}{\zeta(1-\xi)}f^g_{\mu_0^2}\left(\frac{x_F}{\zeta(1-\xi)}\right)\frac{1}{2}\Big[\xi^2+(1-\xi)^2\Big] s(k)s(q)\nonumber\\
&&\times\bigg\{ 
\frac{(p/\zeta-q)^i}{(p/\zeta-q)^2}\bigg[\frac{(p/\zeta-k)^i}{(p/\zeta-k)^2} -\frac{[p/\zeta-(1-\xi)(q-k)]^i}{[p/\zeta-(1-\xi)(q-k)]^2}\bigg]
\nonumber\\
&&
\hspace{0.6cm}
+
\int_t s(t)\frac{[p/\zeta-(1-\xi)(q-k)]^i}{[p/\zeta-(1-\xi)(q-k)]^2}\bigg[ \frac{[p/\zeta-(1-\xi)(q-t)]^i}{[p/\zeta-(1-\xi)(q-t)]^2}-\frac{(p/\zeta-q)^i}{(p/\zeta-q)^2}
\bigg]\bigg\}.
\eeq
Using the same arguments introduced in eq.~\eqref{rewriting}, the final expression for this contribution can be written as 
\beq
\label{g-q-f}
\frac{d\sigma}{d^2p\, d\eta}\bigg|^{\rm g\to q\to H}_{\rm NLO, r}&=&\frac{g^2}{(2\pi)^3}\, S_\perp\int_{x_F}^1\frac{d\zeta}{\zeta^2}\, D^q_{\mu_0^2}(\zeta)\int_{k,q>\mu_0}\int_{\xi_0}d\xi \, \frac{x_F}{\zeta(1-\xi)}\, f^g_{\mu_0^2}\left(\frac{x_F}{\zeta(1-\xi)}\right)\frac{1}{2}\Big[\xi^2+(1-\xi)^2\Big]\, \frac{1}{k^2}
\nonumber\\
&&\times\, 
s(-k+p/\zeta)\, \bigg[1-\frac{k\cdot q}{q^2}\bigg]\, s(-q+p/\zeta)
\nonumber\\
&+& \frac{g^2}{(2\pi)^3}\, S_\perp\int_{x_F}^1\frac{d\zeta}{\zeta^2}\, \frac{1}{(1-\xi)}D^q_{\mu_0^2}\left(\frac{\zeta}{1-\xi}\right)\int_{k,q,t>\mu_0}\int_{\xi_0}d\xi \, \frac{x_F}{\zeta}\, f^g_{\mu_0^2}\left(\frac{x_F}{\zeta}\right)\frac{1}{2}\Big[\xi^2+(1-\xi)^2\Big]\, \frac{1}{k^2}
\nonumber\\
&&\times\, 
s(t)s(p/\zeta+k-t)\, \bigg[1-\frac{k\cdot q}{q^2}\bigg]\, s(p/\zeta+q-t)
\nonumber\\
&+&({\rm Gen.\, NLO})_2\ .
\eeq
\subsubsection{Gluon initiated gluon production}

The partonic level expression for this contribution is given by eq. (4.9) of Ref.~\cite{Altinoluk:2014eka} which, at large $N_c$, reads
\beq
&&
\frac{d\sigma}{d^2p\, d\eta}\bigg|^{\rm g\to g}_{\rm NLO, r}=\frac{g^2}{(2\pi)^3}N_c\int_{\xi_0}d\xi \frac{x_p}{(1-\xi)}f^g_{\mu_0^2}\left(\frac{x_p}{1-\xi}\right)\bigg[ \frac{1-\xi}{\xi}+\frac{\xi}{1-\xi}+\xi(1-\xi)\bigg]\int_{y,\bar y, z}A^i_\xi(y-z)A^i_\xi(\bar y-z)
\nonumber\\
&&\times
\Big\{2 s(y,\bar y)s(\bar y, y)+2s[(1-\xi)y, (1-\xi)\bar y]s[(1-\xi)\bar y, (1-\xi)y]-s(y,z)s[(1-\xi)\bar y+\xi z, y]s[z, (1-\xi)\bar y+\xi z]
\nonumber\\
&&
\hspace{0.4cm}
-s(z,y)s[y, (1-\xi)\bar y+\xi z]s[(1-\xi)\bar y+\xi z, z] - s[\bar y, (1-\xi)y +\xi z]s[(1-\xi)y+\xi z, z]s(z,\bar y)
\nonumber\\
&&
\hspace{0.4cm}
-s[(1-\xi)y+\xi z, \bar y]s[z, (1-\xi)y+\xi z]s(\bar y, z)\Big\}.
\eeq
Again using the translationally invariant dipoles and the definition of the modified WW fields, we get 
\beq
&&
\frac{d\sigma}{d^2p\, d\eta}\bigg|^{\rm g\to g}_{\rm NLO, r}=\frac{g^2}{(2\pi)^3}N_c\int_{\xi_0}d\xi \frac{x_p}{(1-\xi)}f^g_{\mu_0^2}\left(\frac{x_p}{1-\xi}\right)\bigg[ \frac{1-\xi}{\xi}+\frac{\xi}{1-\xi}+\xi(1-\xi)\bigg]\frac{(-1)}{(2\pi)^4}
\int_{y,\bar y, z}\int_{l,m>\mu_0} \int_{k,q,t} \frac{l^i}{l^2}\frac{m^i}{m^2}
\nonumber\\
&&\times
 s(k)s(q)s(t)\Big\{ 
2e^{iy(p-l+k-q)}e^{-i\bar y(p+m+k-q)}e^{iz(l+m)} + 2 e^{iy[p-l+(1-\xi)(k-q)]}e^{-i\bar y[p+m+(1-\xi)(k-q)]}e^{iz(l+m)}\\
&&
-e^{iy(p-l+t-k)}e^{-i\bar y[p+m+(1-\xi)(q-k)]}e^{iz[l+m-t+\xi k+(1-\xi)q]}
-e^{iy(p-l-t+k)}e^{-i\bar y[p+m-(1-\xi)(q-k)]}e^{iz[l+m+t -\xi k-(1-\xi)q]}
\nonumber\\
&&
-e^{iy[p-l+(1-\xi)(q-k)]}e^{-i\bar y(p+m-k+t)}e^{iz[l+m+t-\xi k-(1-\xi)q]}
-e^{iy[p-l+(1-\xi)(k-q)]}e^{-i\bar y(p+m+k-t)}e^{iz[l+m-t+\xi k +(1-\xi)q]}\Big\}.\nonumber
\eeq
After performing the integrals, this contribution can be written as 
\beq
\frac{d\sigma}{d^2p\, d\eta}\bigg|^{\rm g\to g}_{\rm NLO, r}&=&\frac{g^2}{(2\pi)^3}\, S_\perp\int_{\xi_0}d\xi \frac{x_p}{(1-\xi)}f^g_{\mu_0^2}\left(\frac{x_p}{1-\xi}\right)\, 2N_c \bigg[ \frac{1-\xi}{\xi}+\frac{\xi}{1-\xi}+\xi(1-\xi)\bigg]\int_{k,q,t}s(k) \, s(q)\, s(t)
\nonumber\\
&&\times
\bigg\{ \frac{1}{[p-(q-k)]^2}+\frac{1}{[p-(1-\xi)(q-k)]^2}-2\frac{(p-q+t)^i}{(p-q+t)^2}\frac{[p-(1-\xi)(q-k)]^i}{[p-(1-\xi)(q-k)]^2}\bigg\}.
\eeq
Introducing the fragmentation function and using the symmetry between $k,q$ and $t$, it can be organized as 
\beq
&&
\label{Real_g_g_1}
\frac{d\sigma}{d^2p\, d\eta}\bigg|^{\rm g\to g\to H}_{\rm NLO, r}=\frac{g^2}{(2\pi)^3}\, S_\perp
\int_{x_F}^1\frac{d\zeta}{\zeta^2}\, D^g_{\mu_0^2}(\zeta)\int_{k>\mu_0}
\int_{\xi_0}d\xi \frac{x_F}{\zeta(1-\xi)}f^g_{\mu_0^2}\left(\frac{x_F}{\zeta(1-\xi)}\right)\, 2N_c \bigg[ \frac{1-\xi}{\xi}+\frac{\xi}{1-\xi}+\xi(1-\xi)\bigg]
\nonumber\\
&&\times
\int_{q,t}s(k) \, s(q)\, s(t)\bigg\{ \frac{1}{2}\frac{[(q-k)-(q-t)]^2}{[p/\zeta-(q-t)]^2[p/\zeta-(q-k)]^2}+\frac{1}{2}\frac{(1-\xi)^2[(q-k)-(q-t)]^2}{[p/\zeta-(1-\xi)(q-t)]^2[p/\zeta-(1-\xi)(q-k)]^2}\bigg\} 
\nonumber\\
&&+  ({\rm Gen.\, NLO})_3\ ,
\eeq
where the genuine NLO contribution reads 
\beq
&&
({\rm Gen.\, NLO})_3=\frac{g^2}{(2\pi)^3}\, S_\perp
\int_{x_F}^1\frac{d\zeta}{\zeta^2}\, D^g_{\mu_0^2}(\zeta)\int_{k>\mu_0}
\int_{\xi_0}d\xi \frac{x_F}{\zeta(1-\xi)}f^g_{\mu_0^2}\left(\frac{x_F}{\zeta(1-\xi)}\right)\, 2N_c \bigg[ \frac{1-\xi}{\xi}+\frac{\xi}{1-\xi}+\xi(1-\xi)\bigg]
\nonumber\\
&&\times
\int_{q,t}s(k) \, s(q)\, s(t)\bigg[
\frac{[p/\zeta-(q-t)]^i}{[p/\zeta-(q-t)]^2}-\frac{[p/\zeta-(1-\xi)(q-t)]^i}{[p/\zeta-(1-\xi)(q-t)]^2}\bigg]
\bigg[
\frac{[p/\zeta-(q-k)]^i}{[p/\zeta-(q-k)]^2}-\frac{[p/\zeta-(1-\xi)(q-k)]^i}{[p/\zeta-(1-\xi)(q-k)]^2}\bigg].
\eeq
One can again massage the first two terms in eq.~\eqref{Real_g_g_1} using  similar arguments to those introduced in the previous cases and the final result reads 
\beq
\label{g-g-f}
\frac{d\sigma}{d^2p\, d\eta}\bigg|^{\rm g\to g\to H}_{\rm NLO, r}&=&\frac{g^2}{(2\pi)^3}\, S_\perp
\int_{x_F}^1\frac{d\zeta}{\zeta^2}\, D^g_{\mu_0^2}(\zeta)\int_{k>\mu_0}\int_{q,t}
\int_{\xi_0}d\xi \frac{x_F}{\zeta(1-\xi)}f^g_{\mu_0^2}\left(\frac{x_F}{\zeta(1-\xi)}\right)\, 2N_c \bigg[ \frac{1-\xi}{\xi}+\frac{\xi}{1-\xi}+\xi(1-\xi)\bigg]
\nonumber\\
&&
\times\, \frac{1}{k^2}\, s(t)\, s(p/\zeta+k-t)\bigg[1-\frac{k\cdot q}{q^2}\bigg]\, s(p/\zeta+q-t)\nonumber\\
&+&
\frac{g^2}{(2\pi)^3}\, S_\perp
\int_{x_F}^1\frac{d\zeta}{\zeta^2}\, \frac{1}{1-\xi}\, D^g_{\mu_0^2}(\zeta)\int_{k>\mu_0}\int_{q,t}
\int_{\xi_0}d\xi \frac{x_F}{\zeta}f^g_{\mu_0^2}\left(\frac{x_F}{\zeta}\right)\, 2N_c \bigg[ \frac{1-\xi}{\xi}+\frac{\xi}{1-\xi}+\xi(1-\xi)\bigg]
\nonumber\\
&&
\times\, \frac{1}{k^2}\, s(t)\, s(p/\zeta+k-t)\bigg[1-\frac{k\cdot q}{q^2}\bigg]\, s(p/\zeta+q-t)\nonumber\\
&+&({\rm Gen.\, NLO})_3\ .
\eeq
\subsubsection{Quark initiated gluon production}
At large $N_c$, this contribution reads (eq. (4.8) in Ref.~\cite{Altinoluk:2014eka}):
\beq
\frac{d\sigma}{d^2p\, d\eta}\bigg|^{\rm q\to g}_{\rm NLO, r}&=&\frac{g^2}{(2\pi)^3}\int_{\xi_0}d\xi \, \frac{x_p}{\xi}\, f^q_{\mu_0^2}\left(\frac{x_p}{\xi}\right)\frac{N_c}{2}\bigg[\frac{1+(1-\xi)^2}{\xi}\bigg]\int_{z\bar z y}e^{ip(z-\bar z)}A^i_{\xi}(z-y)A^i_{\xi}(\bar z-y)
\nonumber\\
&&\times
\Big\{ s(z,\bar z)s(\bar z, z)+s(\xi z, \xi\bar z)-s[z, (1-\xi)y+\xi \bar z]s(y,z)-s(\bar z,y)s[(1-\xi)y+\xi z, \bar z]\Big\}.
\eeq.
Using the same arguments, we get 
\beq
\frac{d\sigma}{d^2p\, d\eta}\bigg|^{\rm q\to g}_{\rm NLO, r}&=&\frac{g^2}{(2\pi)^3}\int_{\xi_0}d\xi \, \frac{x_p}{\xi}\, f^q_{\mu_0^2}\left(\frac{x_p}{\xi}\right)\frac{N_c}{2}\bigg[\frac{1+(1-\xi)^2}{\xi}\bigg]
\frac{(-1)}{(2\pi)^4}
\int_{z\bar z y}\int_{l,m>\mu_0}\int_{k,q}\frac{l^i}{l^2}\frac{m^i}{m^2}\, s(k)\, s(q)
\nonumber\\
&&
\hspace{-0.7cm}
\times
\Big\{ 
e^{iz(p-l+k-q)}e^{-i\bar z(p+m+k-q)}e^{iy(l+m)}+ e^{iz(p-l+\xi q)}e^{-i\bar z(p+m+\xi q)}e^{iy(l+m)}
\\
&&
-e^{iz(p-l+q-k)}e^{-i\bar z (p+m+\xi q)} e^{iy[l+m+k-(1-\xi)q]}
-e^{iz(p-l+\xi k)}e^{-i\bar z (p+m+k-q)}e^{iy[l+m-q+(1-\xi)k]}\Big\}.\nonumber
\eeq
After performing the integrals, the contribution to the partonic level cross section from this channel reads
\beq
\frac{d\sigma}{d^2p\, d\eta}\bigg|^{\rm q\to g}_{\rm NLO, r}&=&\frac{g^2}{(2\pi)^3}\, S_\perp \int_{\xi_0}d\xi \, \frac{x_p}{\xi}\, f^q_{\mu_0^2}\left(\frac{x_p}{\xi}\right)\frac{N_c}{2}\bigg[\frac{1+(1-\xi)^2}{\xi}\bigg]\int_{k,q>\mu_0} s(k)\, s(q)\nonumber\\
&&\times\bigg\{ 
\frac{1}{[p-(q-k)]^2}+\frac{1}{(p-\xi q)^2}-2\frac{[p-(q-k)]^i}{[p-(q-k)]^2}\frac{(p-\xi q)^i}{(p-\xi q)^2}\bigg\}.
\eeq
After introducing the fragmentation function and symmetrizing our result, we get 
\beq
\label{Real_q_g_1}
\frac{d\sigma}{d^2p\, d\eta}\bigg|^{\rm q\to g\to H}_{\rm NLO, r}&=&\frac{g^2}{(2\pi)^3}\, S_\perp \int_{x_F}^1 \frac{d\zeta}{\zeta^2} \, D^g_{\mu_0^2}(\zeta)\int_{k,q>\mu_0} \int_{\xi_0}d\xi \, \frac{x_F}{\zeta\xi}\, f^q_{\mu_0^2}\left(\frac{x_F}{\zeta\xi}\right)\frac{N_c}{2}\bigg[\frac{1+(1-\xi)^2}{\xi}\bigg]
s(k)s(q)
\\
&&\times\bigg\{ 
\frac{1}{2} \int_t s(t)\frac{[(q-k)-(q-t)]^2}{[p/\zeta-(q-k)]^2[p/\zeta-(q-t)]^2} +\frac{1}{2}\frac{\xi^2(q-k)^2}{(p/\zeta-\xi q)^2(p/\zeta-\xi k)^2}\Bigg\}+ ({\rm Gen.\, NLO})_4\ ,\nonumber
\eeq
with
\beq
&&
({\rm Gen.\, NLO})_4= \frac{g^2}{(2\pi)^3}\, S_\perp \int_{x_F}^1 \frac{d\zeta}{\zeta^2} \, D^g_{\mu_0^2}(\zeta)\int_{k,q>\mu_0} \int_{\xi_0}d\xi \, \frac{x_F}{\zeta\xi}\, f^q_{\mu_0^2}\left(\frac{x_F}{\zeta\xi}\right)\frac{N_c}{2}\bigg[\frac{1+(1-\xi)^2}{\xi}\bigg]
s(k)s(q)\\
&&
\hspace{-0.5cm}
\times
\bigg\{ \int_t s(t)\frac{[p/\zeta-(q-k)]^i}{[p/\zeta-(q-k)]^2}\bigg[\frac{[p/\zeta-(q-t)]^i}{[p/\zeta-(q-t)]^2}-\frac{(p/\zeta-\xi q)^i}{(p/\zeta-\xi q)^2}\bigg]+
 \frac{(p/\zeta-\xi q)^i}{(p/\zeta-\xi q)^2}\bigg[ \frac{(p/\zeta-\xi k)^i}{(p/\zeta-\xi k)^2}-\frac{[p/\zeta-(q-k)]^i}{[p/\zeta-(q-k)]^2}\bigg]\bigg\}.\nonumber
\eeq
Performing the same manipulations for the first two terms in eq.~\eqref{Real_q_g_1}, we get the final expression as 
\beq
\label{q-g-f}
\frac{d\sigma}{d^2p\, d\eta}\bigg|^{\rm q\to g\to H}_{\rm NLO, r}&=&\frac{g^2}{(2\pi)^3}\, S_\perp \int_{x_F}^1 \frac{d\zeta}{\zeta^2} \, D^g_{\mu_0^2}(\zeta)\int_{k>\mu_0} \int_{q,t} \int_{\xi_0}d\xi \, \frac{x_F}{\zeta\xi}\, f^q_{\mu_0^2}\left(\frac{x_F}{\zeta\xi}\right)\frac{N_c}{2}\bigg[\frac{1+(1-\xi)^2}{\xi}\bigg]\frac{1}{k^2}\nonumber\\
&&\times\, s(t)\, s(p/\zeta+k-t)\bigg[1-\frac{k\cdot q}{q^2}\bigg]\, s(p/\zeta+q-t)
\nonumber\\
&+&\frac{g^2}{(2\pi)^3}\, S_\perp \int_{x_F}^1 \frac{d\zeta}{\zeta^2} \, \frac{1}{\xi}D^g_{\mu_0^2}\left(\frac{\zeta}{\xi}\right)\int_{k>\mu_0} \int_{q} \int_{\xi_0}d\xi \, \frac{x_F}{\zeta}\, f^q_{\mu_0^2}\left(\frac{x_F}{\zeta}\right)\frac{N_c}{2}\bigg[\frac{1+(1-\xi)^2}{\xi}\bigg]\frac{1}{k^2}\nonumber\\
&&\times\, s(-k+p/\zeta)\bigg[1-\frac{k\cdot q}{q^2}\bigg]\, s(-q+p/\zeta)\nonumber\\
&+&({\rm Gen.\, NLO})_4\ .
\eeq
\subsubsection{Summing all real contributions}

The definitions of full quark and gluon TMD PDFs and FFs are given in eqs.~\eqref{eq:tmdd1pfull},~\eqref{eq:tmdg1pfull},~\eqref{tmdffkp} and~\eqref{eq:tmdgff1pfull}. Here we ignore the contribution from antiquarks that, as indicated in the main text, can be added straightforwardly.
The contributions to the quark channel arise from the first two terms in eq.~\eqref{q-q-f}, the first term in eq.~\eqref{q-q-f} and the second term in eq.~\eqref{q-g-f}. Adding these terms and using the definitions for the quark TMD PDF and TMD FF, we get 
\beq
\frac{d\sigma}{d^2pd\eta}\bigg|_{\rm NLO, r}^q&=& S_\perp\int_{x_F}^1\frac{d\zeta}{\zeta^2}\int_{k>\mu_0}
\bigg[ D^q_{\mu_0^2}(\zeta)\frac{x_F}{\zeta}{\cal T}_q\left(\frac{x_f}{\zeta}, k^2;k^2,\xi_0\right)+\frac{x_F}{\zeta}f^q_{\mu_0^2}\left(\frac{x_F}{\zeta}\right){\cal F}_q(\zeta, k^2;k^2;\xi_0)\bigg]\nonumber\\
&&\times\,\int_q s(-k+p/\zeta)\, \bigg[1-\frac{k\cdot q}{q^2}\bigg]\, s(-q+p/\zeta).
\eeq
The contributions to the gluon channel arise from the second term in eq.~\eqref{g-q-f}, the first two terms in eq.~\eqref{g-g-f}
and the first term in eq.~\eqref{q-g-f}. Adding those terms and using the definitions for the gluon TMD PDF and TMD FF, we get
\beq
\label{real_NLO_gluon}
\frac{d\sigma}{d^2pd\eta}\bigg|_{\rm NLO, r}^g&=& S_\perp\int_{x_F}^1\frac{d\zeta}{\zeta^2}\int_{k>\mu_0}
\bigg[ D^g_{\mu_0^2}(\zeta)\frac{x_F}{\zeta}{\cal T}_g\left(\frac{x_f}{\zeta}, k^2;k^2,\xi_0\right)+\frac{x_F}{\zeta}f^g_{\mu_0^2}\left(\frac{x_F}{\zeta}\right){\cal F}_g(\zeta, k^2;k^2;\xi_0)\bigg]\nonumber\\
&&\times\,\int_{q,t} s(t) \, s(p/\zeta+k-t)\, \bigg[1-\frac{k\cdot q}{q^2}\bigg]\, s(p/\zeta+q-t).
\eeq
Finally, the sum of the real corrections reads
\beq
\frac{d\sigma}{d^2pd\eta}\bigg|_{\rm NLO}^{\rm real}= \frac{d\sigma}{d^2pd\eta}\bigg|_{\rm NLO, r}^q+\frac{d\sigma}{d^2pd\eta}\bigg|_{\rm NLO, r}^g+\sum_{i=1}^4({\rm Gen. \, NLO})_i\ .
\eeq
\subsection{Virtual corrections at NLO}
\beq
\label{NLO_virtual_sch}
\frac{d\sigma}{d^2p\, d\eta}\bigg|_{\rm NLO}^{\rm virtual}= \frac{d\sigma}{d^2p\, d\eta}\bigg|_{\rm NLO, v}^{\rm q\to q\to H}+\frac{d\sigma}{d^2p\, d\eta}\bigg|_{\rm NLO, v}^{\rm g\to g\to H}+\frac{d\sigma}{d^2p\, d\eta}\bigg|_{\rm NLO, v}^{\rm g\to q \to H}+\frac{d\sigma}{d^2p\, d\eta}\bigg|_{\rm NLO, v}^{\rm g\to \bar q \to H}\ .
\eeq

\subsubsection{Virtual corrections to the quark production}

This is the first term in eq.~\eqref{NLO_virtual_sch} which was discussed earlier in the paper.

\subsubsection{Virtual corrections to the gluon production}

The virtual corrections in the gluon channel  given by the last three terms in eq.~\eqref{NLO_virtual_sch}. The last term is contribution from the antiquark, we will not calculate that contribution explicitly. 
The two terms we need to focus on are the second and the third terms in eq.~\eqref{NLO_virtual_sch}. Let us first consider the second term. This contribution can be divided into two pieces at partonic level. In the large $N_c$ limit, we can write it as (eq. (4.10) in Ref.~\cite{Altinoluk:2014eka})
\beq
\frac{d\sigma}{d^2p\, d\eta}\bigg|^{\rm g\to g}_{\rm NLO, v}=\frac{d\sigma}{d^2p\, d\eta}\bigg|^{\rm g\to g}_{\rm NLO, v - 1}+\frac{d\sigma}{d^2p\, d\eta}\bigg|^{\rm g\to g}_{\rm NLO, v - 2}\ ,
\eeq
where 
\beq
\frac{d\sigma}{d^2p\, d\eta}\bigg|^{\rm g\to g}_{\rm NLO, v - 1}&=&-\frac{g^2}{(2\pi)^3}N_cx_pf^g_{\mu_0^2}(x_p)\int_0^1d\xi \bigg[\frac{1-\xi}{\xi}+\frac{\xi}{1-\xi}+\xi(1-\xi)\bigg]\int_{y,\bar y, z}e^{ip(y-\bar y)}A^i_{\xi}(y-z)A^i_{\xi}(y-z)
\nonumber\\
& \times &
\frac{1}{2}\bigg\{ 2s_A(y,\bar y)-s[y+\xi(y-z),z+\xi(y-z)]s[z+\xi(y-z),\bar y]s[\bar y, y+\xi(y-z)]
\nonumber\\
&&\hspace{0.5cm}
-s[z+\xi(y-z),y+\xi(y-z)]s[\bar y, z+\xi(y-z)]s[y+\xi(y-z),\bar y]\Big\}
\eeq
and 
\beq
\frac{d\sigma}{d^2p\, d\eta}\bigg|^{\rm g\to g}_{\rm NLO, v - 2}&=&-\frac{g^2}{(2\pi)^3}N_cx_pf^g_{\mu_0^2}(x_p)\int_0^1d\xi \bigg[\frac{1-\xi}{\xi}+\frac{\xi}{1-\xi}+\xi(1-\xi)\bigg]\int_{y,\bar y, z}e^{ip(y-\bar y)}A^i_{\xi, x_p}(\bar y-z)A^i_{\xi,x_p}(\bar y-z)
\nonumber\\
& \times &
\frac{1}{2}\bigg\{ 2s_A(y,\bar y)-s[\bar y+\xi(\bar y -z), z+\xi(\bar y - z)]s[z+\xi(\bar y -z), y] s[y, \bar y + \xi(\bar y-z)]
\nonumber\\
&&\hspace{0.5cm}
-s[z+\xi(\bar y-z),\bar y+\xi(\bar y-z)]s[y, z+\xi(\bar y-z)]s[\bar y+\xi(\bar y-z), y]\Big\}.
\eeq
Following the same procedure indicated previously, these two contributions can be written as 
\beq
&&\frac{d\sigma}{d^2p\, d\eta}\bigg|^{\rm g\to g}_{\rm NLO, v - 1}=-\frac{g^2}{(2\pi)^3}\, N_c\, x_p\, f^g_{\mu_0^2}(x_p)\int_{\xi_0}^{1-\xi_0}d\xi \bigg[\frac{1-\xi}{\xi}+\frac{\xi}{1-\xi}+\xi(1-\xi)\bigg]\frac{(-1)}{(2\pi)^4}\int_{y,\bar y, z}\int_{l,m>\mu_0}\int_{k,q,t}\frac{l^i}{l^2}\frac{m^i}{m^2}\nonumber\\
&&\hskip 0.6cm \times s(k)s(q)s(t)\frac{1}{2}\Big\{ 
2e^{iy(p-l-m+k-q)}e^{-i\bar y (p+k-q)}e^{iz(l+m)}-e^{iy[p-l-m+k+\xi q-(1+\xi)t]}e^{-i\bar y (p+q-t)}e^{iz[l+m-k+(1-\xi)q+\xi t]}
\nonumber\\
&&
\hspace{3.2cm}
-\, 
e^{iy[p-l-m-k-\xi q+(1+\xi)t]}e^{-i\bar y(p-q+t)}e^{iz[l+m+k-(1-\xi)q-\xi t]}\Big\}
\eeq
and 
\beq
&&\frac{d\sigma}{d^2p\, d\eta}\bigg|^{\rm g\to g}_{\rm NLO, v - 2}=-\frac{g^2}{(2\pi)^3}\, N_c\, x_p\, f^g_{\mu_0^2}(x_p)\int_{\xi_0}^{1-\xi_0}d\xi \bigg[\frac{1-\xi}{\xi}+\frac{\xi}{1-\xi}+\xi(1-\xi)\bigg]\frac{(-1)}{(2\pi)^4}\int_{y,\bar y, z}\int_{l,m>\mu_0}\int_{k,q,t}\frac{l^i}{l^2}\frac{m^i}{m^2}\nonumber\\
&&\hskip 0.6cm\times
s(k)s(q)s(t)\frac{1}{2} \Big\{ 2e^{iy(p+k-q)}e^{-i\bar y (p+l+m+k-q)}e^{iz(l+m)}-e^{iy(p-q+t)}e^{-i\bar y [p+l+m-k-\xi q+(1+\xi)t]}e^{iz[l+m-k+(1-\xi)q+\xi t]}
\nonumber\\
&&
\hspace{3.2cm}
- \, e^{iy(p+q-t)}e^{-i\bar y [p+l+m+k+\xi q-(1+\xi)t]}e^{iz[l+m+k-(1-\xi)q-\xi t]}\Big\}.
\eeq
After performing the integrals, these two contributions can be combined with some change of variables. The final result at partonic level reads 
\beq
\frac{d\sigma}{d^2p\, d\eta}\bigg|^{\rm g\to g}_{\rm NLO, v}&=&-\frac{g^2}{(2\pi)^3}\, S_\perp x_pf^g_{\mu_0^2}(x_p)\int_{\xi_0}^{1-\xi_0}d\xi \, 2N_c\, \bigg[ \frac{1-\xi}{\xi}+\frac{\xi}{1-\xi}+\xi(1-\xi)\bigg]
\nonumber\\
&&\times\, 
\int_{l>\mu_0}\int_{q,t}s(q)s(t-p)s(t)
\bigg\{ \frac{1}{l^2}+\frac{l^i}{l^2}\frac{[(1-\xi)p-l-q-t]^i}{[(1-\xi)p-l-q-t]^2}\bigg\}.
\eeq
Upon introducing the fragmentation function, this contribution becomes
\beq
\label{virtual_g_g_f_1}
\frac{d\sigma}{d^2p\, d\eta}\bigg|^{\rm g\to g\to H}_{\rm NLO, v}&=&-\frac{g^2}{(2\pi)^3}\, S_\perp 
\int_{x_F}^1\frac{d\zeta}{\zeta^2}D^g_{\mu_0^2}(\zeta)\int_{\xi_0}^{1-\xi_0}d\xi \frac{x_F}{\zeta}f^g_{\mu_0^2}\left(\frac{x_F}{\zeta}\right) \, 2N_c\, \bigg[ \frac{1-\xi}{\xi}+\frac{\xi}{1-\xi}+\xi(1-\xi)\bigg]
\nonumber\\
&&\times\,
\int_{l>\mu_0}\int_{q,t}s(q)s(t-p/\zeta)s(t)
 \bigg\{ \frac{1}{l^2}+\frac{l^i}{l^2}\frac{[(1-\xi)p/\zeta-l-q-t]^i}{[(1-\xi)p/\zeta-l-q-t]^2}\bigg\}\\
 &=&-\frac{g^2}{(2\pi)^3}\, S_\perp 
\int_{x_F}^1\frac{d\zeta}{\zeta^2}D^g_{\mu_0^2}(\zeta)\int_{\xi_0}^{1-\xi_0}d\xi \frac{x_F}{\zeta}f^g_{\mu_0^2}\left(\frac{x_F}{\zeta}\right) \, 2N_c\, \int_{l>\mu_0}\int_{q,t}s(q)s(t-p/\zeta)s(t)\nonumber\\
&&\hspace{0.2cm}\times\,\Bigg\{
\Bigg[ \frac{1-\xi}{\xi}+\frac{\xi}{1-\xi}+\xi(1-\xi)\bigg]
 \bigg\{ \left[\frac{1}{l^2}+\frac{l^i}{l^2}\frac{[p/\zeta-l-q-t]^i}{[p/\zeta-l-q-t]^2}\right]\nonumber \\
&& \hskip 0.8cm+
 \left[\frac{l^i}{l^2}\frac{[(1-\xi)p/\zeta-l-q-t]^i}{[(1-\xi)p/\zeta-l-q-t]^2}-\frac{l^i}{l^2}\frac{[p/\zeta-l-q-t]^i}{[p/\zeta-l-q-t]^2}\right]
 \bigg\}\Bigg\}.\nonumber
\eeq
 Let us now consider the third term in eq.~\eqref{NLO_virtual_sch} (the last term in eq.~\eqref{NLO_virtual_sch} corresponds to the antiquark and we will not consider it explicitly). This term can also be grouped into two contributions 
\beq
\frac{d\sigma}{d^2p\, d\eta}\bigg|^{\rm g\to q}_{\rm NLO, v}=\frac{d\sigma}{d^2p\, d\eta}\bigg|^{\rm g\to q}_{\rm NLO, v - 1}+\frac{d\sigma}{d^2p\, d\eta}\bigg|^{\rm g\to q}_{\rm NLO, v - 2}\ ,
\eeq
and each of them are given by (eq. (4.11) in Ref.~\cite{Altinoluk:2014eka})
\beq
\frac{d\sigma}{d^2p\, d\eta}\bigg|^{\rm g\to q}_{\rm NLO, v - 1}&=&-\frac{g^2}{(2\pi)^3}x_pf^g_{\mu_0^2}(x_p)\int_{\xi_0}d\xi\frac{1}{2}\Big[\xi^2+(1-\xi)^2\Big]
\int_{y,\bar y, z} e^{ip(y-\bar y)}A^i_{\xi}(y-z)A^i_{\xi}(y-z)
\nonumber\\
&&\times
\Big\{s_A(y,\bar y) - s[y+\xi(y-z), \bar y]s[\bar y, z+\xi(y-z)]\Big\}
\eeq
and
\beq
\frac{d\sigma}{d^2p\, d\eta}\bigg|^{\rm g\to q}_{\rm NLO, v - 2}&=&-\frac{g^2}{(2\pi)^3}x_pf^g_{\mu_0^2}(x_p)\int_{\xi_0}d\xi\frac{1}{2}\Big[\xi^2+(1-\xi)^2\Big]
\int_{y,\bar y, z} e^{ip(y-\bar y)}A^i_{\xi}(\bar y -z)A^i_{\xi}(\bar y-z)
\nonumber\\
&&\times
\Big\{s_A(y,\bar y) - s[\bar y+\xi(\bar y-z), y]s[y, z+\xi(\bar y-z)]\Big\}.
\eeq
Within our approximations, these two contributions can be written  
\beq
&&
\frac{d\sigma}{d^2p\, d\eta}\bigg|^{\rm g\to q}_{\rm NLO, v - 1}=-\frac{g^2}{(2\pi)^3}x_pf^g_{\mu_0^2}(x_p)\int_{\xi_0}d\xi\, \frac{1}{2}\, \Big[\xi^2+(1-\xi)^2\Big]
\frac{(-1)}{(2\pi)^4}
\int_{y,\bar y, z} 
\int_{l,m>\mu_0}\int_{k,q}\frac{l^i}{l^2}\frac{m^i}{m^2}\, s(k)s(q)
%e^{ip(y-\bar y)}A^i_{\xi}(y-z)A^i_{\xi}(y-z)
\nonumber\\
&&\times
\Big\{
e^{iy(p-l-m+k+q)}e^{-i\bar y (p+k-q)}e^{iz(l+m)}
-
e^{iy[p-l-m+(1+\xi)k-\xi q]}e^{-i\bar y (p+k-q)} e^{iz[l+m-\xi k-(1-\xi) q]}
\Big\}
\eeq
and 
\beq
&&
\frac{d\sigma}{d^2p\, d\eta}\bigg|^{\rm g\to q}_{\rm NLO, v - 1}=-\frac{g^2}{(2\pi)^3}x_pf^g_{\mu_0^2}(x_p)\int_{\xi_0}d\xi\, \frac{1}{2}\, \Big[\xi^2+(1-\xi)^2\Big]
\frac{(-1)}{(2\pi)^4}
\int_{y,\bar y, z} 
\int_{l,m>\mu_0}\int_{k,q}\frac{l^i}{l^2}\frac{m^i}{m^2}\, s(k)s(q)
%e^{ip(y-\bar y)}A^i_{\xi}(y-z)A^i_{\xi}(y-z)
\nonumber\\
&&\times
\Big\{
e^{iy(p+k-q)}e^{-i\bar y (p+l+m+k-q)}e^{iz(l+m)}
-
e^{iy(p-k+q)}e^{-i\bar y [p+l+m-(1+\xi)k+\xi q]} e^{iz[l+m-\xi k-(1-\xi) q]}
\Big\}.
\eeq
These two contributions can be combined after performing the integrals and the final result at partonic level reads
\beq
\frac{d\sigma}{d^2p\, d\eta}\bigg|^{\rm g\to q}_{\rm NLO, v}&=&-\frac{g^2}{(2\pi)^3}x_pf^g_{\mu_0^2}(x_p)\int_{\xi_0}d\xi\, \frac{1}{2}\, \Big[\xi^2+(1-\xi)^2\Big]
\int_{l>\mu_0}\int_q s(p-q)s(q)
\nonumber\\
&&\times\, 
\bigg\{ \frac{2}{l^2}+\frac{l^i}{l^2}\frac{[\xi p-l-q]^i}{[\xi p-l-q]^2}+\frac{l^i}{l^2}\frac{[(1-\xi)p-l-q]^i}{[(1-\xi)p-l-q]^2}\bigg\}.
\eeq
Introducing the fragmentation function and reorganizing the terms, we get 
\beq
\label{virtual_g_q_f_1}
\frac{d\sigma}{d^2p\, d\eta}\bigg|^{\rm g\to q\to H}_{\rm NLO, v}&=&-\frac{g^2}{(2\pi)^3}\, S_\perp \int_{x_F}^1\frac{d\zeta}{\zeta^2} D^g_{\mu_0^2}(\zeta)\int_{\xi_0}^1 \frac{x_F}{\zeta}f^g_{\mu_0^2}\left(\frac{x_F}{\zeta}\right)\, \frac{1}{2}\, \Big[\xi^2+(1-\xi)^2\Big]
\int_{l>\mu_0}\int_{q,t} s(t)s(t-p/\zeta)s(q)
\nonumber\\
&&\times\, 
\bigg\{ \frac{2}{l^2}+\frac{l^i}{l^2}\frac{[\xi p/\zeta-l-t]^i}{[\xi p/\zeta-l-t]^2}+\frac{l^i}{l^2}\frac{[(1-\xi)p/\zeta-l-t]^i}{[(1-\xi)p/\zeta-l-t]^2}\bigg\}.
\eeq
Note that eqs.~\eqref{virtual_g_g_f_1}  and~\eqref{virtual_g_q_f_1} are the final expressions for the virtual contributions in the gluon channel. In the rest of the discussion we will follow the same arguments adopted in the analysis of the virtual contributions in the quark channel. Namely, we will separate each of these two contributions into a piece containing a large logarithm which will be included in the evolution of the TMDs and a piece without any large logarithms which therefore will be referred to as new genuine NLO corrections.  

Let us start with eq.~\eqref{virtual_g_g_f_1}. The last term in this equation is finite - it does not contain either transverse or longitudinal logs, and we set it aside. Just like in eq.~\eqref{eq:myvirrtual2}, we can perform the angular integration over the angle of vector $l$ in the first tern in eq.~\eqref{virtual_g_g_f_1}, 

\beq
\label{virtual_g_g_f_2}
\frac{d\sigma}{d^2p\, d\eta}\bigg|^{\rm g\to g\to H(1)}_{\rm NLO, v}&=&- 2 \frac{g^2}{(2\pi)^3}\, S_\perp 
\int_{x_F}^1\frac{d\zeta}{\zeta^2}D^g_{\mu_0^2}(\zeta) \int_{q,t}\int_{\mu_0^2}^{[q+t-p/\zeta]^2}\frac{d^2l}{l^2}
\int_{\xi_0}^1d\xi \, \frac{x_F}{\zeta} \, f^g_{\mu_0^2}\left(\frac{x_F}{\zeta}\right) \, 
\nonumber\\
&&\times\,
N_c\, \bigg[ \frac{1-\xi}{\xi}+\frac{\xi}{1-\xi}+\xi(1-\xi)\bigg]
s(q)s(t-p/\zeta)s(t).
\eeq
Again, following the same arguments in eq.~\eqref{eq:myvirrtual2} we divide the integration region over $l$ into two pieces as
\beq
\label{virtual_g_g_f_3}
\frac{d\sigma}{d^2p\, d\eta}\bigg|^{\rm g\to g\to H(1)}_{\rm NLO, v}&=&- 2 \frac{g^2}{(2\pi)^3}\, S_\perp 
\int_{x_F}^1\frac{d\zeta}{\zeta^2}D^g_{\mu_0^2}(\zeta) \int_{q,t}\bigg[\int_{\mu_0^2}^{\mu^2}+ \int_{\mu^2}^{[q+t-p/\zeta]^2}\bigg]\frac{d^2l}{l^2}
\int_{\xi_0}^1d\xi \, \frac{x_F}{\zeta} \, f^g_{\mu_0^2}\left(\frac{x_F}{\zeta}\right) \, 
\nonumber\\
&&\times\,
N_c\, \bigg[ \frac{1-\xi}{\xi}+\frac{\xi}{1-\xi}+\xi(1-\xi)\bigg]
s(q)s(t-p/\zeta)s(t).
\eeq
The first integral over $l$ in eq.~\eqref{virtual_g_g_f_3} will give a large log which will contribute to the evolution of the gluon TMD PDFs and TMD FFs, the second integral does not give a large log and therefore it is a genuine NLO correction. We write it as 
\beq
\label{virtual_g_g_f_4}
\frac{d\sigma}{d^2p\, d\eta}\bigg|^{\rm g\to g\to H}_{\rm NLO, v}&=&- 2 \frac{g^2}{(2\pi)^3}\, S_\perp 
\int_{x_F}^1\frac{d\zeta}{\zeta^2}D^g_{\mu_0^2}(\zeta) \int_{t} \, \int_{\mu_0^2}^{\mu^2}\frac{d^2l}{l^2}
\int_{\xi_0}^1d\xi \, \frac{x_F}{\zeta} \, f^g_{\mu_0^2}\left(\frac{x_F}{\zeta}\right) \, 
\nonumber\\
&&\times\,
N_c\, \bigg[ \frac{1-\xi}{\xi}+\frac{\xi}{1-\xi}+\xi(1-\xi)\bigg]
s(t-p/\zeta)s(t) +({\rm Gen. NLO})_5\ ,
\eeq
with 
\beq
({\rm Gen. NLO})_5&=& - 2 \frac{g^2N_c}{(2\pi)^3}\, S_\perp 
\int_{x_F}^1\frac{d\zeta}{\zeta^2}D^g_{\mu_0^2}(\zeta) \int_{q,t} \,  \int_{\mu^2}^{[q+t-(1-p/\zeta]^2}\bigg]\frac{d^2l}{l^2}
\int_{\xi_0}^1d\xi \, \frac{x_F}{\zeta} \, f^g_{\mu_0^2}\left(\frac{x_F}{\zeta}\right) \, 
\nonumber\\
&&\times\,
\, \bigg[ \frac{1-\xi}{\xi}+\frac{\xi}{1-\xi}+\xi(1-\xi)\bigg]
s(q)s(t-p/\zeta)s(t) \\
 &-&2\frac{g^2N_c}{(2\pi)^3}\, S_\perp 
\int_{x_F}^1\frac{d\zeta}{\zeta^2}D^g_{\mu_0^2}(\zeta)\int_{\xi_0}^1d\xi \frac{x_F}{\zeta}f^g_{\mu_0^2}\left(\frac{x_F}{\zeta}\right) \, \, \bigg[ \frac{1-\xi}{\xi}+\frac{\xi}{1-\xi}+\xi(1-\xi)\bigg]
\nonumber\\
&&\times\,
\int_{l>\mu_0}\int_{q,t}s(q)s(t-p/\zeta)s(t)
 \bigg\{ \left[\frac{l^i}{l^2}\frac{[(1-\xi)p/\zeta-l-q-t]^i}{[(1-\xi)p/\zeta-l-q-t]^2}-\frac{l^i}{l^2}\frac{[p/\zeta-l-q-t]^i}{[p/\zeta-l-q-t]^2}\right]
 \bigg\}.\nonumber
\eeq
%+
Let us now consider eq.~\eqref{virtual_g_q_f_1}. We write it as 
\beq
\label{virtual_g_q_f_2}
\frac{d\sigma}{d^2p\, d\eta}\bigg|^{\rm g\to q\to H}_{\rm NLO, v}&=&-\frac{g^2}{(2\pi)^3}\, S_\perp \int_{x_F}^1\frac{d\zeta}{\zeta^2} D^g_{\mu_0^2}(\zeta)\int_{\xi_0}^1 \frac{x_F}{\zeta}f^g_{\mu_0^2}\left(\frac{x_F}{\zeta}\right)\, \frac{1}{2}\, \Big[\xi^2+(1-\xi)^2\Big]
\int_{l>\mu_0}\int_{q,t} s(t)s(t-p/\zeta)s(q)
\nonumber\\
&&\times\, 
\bigg\{ \bigg[\frac{1}{l^2}+\frac{l^i}{l^2}\frac{[\xi p/\zeta-l-t]^i}{[\xi p/\zeta-l-t]^2}\bigg]+\bigg[\frac{1}{l^2}+\frac{l^i}{l^2}\frac{[(1-\xi)p/\zeta-l-t]^i}{[(1-\xi)p/\zeta-l-t]^2}\bigg]\bigg\}.
\eeq
We can now perform the angular integration over the angle of vector $l$, separately for the first two terms and the last two terms. The result reads 
\beq
\label{virtual_g_q_f_3}
\frac{d\sigma}{d^2p\, d\eta}\bigg|^{\rm g\to q\to H}_{\rm NLO, v}&=&-\frac{g^2}{(2\pi)^3}\, S_\perp \int_{x_F}^1\frac{d\zeta}{\zeta^2} D^g_{\mu_0^2}(\zeta)
\int_t\bigg[\int_{\mu_0^2}^{[t-(1-\xi)p/\zeta]^2}+\int_{\mu_0^2}^{[t-\xi p/\zeta]^2}\bigg] \frac{d^2l}{l^2}
\int_{\xi_0}^1 \frac{x_F}{\zeta}f^g_{\mu_0^2}\left(\frac{x_F}{\zeta}\right)\, \nonumber\\
&&\times\, \frac{1}{2}\, \Big[\xi^2+(1-\xi)^2\Big]\,  s(t)s(t-p/\zeta).
\eeq
Following the same arguments, we divide the integration region over $l$ into two pieces: 
\beq
\label{virtual_g_q_f_4}
\frac{d\sigma}{d^2p\, d\eta}\bigg|^{\rm g\to q\to H}_{\rm NLO, v}&=&-\frac{g^2}{(2\pi)^3}\, S_\perp \int_{x_F}^1\frac{d\zeta}{\zeta^2} D^g_{\mu_0^2}(\zeta)
\int_t\bigg[\int_{\mu_0^2}^{\mu^2}+\int_{\mu^2}^{[t-(1-\xi)p/\zeta]^2}+\int_{\mu_0^2}^{\mu^2}+\int_{\mu^2}^{[t-\xi p/\zeta]^2}\bigg] \frac{d^2l}{l^2}
\int_{\xi_0}^1 \frac{x_F}{\zeta}f^g_{\mu_0^2}\left(\frac{x_F}{\zeta}\right)\, \nonumber\\
&&\times\, \frac{1}{2}\, \Big[\xi^2+(1-\xi)^2\Big]\,  s(t)s(t-p/\zeta).
\eeq
The first and the third terms in eq.~\eqref{virtual_g_q_f_4} contain large logarithms and they contribute to the evolution of the TMDs while the second and fourth terms are genuine NLO corrections. Thus the final expression can be written as 
\beq
\label{virtual_g_q_f_5}
\frac{d\sigma}{d^2p\, d\eta}\bigg|^{\rm g\to q\to H}_{\rm NLO, v}&=&-2\, \frac{g^2}{(2\pi)^3}\, S_\perp \int_{x_F}^1\frac{d\zeta}{\zeta^2} D^g_{\mu_0^2}(\zeta)
\int_t\int_{\mu_0^2}^{\mu^2} \frac{d^2l}{l^2}
\int_{\xi_0}^1 \frac{x_F}{\zeta}f^g_{\mu_0^2}\left(\frac{x_F}{\zeta}\right)\, 
\frac{1}{2}\, \Big[\xi^2+(1-\xi)^2\Big]\,  s(t)s(t-p/\zeta)
\nonumber\\
&+&  ({\rm Gen. NLO})_6\ ,
\eeq
where the genuine NLO correction reads 
\beq
({\rm Gen. NLO})_6&=& -\, \frac{g^2}{(2\pi)^3}\, S_\perp \int_{x_F}^1\frac{d\zeta}{\zeta^2} D^g_{\mu_0^2}(\zeta)
\int_t \bigg[\int_{\mu^2}^{[t-(1-\xi)p/\zeta]^2}+\int_{\mu^2}^{[t-\xi p/\zeta]^2}\bigg] \frac{d^2l}{l^2}
\int_{\xi_0}^1 \frac{x_F}{\zeta}f^g_{\mu_0^2}\left(\frac{x_F}{\zeta}\right)\, 
\nonumber\\
&&\times\, 
\frac{1}{2}\, \Big[\xi^2+(1-\xi)^2\Big]\,  s(t)s(t-p/\zeta).
\eeq

\subsection{Combining all contributions: gluon channel}
In the gluon channel, the leading order expression reads 
\beq
\label{G_LO}
\frac{d\sigma}{d^2p\, d\eta}\bigg|^{\rm g\to H}_{\rm LO}&=& S_\perp\int_{x_F}^1\frac{d\zeta}{\zeta^2}\, D^g_{\mu_0^2}\frac{x_F}{\zeta}f^g_{\mu_0^2}\left(\frac{x_F}{\zeta}\right)\int_t\, s(t)s(t-p/\zeta).
\eeq
Combining the LO expression (eq.~\eqref{G_LO}) with the virtual corrections (eqs.~\eqref{virtual_g_g_f_4} and ~\eqref{virtual_g_q_f_5}) in the gluon channel, we get 
\beq
\label{LO+Virtual_gluon}
&&
\hspace{-2cm}
\frac{d\sigma}{d^2p\, d\eta}\bigg|^{\rm g\to H}_{\rm LO}+\frac{d\sigma}{d^2p\, d\eta}\bigg|^{\rm g\to g\to H}_{\rm NLO, v}+\frac{d\sigma}{d^2p\, d\eta}\bigg|^{\rm g\to q\to H}_{\rm NLO, v}+\frac{d\sigma}{d^2p\, d\eta}\bigg|^{\rm g\to \bar q\to H}_{\rm NLO, v}=S_\perp\int_{x_F}^1
\frac{d\zeta}{\zeta^2}\int_t\int_0^{\mu_0^2}d^2k
\nonumber\\
&&
\hspace{1.8cm}
\times\, \bigg[ D^g_{\mu_0^2}(\zeta)\frac{x_F}{\zeta}{\cal T}_g\left( \frac{x_F}{\zeta},k^2;\mu^2;\xi_0\right)+{\cal F}_g(\zeta,k^2;\mu^2;\xi_0)\frac{x_F}{\zeta}f^g_{\mu_0^2}\left(\frac{x_F}{\zeta}\right)\bigg]\, s(t)s(t-p/\zeta)
\nonumber\\
&&
\hspace{1.8cm}
+ \; ({\rm Gen. NLO})_5+({\rm Gen. NLO})_6\ .
\eeq
% %
Here, we have used the evolution equations for the TMDs over the resolution scale given in eqs.~\eqref{eq:pdftmdgmufull} and~\eqref{eq:pdftmdgffmufull}. Note that we have not calculated the last term on the left hand side of eq.~\eqref{LO+Virtual_gluon}. We can rewrite eq.~\eqref{LO+Virtual_gluon} in a more convenient way (by following the discussions in the quark channel):
\beq
\label{LO+Virtual_gluon_1}
&&
\hspace{-1cm}
\frac{d\sigma}{d^2p\, d\eta}\bigg|^{\rm g\to H}_{\rm LO}+\frac{d\sigma}{d^2p\, d\eta}\bigg|^{\rm g\to g\to H}_{\rm NLO, v}+\frac{d\sigma}{d^2p\, d\eta}\bigg|^{\rm g\to q\to H}_{\rm NLO, v}+\frac{d\sigma}{d^2p\, d\eta}\bigg|^{\rm g\to \bar q\to H}_{\rm NLO, v}\simeq S_\perp\int_{x_F}^1
\frac{d\zeta}{\zeta^2}\int_{t,q}\int_0^{\mu_0^2}d^2k\int_0^{\mu_0^2}d^2l
\nonumber\\
&&
\hspace{0.8cm}
\times\, {\cal F}_g(\zeta,l^2;\mu^2;\xi_0)\frac{x_F}{\zeta} {\cal T}_g\left( \frac{x_F}{\zeta},k^2;\mu^2;\xi_0\right)
s(t)\, s\big(p/\zeta-t+(k+l)\big) \bigg[1-\frac{(k+l)\cdot q}{q^2}\bigg]\, s(p/\zeta-t+q)
\nonumber\\
&&
\hspace{0.8cm}+ \; ({\rm Gen. NLO})_5+({\rm Gen. NLO})_6\ .
\eeq
Finally, combining this result with the real NLO correction given in eq.~\eqref{real_NLO_gluon}, we get 
\beq
\label{LO+Virtual_gluon_1+real}
&&
\hspace{-1cm}
\frac{d\sigma}{d^2p\, d\eta}\bigg|^{\rm g\to H}_{\rm LO}+\frac{d\sigma}{d^2p\, d\eta}\bigg|^{\rm g}_{\rm NLO, r}+\frac{d\sigma}{d^2p\, d\eta}\bigg|^{\rm g\to g\to H}_{\rm NLO, v}+\frac{d\sigma}{d^2p\, d\eta}\bigg|^{\rm g\to q\to H}_{\rm NLO, v}+\frac{d\sigma}{d^2p\, d\eta}\bigg|^{\rm g\to \bar q\to H}_{\rm NLO, v}\simeq S_\perp\int_{x_F}^1
\frac{d\zeta}{\zeta^2}\int_{t,q}\int d^2k\int d^2l
\nonumber\\
&&
\hspace{0.8cm}
\times\, {\cal F}_g(\zeta,l^2;\mu^2;\xi_0)\frac{x_F}{\zeta} {\cal T}_g\left( \frac{x_F}{\zeta},k^2;\mu^2;\xi_0\right)
s(t)\, s\big(p/\zeta-t+(k+l)\big) \bigg[1-\frac{(k+l)\cdot q}{q^2}\bigg]\, s(p/\zeta-t+q)
\nonumber\\
&&
\hspace{0.8cm}+ \; ({\rm Gen. NLO})_5+({\rm Gen. NLO})_6\ .
\eeq

\end{document}